\PassOptionsToPackage{noend, linesnumbered}{algorithm2e}
\documentclass[12pt]{l4dc2023}

\let\vec\relax
\title{Multi-Agent Reinforcement Learning with Reward Delays }
\usepackage{times}

\usepackage{bm}
\usepackage{float}
\usepackage{hyperref}
\usepackage[normalem]{ulem}

\usepackage{accents}
\usepackage{comment}
\SetKwInput{KwInit}{Init}

\newtheorem{mylemma}{Lemma}
\newtheorem{assumption}{Assumption}
\newtheorem{mydefinition}{Definition}

\newcommand{\myM}[4]{{\mathcal{M}}_{#1, #2}^{#4}(#3)} % M_{m,h}^k(s)
\newcommand{\myMm}[4]{{\mathcal{M}}_{#1, #2}^{-,#4}(#3)} % M_{m,h}^k(s)
\newcommand{\myMp}[4]{{\mathcal{M}}_{#1, #2}^{+,#4}(#3)} % M_{m,h}^k(s)
\newcommand{\myaMp}[3]{{\mathcal{M}}_{#1, #2}^{+}(#3)} % M_{m,h}^k(s)
\newcommand{\myaMm}[3]{\mathcal{M}_{#1, #2}^{-}(#3)}         % M_{h}(s), TBD  
\newcommand{\myaM}[3]{\mathcal{M}_{#1, #2}^{}(#3)} 
\newcommand{\myO}[4]{\mathcal{O}_{#1, #2}^{#4}(#3)} % M_{m,h}^k(s)
\newcommand{\mysO}[1]{\mathcal{O}_{#1}}
\newcommand{\myF}[4]{\mathcal{F}_{#1, #2}^{#4}(#3)} % F_{m,h}^k(s)
\newcommand{\myaF}[3]{\mathcal{F}_{#1,#2}(#3)}         % F_{h}(s), TBD
\newcommand{\mysM}[1]{\mathcal{M}_{#1}}             % M_k
\newcommand{\mysMm}[1]{\mathcal{M}^{-}_{#1}}             % M_k
\newcommand{\mysMp}[1]{\mathcal{M}^{+}_{#1}}             % M_k
\newcommand{\mysF}[1]{\mathcal{F}_{#1}}             % F_k

\newcommand{\myd}[4]{d_{#1, #2}^{#4}(#3)}                           % d_{m,h}^k
\newcommand{\mydpr}[4]{{d'}_{#1, #2}^{#4}(#3)}                           % d_{m,h}^k
\newcommand{\mysdpr}[1]{{d'}_{#1}}                           % d_{m,h}^k
\newcommand{\myphi}[5]{\phi_{#1, #2}^{#4,#5}(#3)}                  % \phi_{m,h}^k(s)                 
\newcommand{\myT}[4]{\mathcal{T}_{#1, #2}^{#4}(#3)} 
\newcommand{\myTtil}[4]{\mathcal{T}_{#1, #2}^{#4}(#3)} % T_{m,h}^i(s)
\newcommand{\mysTtil}[1]{\mathcal{T}_{#1}} % T_{m,h}^i(s)
\newcommand{\mybetaOver}[4]{\overline{\beta}_{#1, #2}^{#4}(#3)} % \overline{beta}_{m,h,s}^i
\newcommand{\mybetaUnd}[4]{\underline{\beta}_{#1, #2}^{#4}(#3)} % \underline{beta}_{m,h,s}^i
\newcommand{\mysphi}[2]{\phi_{#1,#2}}                              % \phi_k                 
\newcommand{\mysT}[1]{\mathcal{T}_{#1}}                         % T_i

\newcommand{\mysbetaOver}[1]{\overline{\beta}_{#1}}             % \overline{beta}_i
\newcommand{\mysbetaUnd}[1]{\underline{\beta}_{#1}} 
\newcommand{\myn}[4]{\underline{n}_{#1, #2}^{#4}(#3)}                       % n_{m,h}^k(s)
\newcommand{\mynpr}[4]{\overline{n}_{#1, #2}^{#4}(#3)}                  % n'_{m,h}^k(s)
\newcommand{\myk}[3]{k_{#1}^{#3}(#2)}                           % k_{h}^i(s)
\newcommand{\mysn}[1]{\underline{n}_{#1}}                                   % \mysn{k}
\newcommand{\mysnpr}[1]{\overline{n}_{#1}}                              % \mysnpr{k}
\newcommand{\mysk}[1]{k_{#1}}                                   % k_i

\newcommand{\myeta}[4]{\eta_{#1, #2}^{#4}(#3)}                  % \eta_{m,h,s}^i
\newcommand{\mygamma}[4]{\gamma_{#1, #2}^{#4}(#3)}              % \gamma_{m,h,s}^i
\newcommand{\mye}[4]{e_{#1, #2}^{#4}(#3)}
\newcommand{\myse}[1]{e_{#1}}
\newcommand{\myl}[5]{l_{#1, #2}^{#4}(#3, #5)}                   % l_{m,h,s}^i(a)
\newcommand{\mylpr}[5]{{l'}_{#1, #2}^{#4}(#3, #5)}                   % l_{m,h,s}^i(a)
\newcommand{\myslpr}[2]{{l'}_{#1}(#2)}                   % l_{m,h,s}^i(a)
\newcommand{\mylhat}[5]{\hat{l}_{#1, #2}^{#4}(#3, #5)}          % \hat{l}_{m,h,s}^i(a)
\newcommand{\mylbarpr}[5]{\bar{l'}_{#1, #2}^{#4}(#3, #5)}          % \hat{l}_{m,h,s}^i(a)
\newcommand{\myslbarpr}[2]{\bar{l'}_{#1}(#2)}          % \hat{l}_{m,h,s}^i(a)
\newcommand{\mylbar}[5]{\bar{l}_{#1, #2}^{#4}(#3, #5)}          % \bar{l}_{m,h,s}^i(a)
\newcommand{\myLhat}[5]{\hat{L}_{#1, #2}^{#4}(#3, #5)}          % \hat{L}_{m,h,s}^i(a)
\newcommand{\myR}[4]{R_{#1, #2}^{#4}(#3)}                       % R_{m,h,s}^i
\newcommand{\myVOver}[4]{\overline{V}_{#1, #2}^{#4}(#3)}
\newcommand{\myVOvertil}[4]{\tilde{V}_{#1, #2}^{#4}(#3)}
\newcommand{\myVUnd}[4]{\underline{V}_{#1, #2}^{#4}(#3)}
\newcommand{\myVUndtil}[4]{\undertilde{V}_{#1, #2}^{#4}(#3)}
\newcommand{\mysl}[2]{l_{#1}(#2)}                               % l_i(a)
\newcommand{\myslhat}[2]{\hat{l}_{#1}(#2)}                      % \mysLhat{i}{a}
\newcommand{\myslbar}[2]{\bar{l}_{#1}(#2)}                      % \bar{l}_i(a)
\newcommand{\mysLhat}[2]{\hat{L}_{#1}(#2)}                      % \mysLhat{i}{a}
\newcommand{\mysLtil}[2]{\tilde{L}_{#1}(#2)}                    % \tilde{L}_i(a)
\newcommand{\mysR}[1]{R_{#1}}                                   % R_i
\newcommand{\gap}[1]{\text{CCE-gap}(#1)}
\usepackage{enumitem}
\setlist[itemize]{leftmargin=*}
\author{%
 \Name{Yuyang Zhang} \Email{yuyangzhang@g.harvard.edu}\\
 \Name{Runyu Zhang} \Email{runyuzhang@fas.harvard.edu}\\
 \addr Harvard University, School of Engineering and Applied Science
 \AND
 \Name{Yuantao Gu}
 \Email{gyt@tsinghua.edu.cn}\\
 \addr Tsinghua University, Department of Electronic Engineering
 \AND
 \Name{Na Li}
 \Email{nali@seas.harvard.edu}\\
 \addr Harvard University, School of Engineering and Applied Science
}

\usepackage{xcolor}

\begin{document}

\maketitle
\vspace{-12pt}
\begin{abstract}%
    This paper considers multi-agent reinforcement learning (MARL) where the rewards are received after delays and the delay time varies across agents and across time steps. Based on the V-learning framework, this paper proposes MARL algorithms that efficiently deal with reward delays. When the delays are finite, our algorithm reaches a coarse correlated equilibrium (CCE) with rate $\tilde{\mathcal{O}}(\frac{H^3\sqrt{S\mathcal{T}_K}}{K}+\frac{H^3\sqrt{SA}}{\sqrt{K}})$ where $K$ is the number of episodes, $H$ is the planning horizon, $S$ is the size of the state space, $A$ is the size of the largest action space, and $\mathcal{T}_K$ is the measure of total delay formally defined in the paper. Moreover, our algorithm is extended to cases with infinite delays through a reward skipping scheme. It achieves convergence rate similar to the finite delay case.
\end{abstract}

\begin{keywords}%
  Reward Delays, Markov Games, Multi-Agent Reinforcement Learning
\end{keywords}

\section{Introduction}

% MARL intro

% MARL state of the art currently (V-learning etc.)

% Reward delay and its lack of theoretical guarantees

% Our contribution

% Other related works

Multi-agent reinforcement learning (MARL) finds extensive applications such as recommendation systems \citep{intro_recom}, medical treatments \citep{intro_med, intro_med2}, multi-agent robotics systems \citep{intro_rob, intro_rob2, intro_rob3}, autonomous driving \citep{intro_aud}, etc. In these multi-agent problems, individuals aim to learn to interact with the environment under the influence of other agents.
%Despite all the empirical success, many MARL algorithms with theoretical guarantees fail to perform well when put into practice. One major challenge that lies between theories and applications is the reward delay, or feedback delay in other literature.

Motivated by the empirical success of MARL, there is a recent surge of studies on MARL algorithms with theoretical convergence guarantees such as V-learning, V-learning OMD, SPoCMAR, etc \citep{alg_ma_VL, alg_ma_VL2, alg_ma2, alg_ma3}. In these algorithms, agents rely heavily on real-time observations of reward values to update their policies or value functions. However, in real-life MARL applications, rewards generally come with delays. One example is the medical treatment process \citep{intro_med}, where the effectiveness of a treatment strategy cannot be observed immediately. It generally takes a long time for a patient to respond and recover. Similar reward delays also widely exist in recommendation systems \citep{intro_recom_delay}, autonomous driving \citep{intro_autodrive}, neuroscience \citep{intro_neuro}, etc. Another example is the reward delays due to communication latency in all kinds of distributed systems \citep{intro_rob3_lqr, intro_distri, intro_distri2} where even infinite delays are common due to packet loss and network failure. Reward delays in these applications are typically time-varying, depending on factors including the patient's physiological state, the status of communication channels, etc. All the examples suggest that it is crucial to understand how reward delays affect the learning process and how to design MARL algorithms that could accommodate the delays efficiently.

In existing empirical work on MARL with reward delays, different approaches are proposed to handle delays, %guide the agents in the absence of the rewards,
including but not restricted to learning temporal structures \citep{rw_rdelay3}, predicting strategic interactions \citep{rw_rdelay2}, evaluating curiosity \citep{rw_rdelay}, and predicting the environment \citep{rw_sdelay2} with neural networks. However, from the theoretical perspective, few results are known for MARL. We acknowledge the lines of work studying state or action delays in MARL \citep{rw_sdelay3, rw_sdelay4, rw_sdelay5}, but the settings are different and out of the scope of this paper. Other related settings include single-agent reinforcement learning (SARL) and multi-arm bandit (MAB). For SARL, recent work \citep{rw_mdp4, rw_mdp5} studies adversarial reward delays. Unfortunately, their methods suffer from the curse of dimensionality when directly extended to MARL. Other work \citep{rw_sdelay, rw_sdelay0} only focuses on constant reward delays. For MAB, \citet{rw_bandit, rw_bandit2, rw_bandit4} tackle adversarial reward delays while \citet{rw_bandit5, rw_bandit6} focus on constant delays. 
%\lina{one possible question: If they consider delays on state/action, does it mean they also have delay in the reward?  Should their case include our case?  }
% \lina{One thought: I think it would be useful to have a subsection "Related work" in Appendix where we can discuss the literature in more details. Reviewers will feel our literature review is too short for a RL paper. }

\vspace{5pt}\noindent\textbf{Our Contributions.} In this paper, we focus on a specific MARL model, the general-sum Markov games \citep{intro_stochaticgame,intro_generalsum}. 
We propose the delay-adaptive multi-agent V-learning (DA-MAVL) to learn coarse-correlated equilibria (CCEs) under time-varying reward delays, where the learning of the agents can be finished in a fully decentralized manner. Namely, every agent runs its own learning algorithm without communicating with others. Note that this is nontrivial because different agents may receive the same reward at different episodes due to the heterogeneous delays among them. Without careful design, fully distributed learning might lead to misalignment and divergent behavior. Our DA-MAVL algorithm circumvents this problem by carefully selecting proper reward information for learning and therefore aligning the behaviour of the agents. %\cathy{TBD: Needs revision}
% It also comes with batched and dedicated subroutines whose parameters are crafted to utilize the reward information more efficiently.

For finite delays, our algorithm achieves the CCE-gap as small as $\tilde{\mathcal{O}}(\frac{H^3\sqrt{S\mysT{K}}}{K}+\frac{H^3\sqrt{SA}}{\sqrt{K}})$ with samples from $K$ episodes (Theorem~\ref{thm:comp}). Here $H$ is the planning horizon, $S$ is the size of the state space, $A=\max_m|\mathcal{A}_m|$ is the largest size of one agent's action space, and $\mysT{K}$ can be seen as a measure of the total delay. %\footnote{We formally define the term $\mysT{K}$ in Equation~\eqref{def:holding} and \eqref{eq:holdmax}.}
In the worst case, $\sqrt{\mysT{K}}$ is the order of $\mathcal{O}(\sqrt{Kd_{max}})$, where $d_{max}$ is the largest possible delay. This implies that the CCE-gap is as small as $\tilde{\mathcal{O}}(\frac{1}{\sqrt{K}})$. This dependence of $K$ matches the original result of V-learning \citep{alg_ma_VL, alg_ma_VL2}, indicating that DA-MAVL successfully aligns the behaviour of the agents. Moreover, both terms are independent of the number of agents, meaning that DA-MAVL scales nicely with the system size. Our proposed DA-MAVL algorithm can be extended to settings with infinite delays. With a novel skipping metric inspired by \citet{rw_bandit2}, our algorithm can skip the infinite delays without prior knowledge of the delay sequence and achieve the CCE-gap similarly to the finite delay case (Theorem~\ref{thm:compskip}). 
To the best of our knowledge, our results give the first convergence rate guarantee for general-sum MGs under time-varying reward delays. 

Due to the space limit, we defer related work, some of the algorithms, proofs, and simulation settings to the appendix of our full paper \citep{fullreport}.

\section{Problem Setup \& Preliminary}
% \lina{the original notation explanation of $\arg$ indices are confusing without any context. I temporally remove it. it would be better to define it when you first time use it.}

%\noindent\textit{Notations:} 
%and $-m$ means $[m]-\left\{i\right\}$
%and use $-m$ to denote all agents other than agent $m$. 

% $\arg$ will be specified in the context.
% For any set $\mathcal{M} = \{(n_1, \dots), \dots, (n_{|\mathcal{M}|}, \dots)\}$ whose elements are ordered tuples, we define $\arg\min \{\mathcal{M}\} = \min_i n_i$ as the minimal first entry of the tuples.

\subsection{Markov Games with Reward Delays}\label{sec:mg}
%\lina{Apologize that I think markov games $MG$ might be more aligned with the current literature and it is actually more informative than stochastic games.}
We study general-sum Markov games (MGs, also called stochastic games in \cite{intro_stochaticgame})  with reward delays. In its episodic and tabular form, an MG can be defined by the following tuple:
\begin{align} \label{eq:SG}
    \mathcal{MG} \Big(H, \mathcal{S}, &\{\mathcal{A}_m\}_{m\in[M]}, \{\mathbb{P}_h\}_{h\in[H]}, \{r_{m,h}\}_{m\in[M], h\in[H]},\{\myd{m}{h}{s}{n}\}_{m\in [M], h\in [H], s\in\mathcal{S}, n\in [K]}\Big).
\end{align}
Here we use $[i]$ to denote set $\{1, \dots, i\}$ for any integer $i$. In the subscripts, $m\in[M]$ stands for the agents, $k\in[K]$ stands for the episode, $h\in[H]$ stands for the time step over the finite horizon.  $\mathcal{S}$ is a global state space with cardinality $S = |\mathcal{S}|$. $\mathcal{A}_m$ is the action space of agent $m$. Define $A = \max_{m\in[M]} |\mathcal{A}_m|$. The joint action space is given by $\bm{\mathcal{A}} = \mathcal{A}_1\times \dots\times \mathcal{A}_M$, and the joint action is given as $\bm{a} = (a_1, \dots, a_M)$. $\mathbb{P}_h(s'|s,\bm{a})$ with $s, s'\in \mathcal{S}, \bm{a}\in \bm{\mathcal{A}}$ is the transition function for step $h$. $r_{m,h}(s,\bm{a})$ is a deterministic reward for agent $m$ at step $h$ when the current state and joint action are $s$ and $\bm{a}$ respectively. The sequence $\{\myd{m}{h}{s}{n}\}_{m\in [M], h\in [H], s\in\mathcal{S}, n\in \mathbb{N}}$ represents reward delays which will be detailed in later paragraphs. Without loss of generality, we assume every episode $k$ starts from a fixed initial state $s_1$.\footnote{For any MG with initial distribution $\mu$, one can always add a step with only one state as the first time step and let the transition function be $\mu$ for all actions. This leads to an equivalent MG with a fixed initial state.} At every step $h$, every agent observes the current state $s^k_h$, takes action $a^k_{m,h}$. The environment transits to the next state $s^k_{h+1}$ according to $\mathbb{P}_h$ until step $H+1$ is reached. 

\vspace{5pt}
\noindent{\textbf{Visits \& happening order:}} When the agents visit state $s$ at step $h$ for the $n$-th time, we say that the $n$-th visit of $(h,s)$ happens, and $n$ is the happening order of this visit. 

\vspace{5pt}
\noindent{\textbf{Reward delays:}} We allow the reward delays to be heterogeneous among different agents $m$, different visits $(h,s)$ and different happening orders $n$. In specific, for the $n$-th visit of $(h,s)$ which happens at episode $k$, agent $m$ will receive its reward $r_{m,h}(s, \bm{a})$ by the end of episode $k+\myd{m}{h}{s}{n}$. When $\myd{m}{h}{s}{n}=0$ for all $m\in[M],h\in[H],s\in \mathcal{S},n\in\mathbb{N}$, our setting reduces to a classic MG. 
%by the end of episode $k$.

\vspace{5pt}
\noindent{\textbf{(Un)received visits:}} When the reward of a visit has been received, we call the visit a received visit; otherwise, we call it an unreceived visit. It is worth mentioning that visits that happen early are not necessarily received early.

\noindent{\textbf{(Un)usable visits:}} We denote the episode when the $n$-th visit of $(h,s)$ happens as $\myk{h}{s}{n}$. At the beginning of episode $\myk{h}{s}{n}$,\footnote{Without causing any confusion, we will use ``at the beginning of episode $k$'' and ``by the end of episode $k-1$'' interchangeably.} for agent $m$, some of the first $n-1$ visits of $(h,s)$ may not be received because of the reward delays. In this case, we define index $e_{m,h}^n(s)$ as the earliest unreceived visit:
% \vspace{-6pt}
\begin{equation}\label{eq:mye}
    e_{m,h}^n(s) := \min \Big\{j:\myd{m}{h}{s}{j} + \myk{h}{s}{j} > \myk{h}{s}{n}-1,  j\in[n-1]\Big\}.
\end{equation}
If all of the first $n-1$ visits have been received, we define
% \vspace{-6pt}
\begin{equation}\label{eq:mye_2}
    e_{m,h}^n(s) := n.
\end{equation}
It means that all the visits of $(h,s)$ that happen earlier than the $e_{m,h}^n(s)$-th visit have been received at the beginning of episode $\myk{h}{s}{n}$; but the $e_{m,h}^n(s)$-th visit has not been received yet. 
%$then l^n is n”, “if there is some visit that has not been received, then it is the earliest unreceived visit”.
We call a received visit as \emph{usable} if all visits happening earlier have all been received. The rest of the received visits are called \emph{unusable}. At the beginning of episode $\myk{h}{s}{n}$, the usable visits of $(h,s)$ have happening orders $1, 2, \dots,  e_{m,h}^n(s)-1$. The unusable visit of $(h,s)$, if $e_{m,h}^n(s) < n$, have happening orders $e_{m,h}^n(s), \dots, n-1$.

In our algorithms, to ensure that the agents are aligned, we only use the usable visits. Consequently, the performance of our algorithms strongly relates to the number of unusable and unreceived visits, for which we define a counting sequence $\{\myT{m}{h}{s}{n}\}_{m\in[M], h\in[H], s\in \mathcal{S}, n\in[K]}$:
\vspace{-6pt}
\begin{equation}\label{def:holding}\begin{split}
    & \myT{m}{h}{s}{n} := \sum_{i=1}^{n} \left(i -e_{m,h}^i(s)\right) = \sum_{i=1}^{n} \left(i - \min\Big\{j: \myd{m}{h}{s}{j} + \myk{h}{s}{j} > \myk{h}{s}{i}-1\Big\}\right).\\
\end{split}\end{equation}
$\myT{m}{h}{s}{n}$ counts the accumulated number of unusable and unreceived visits of $(h,s)$ till the $n$-th visit. Note that in the classic MG setting without reward delays, we have $\myT{m}{h}{s}{n}=0$ for all $m\in[M],h\in[H],s\in \mathcal{S},n\in\mathbb{N}$. 
% We also let $\mysT{n}$ denote the maximum of $\myT{m}{h}{s}{n}$ over all $(m,h,s)$:
% \begin{equation}\label{def:holdmax}
%     \mysT{K} := \max_{m,h,s} \myT{m}{h}{s}{n}.
% \end{equation}

% We use the sequence $\{\myD{k}\}_{k\in[K]}$ to denote the total delays for agent $m$,
% \begin{equation}
%     \myD{k} = \max_{m\in[M],h\in[H]} \sum_{\kappa=1}^{k-1} \myd{m}{h}{\kappa}{}
% \end{equation}
% \lina{ use $\kappa$ to stands for the episode?}
% Performance of our algorithm strongly relates to the number of unusable and unreceived visits. We define the holding sum $\{\myT{m}{h}{s}{n}\}_{m\in[M], h\in[H], s\in \mathcal{S}, k\in[K]}$ to give a count of this:
% \begin{equation}\begin{split}
%     & \myT{m}{h}{s}{n} = \sum_{i=1}^n i - \mathop{\arg\min}\limits_j \Big[ \myd{m}{h}{s}{j} + \myk{h}{s}{j} \geq \myk{h}{s}{i}\Big]\\
% \end{split}\end{equation}
% $\myD{k}$ is  the maximal summation of delays during the first $k$ episodes over each time step $h$ and agent $m$. 
%Note that this term is deterministic, because the delays $\{\myd{m}{h}{t}\}_{m\in[M], h\in[H], k\in[K]}$ are chosen by an adversarial before the algorithm begins.\lina{I am trying to avoid discussing randomness/adversarial etc. For control community, just say delay is ok so far. For the performance analysis, we might want to make it more clear. But saying `adversarial' here is confusing: if it is `adversarial', should I just make the delay be infinite to make no method is going to work. That's why I don't want to discuss this now.}

\subsection{Learning Objective - Coarse Correlated Equilibrium}

%\lina{Should start describing the general goal. like: }
%\cathy{before this, need to define policy and correlated-policy first}

Agent $m$'s policy is denoted as $\pi_m = \{\pi_{m,h}\}_{h\in[H]}$. The policy at step $h$ is $\pi_{m,h}: \Omega\times (\mathcal{S}\times \mathcal{A})^{h-1}\times \mathcal{S} \to \Delta_{\mathcal{A}_m}$, where $\pi_{m,h}$ maps a random sample $\omega_h$ from probability space $\Omega$ and a trajectory $(s_1, \bm{a}_1, \dots, s_h)$ to a point in probability simplex $\Delta_{\mathcal{A}_m}$.
An important subclass of policy is the \emph{independent Markov policy}, with $\pi_{m,h}: \mathcal{S}\to \Delta_{\mathcal{A}_m}$ maps the current state to a point in probability simplex $\Delta_{\mathcal{A}_m}$.

A \emph{joint policy} $\pi$ is a set of policies $\{\pi_m\}_{m\in[M]}$ of all agents. If the random samples $\{\omega_h \in \Omega\}_{h\in[H]}$ are shared among all agents, policies of all agents are correlated. In this case, we denote the joint policy $\pi$ as $\pi = \pi_1\!\odot\!\pi_2\!\odot\!\dots\!\odot\! \pi_M$, and call $\pi$ as a correlated policy. We also use $\pi_{-m} = \pi_1\!\odot\!\dots\!\pi_{m-1}\odot\!\pi_{m+1}\dots\!\odot\! \pi_M$ to denote the policy excluding agent $m$. 
If the randomness of $\pi_m$ is independent of other policies $\pi_{-m}$,
% , for example, the random sample $\omega \in \Omega$ has a special form $\omega = (\omega_m, \omega_{-m})$ and $\pi_m$ only depends on $\omega_m$, $\pi_{-m}$ depends $\omega_{-m}$, where $\omega_m$ and $\omega_{-m}$ are independent of each other. In this case, 
i.e., the random samples $\{\omega_h \in \Omega\}_{h\in[H]}$ are shared among agents except agent $m$, we denote the joint policy as $\pi = \pi_m \times \pi_{-m}$.

%At each step $h$, every agent $m$ takes an action $a_{m,h}$ according to its policy $\pi_m$. We denote the joint policy as $\pi:=(\pi_1,\ldots, \pi_m)$. We also define $\pi_{-m} = \{\pi_{m'}\}_{m'\in[M]\backslash{m}}$ to be the policy without agent $m$.
%To evaluate the performance of an output policy $\pi$, 
For a joint policy $\pi$, we define its value function for agent $m$ as:
% \vspace{-9pt}
\begin{equation}
   V^{\pi}_{m,h}(s_h) := \mathbb{E}_{\pi} \Big[ \sum\limits_{h'=h}^H r_{m,h'}(s_{h'}, \bm{a}_{h'}) | s_h\Big], \ \ \forall m\in [M] .
\end{equation}
Given policy $\pi_{-m}$, the best response for agent $m$ is defined as the best policy that maximizes the value function for agent $m$, i.e., $\pi_m^\dag = \mathop{\arg\max}_{\pi_m} V_{m,1}^{\pi_m\times\pi_{-m}}$. For notation simplicity, we denote the value function of the best response as $V^{\dag, \pi_{-m}}_{m,h} = V^{\pi_m^\dag, \pi_{-m}}_{m,h}$.
Our objective is to find a joint policy $\pi$ that is an $\epsilon$-coarse correlated equilibrium (CCE) defined as follows:
\begin{mydefinition}[Coarse Correlated Equilibrium (CCE \citep{intro_CCE_def})]
    We define the CCE-gap of a joint policy $\pi$ as:
     \vspace{-6pt}
    \begin{equation}
        \text{CCE-gap}(\pi) := \max\limits_{m\in[M]} (V^{\dagger,\pi_{-m}}_{m,1} - V^{\pi}_{m,1})(s_1).
    \end{equation}
    A joint policy $\pi$ is a CCE if the CCE-gap is zero:
    \vspace{-6pt}
    \begin{equation}
        \gap{\pi} = 0.
    \end{equation}
    A joint policy $\pi$ is an $\epsilon$-CCE if the CCE-gap satisfies:
     \vspace{-6pt}
    \begin{equation}
        \gap{\pi} \leq \epsilon.
    \end{equation}
\end{mydefinition}
When the agents reach a CCE, they have no incentive to deviate to any independent policy. 

%\lina{need to define what is a `markovian policy'. It is unclear what policies we are considering here, from where to where? agents' dependence? Please take a look at Chi Jin's paper when he introduce policies}
%At step $h$ of episode $k$, every agent $m$ learns a markovian policy $\hat{\pi}^k_{m,h}$. Based on all the policies $\{\hat{\pi}^K_{m,h}\}_{h\in[H]}$, \dots,$\{\hat{\pi}^K_{m,h}\}_{h\in[H]}$, every agent $m$ outputs its policy $\pi_m = \{\pi^K_{m,h}\}_{h\in[H]}$ (which may be non-Markovian). Note that the non-markovian policies of any two agents may be correlated. \lina{again, this is confusing. This paragraph should move after we describe the general goal.} \lina{depending how to define the general goal at the beginning, you may not need this paragaraph here.}

\section{Delay-Adaptive Multi-Agent V-Learning}\label{sec:damavl}

In this section, we present our main algorithm: Delay-Adaptive Multi-Agent V-Learning (DA-MAVL). Similar to V-learning in \cite{alg_ma_VL, alg_ma_VL2}, DA-MAVL contains two consecutive algorithms - i) the training algorithm (Algorithm \ref{alg:damavl}), where the agents learn and store a set of independent Markov policies \{$\hat{\pi}^k_{m,h}\}_{m\in[M], h\in[H], k\in[K]}$,  and ii) the output algorithm (Algorithm \ref{alg:damavl_output}) that constructs the final output policies (which can be correlated and non-Markov) $\{\pi_m\}_{m\in[M]}$ from the set of independent Markov policies \{$\hat{\pi}^k_{m,h}\}_{m\in[M], h\in[H], k\in[K]}$.
%
%At step $h$ of episode $k$, every agent $m$ learns a markovian policy $\hat{\pi}^k_{m,h}$. Based on all the policies $\{\hat{\pi}^K_{m,h}\}_{h\in[H]}$, \dots,$\{\hat{\pi}^K_{m,h}\}_{h\in[H]}$, every agent $m$ outputs its policy $\pi_m = \{\pi^K_{m,h}\}_{h\in[H]}$ (which may be non-Markovian). Note that the non-markovian policies of any two agents may be correlated.

The training algorithm is \textit{fully decentralized}, i.e., the agents update their own policies with their own delayed reward information without communication with each other. The algorithm framework resembles the V-learning algorithm but comes with a mechanism that carefully chooses \emph{usable} visits for learning. This mechanism enables agents to align their behaviour under the influence of heterogeneous reward delays and leads the algorithm toward convergence (see more discussions at the end of next subsection). 
% \lina{The discussion on explaining this is non-trivial task should be mentioned in the intro which could help people understand the importance of the work. Now in the intro, it only mentions that the "delays are typically heterogeneous" in the second paragraph. After pragraph 3, it should discuss the technical difficulty induced by the heterogeneous delays. Or in the contribution section, highlight that our alg doesn't need communication among agents. This is non-trivial because without communication, it is more challenging to handle the heterogeneous delays... You can try first to see whether you can revise the intro do highlight the main novelty. }
% We would also like to point out that when there's no delay, our algorithm is an exact equivalence to V-learning.

%It avoids negative effects of delays by carefully selecting suitable visits for learning, and processes information in a batched way. Theoretical guarantee of the algorithm is given together with the proof sketch.

Recall that $\myk{h}{s}{n}$ is the episode when the agents visit $(h, s)$ for the $n$-th time. For agent $m$, we also define $\mynpr{m}{h}{s}{k}$ as the count of happened visits of $(h, s)$ and define $\myn{m}{h}{s}{k}$ as the count of
usable visits of $(h, s)$ at the beginning of episode $k$.

\subsection{The Training Algorithm} \label{sec:alg}
We now present the training algorithm of DA-MAVL for agent $m$ (Algorithm~\ref{alg:damavl}). The algorithm contains three major processes, which we name as `Preparation', `Learning' and `Sampling'. At each episode $k$, for every time step $h$, the three processes are carried out iteratively:%The delayed rewards are received at the beginning of every episode $k$.  For every agent $m$, at the beginning of episode $k$, delayed rewards are received. At step $h$ of episode $k$, where state $s$ is visited, three major process are carried out:
\begin{itemize}
    \item In the `Preparation' process, we keep track of three important sets, namely the set of \textit{visits to be used} $\myaF{m}{h}{s}$ (including all usable visits that have not been used previously), the set of \textit{unusable visits} $\myaMp{m}{h}{s}$ and the set of \textit{unreceived visits} $\myaMm{m}{h}{s}$. Usable Visits in $\myaF{m}{h}{s}$ will be fed into later processes and will no longer be used again in future episodes. Unusable and unreceived visits in $\myaM{m}{h}{s}=\myaMp{m}{h}{s} \cup \myaMm{m}{h}{s}$ are stored in memory until they become usable.
    
    For set $\mathcal{M} = \myM{m}{h}{s}{}$ (or $\mathcal{M} = \myaMm{m}{h}{s}{}$), whose entries are tuples $(i,a,\hat{\pi},\overline{V}',\underline{V}',r)$ (or $(i,a,\hat{\pi},\overline{V}',\underline{V}')$) indexed by the first element $i$, we define $\arg \{\mathcal{M}\} := \{i\}$ as the set of indices. 
    
    %two important sets $\myaM{m}{h}{s}=\myaMp{m}{h}{s} \cup \myaMm{m}{h}{s}$ and $\myaF{m}{h}{s}$ are maintained. The former contains information of \textbf{unreceived visits} (in subset $\myaMm{m}{h}{s}$) and \textbf{unusable visits} (in subset $\myaMp{m}{h}{s}$). The latter contains information of \textbf{usable visits}. Note that visits in the two sets are arranged in the order that they happened.
    \item In the `Learning' process, visits in $\myaF{m}{h}{s}$ are fed into subroutines `VALUE\_UPDATE' and `POLICY\_OPT' (Algorithm~\ref{alg:damavl_sub1} and Algorithm \ref{alg:damavl_sub2} in Appendix~\ref{sec:sub_damavl} in \cite{fullreport}) \emph{consecutively in their happening orders}. Subroutine `VALUE\_UPDATE' updates an ``optimistic'' value estimate  $\myVOver{m}{h}{s}{}$ by using all visits in $\myaF{m}{h}{s}$ with parameters $\{\alpha_i\}_i$ and $\{\mybetaOver{m}{h}{s}{i}\}_i$, where $\alpha_i$ can be viewed as the learning rate and $\mybetaOver{m}{h}{s}{i}$ is a bonus term. 
    Subroutine `POLICY\_OPT' runs an adversarial-bandit-type algorithm (similar to the algorithm in \citet{rw_bandit2}) to update the policy, where the bandit loss is calculated using the optimistic value estimates $\myVOver{m}{h}{s}{}$. 
    
    Note that in Algorithm~\ref{alg:damavl} and Subroutine `VALUE\_UPDATE', we also introduce a pessimistic value estimate $\myVUnd{m}{h}{s}{}$. This pessimistic estimate is an auxiliary variable that is not needed for running the algorithm but is used in the proof.
    %The two subroutines perform value update and policy optimization on $\myaF{m}{h}{s}$ in a batched and sequential fashion. Namely, visits in $\myaF{m}{h}{s}$ are arranged in their happening order and used sequentially by the subroutines. The updated value estimates and the policies of $(h,s)$ are then returned. 
    \item In the `Sampling' process, every agent chooses its action based on the updated policy, and the next state is sampled. Finally, every agent stores related information, receives delayed rewards and moves on to the next step $h+1$.
\end{itemize}

\noindent\textbf{Discussions - The role of usable visits.} As previously mentioned, one key challenge for the decentralized learning algorithm is to avoid misalignment due to heterogeneous reward delays among different agents. Our algorithm addresses this challenge by only using usable visits for learning in subroutines `VALUE\_UPDATE' and `POLICY\_OPT'. The main intuition is to ensure that the \emph{happening order} of the visits is also \emph{the order in which they are used in the subroutines}. Consequently, although rewards of the visits might be received and used in different episodes for different agents, the order in which they are used remains the same among agents. This design leads to cooperative policies among the agents without any communication in the training algorithm. % has the irreplaceable advantage of not requiring any communication during the training process.

To better understand the role of usable visits, we also compare our algorithm numerically with the naive algorithm, where visits are immediately fed into subroutines once they are received (see Appendix~\ref{sec:sub_naive} in \cite{fullreport} for details). In the naive algorithm, the reward of the same visit may be used in different orders among agents, which causes extra misalignment among the agents. The numerical results are discussed in Section~\ref{sec:simulation}, where we indeed observe that with the notion of usable visits, our algorithm outperforms the naive algorithm. 
% Additionally, we also tried to analyze the performance of the naive method, yet failed to give a rigorous bound due to the misalignment caused by the heterogeneity of the delay. 
However, it remains an open question to prove or to disapprove whether the naive method would converge to a CCE. 
%unclear whether this is a proof artifact or there actually exists counterexamples where the naive method fails to converge. We leave it as an interesting open question for future work.

We also note that the notion of usable visits alone is not sufficient to fully align all agents, nor does it reduce the problem to MARL without reward delays. This is because different agents still have different amount of information in the episodes. This information mismatch is further addressed by a critical modification in Algorithm~\ref{alg:damavl_output} in the following subsection.

\begin{algorithm2e}[H]
    \caption{DA-MAVL Training for Agent $m$}
    \label{alg:damavl}
    \KwInit{$\forall (h,s)$, $\mynpr{m}{h}{s}{0} \gets 0$, $\myn{m}{h}{s}{0} \gets 0$, $\myT{m}{h}{s}{0} \gets 0$, $\myaF{m}{h}{s}\gets \emptyset$, $\myaM{m}{h}{s}\gets \emptyset$\;}

    \For(){Episode\ \  $k = 1, \dots, K$}{
        Receive initial state $s^k_1$\;
        \For(){Step\ \  $h = 1, \dots, H$}{
            \textcolor{cyan}{// Preparation}\\
            $s \gets s^k_h$\;

            \For(){$(i,a,\hat{\pi},\overline{V}', \underline{V}',r) \in \myaMp{m}{h}{s}$}{
                \If{$\forall j<i, j \notin \arg \{\myaMm{m}{h}{s}\}$}{
                    Save $(i,a,\hat{\pi},\overline{V}', \underline{V}',r)$ to $\myaF{m}{h}{s}$;
                    Remove $(i,a,\hat{\pi},\overline{V}', \underline{V}',r)$ from $\myaMp{m}{h}{s}$\;
                }
            }
            $\mysnpr{} \gets \mynpr{m}{h}{s}{k} = \mynpr{m}{h}{s}{k-1} + 1$;
            $\mysn{} \gets \myn{m}{h}{s}{k} = \myn{m}{h}{s}{k-1} + |\myaF{m}{h}{s}|$\;
            
            % $\mytau{m}{h}{s}{n'} \gets |\myM{m}{h}{s}{}|$\;
            $\myT{m}{h}{s}{\mysnpr{}} \gets \myT{m}{h}{s}{\mysnpr{}-1} + |\myM{m}{h}{s}{}|$\;
            \textcolor{cyan}{// Learning}\\
            $\myVOver{m}{h}{s}{k}, \myVUnd{m}{h}{s}{k} \gets \text{VALUE\_UPDATE}_{m,h,s}\big(\myaF{m}{h}{s}, \mysn{}\big)$\;
            $\hat{\pi}_{m,h}^k(\cdot|s) \gets \text{POLICY\_OPT}_{m,h,s}\big(\myaF{m}{h}{s}, \mysnpr{}\big)$\;   
            \textcolor{cyan}{// Sampling}\\
            Take action $a_{m,h}^k \sim \hat{\pi}_{m,h}^k(\cdot | s)$;\ \ Observe next state $s_{h+1}^k$\;
            \For(){$s'\in \mathcal{S}\backslash s$}{
                $\mynpr{m}{h}{s'}{k} \gets \mynpr{m}{h}{s'}{k-1}$; $\myn{m}{h}{s'}{k} \gets \myn{m}{h}{s'}{k-1}$\;
                $\myVOver{m}{h}{s'}{k} \gets \myVOver{m}{h}{s'}{k-1}$; $\myVUnd{m}{h}{s'}{k} \gets \myVUnd{m}{h}{s'}{k-1}$; $\hat{\pi}^k_{m,h}(\cdot|s')\gets \hat{\pi}^{k-1}_{m,h}(\cdot|s')$\;
            }
            }
        \For(){Step $h = 1, \dots, H$}{

            Save $\big(\mynpr{m}{h}{s^k_h}{k}, a_{m,h}^k, \hat{\pi}_h^k(a_{m,h}^k|s^k_h), 
            \myVOver{m}{h+1}{s^k_{h+1}}{k}, \myVUnd{m}{h+1}{s^k_{h+1}}{k}\big)$ to $\myaMm{m}{h}{s^k_h}$\;
            % Save $\mytau{m}{h}{s}{n'}$ to $\mathcal{M}em_{m,h}(s)$\;
        }
        Receive delayed rewards for all states $s$\;
        \For(){Delayed Reward $(m, h, s, i, r)$}{ 
            Extract and remove $\big(i, a, \hat{\pi}, \overline{V}', \underline{V}'\big)$ from $\myaMm{m}{h}{s}$\;
            Save $\big(i, a, \hat{\pi}, \overline{V}', \underline{V}', r\big)$ to $\myaMp{m}{h}{s}$\;
        }
    }
\end{algorithm2e}

\subsection{Execution of the Output Policy}
\begin{algorithm2e}[H]
    \caption{DA-MAVL Output for Policy $\pi_{m}$}
    \label{alg:damavl_output}
    Sample $k \sim \text{Uniform}([K])$\;
    \For(){step\ \ $h=1, \dots, H$}{
        Observe current state $s_h$; $n \gets \max_m\myn{m}{h}{s_h}{k}$\;
        Sample $i$ from $[n]$ with probability $\alpha_n^i$; $k \gets \myk{h}{s_h}{i}$\;
        Take action $a_{m,h} \sim \hat{\pi}_{m,h}^k(\cdot|s_h)$\;
    }
\end{algorithm2e} 
% \lina{What is the input to the policy $\pi_m$? }
%\cathy{TODO: In the V-learning paper, the definition of $\hat{\pi}$ and $\pi$ is actually flipped. Shall we be consistent with their paper?}
Algorithm~\ref{alg:damavl} outputs a set of independent Markov policies $\{\hat{\pi}^k_{m,h}\}_{m\in[M], h\in[H], k\in[K]}$. Based on this policy set, we now construct joint policy $\pi = \{\pi_m\}_{m\in[M]}$ as the output of DA-MAVL. The policy is defined by its execution in Algorithm \ref{alg:damavl_output}. Notice that all random samples (line 1 and line 4) are shared across all agents. 

This algorithm follows V-learning in \citet{alg_ma_VL, alg_ma_VL2} except for the critical modification in line 3. Intuitively speaking, choosing $n=\max_m \myn{m}{h}{s}{k}$ ensures that agent $m$ is aware of the extra information that the most informed agent possesses, and therefore guarantees that the output policy of agent $m$ is compatible with that of the most informed agent. Technically speaking, it ensures the optimistic value estimates in Algorithm~\ref{alg:damavl} upper bound the policy performance.

\section{Performance Guarantee and Proof Sketch}
Recall that the counting sequence $\{\myT{m}{h}{s}{n}\}_{m\in[M], h\in[H], s\in \mathcal{S}, n\in[K]}$ (Equation \eqref{def:holding}) is agent $m$'s accumulated count of unusable and unreceived visits till the $n$-th visit of $(h,s)$. Also, recall that $\myn{m}{h}{s}{k}$ is the count of
usable visits of $(h, s)$ at the beginning of episode $k$. Using the two notations, we define $\mysT{K} := \max_{m,h} \sum_{s\in\mathcal{S}}\myT{m}{h}{s}{\myn{m}{h}{s}{K}}$ which will be used in bounding the CCE-gap after $K$ episodes. We also assume that the reward delays of the MG are upper bounded by some constant.
\begin{assumption}\label{ass:dmax}
    The delays are bounded by $d_{max}$, that is, $
       \max\limits_{m\in[M], h\in[H], s\in \mathcal{S}, n\in [K]} \myd{m}{h}{s}{n} \leq d_{max}.
    $
\end{assumption}
Now we are ready to present the performance guarantee for DA-MAVL:

\begin{theorem}\label{thm:comp}
 Under Assumption~\ref{ass:dmax}, for any $\delta\in (0, 1), K\geq d_{max}^2S\iota^3$ where $\iota = \log ( 4MHSAK /\delta)$, suppose Algorithm~\ref{alg:damavl} is run for $K$ episodes, then the following equation holds for the output policy $\pi$ of Algorithm \ref{alg:damavl_output} with probability at least $1-\delta$ 
%Suppose Assumption~\ref{ass:dmax} holds. For $\forall \delta\in (0, 1)$ and $\forall K \geq d_{max}^2S\iota^3$, let $\log ( 4MHSAK /\delta)$. Let policy $\pi$ be the output of Algorithm~\ref{alg:damavl_output} after running Algorithm~\ref{alg:damavl} for $K$ episodes. The following equation holds with probability at least $1-\delta$: 
    \begin{equation}\begin{split}
        % &\max\limits_{m\in[M]} \Big( V_{m,1}^{\dag, \pi_{-m}} - V_{m,1}^{\pi} \Big)(s_1) \lesssim H^3 S\sqrt{\mysT{K}/K^2}\iota^2 + H^3\sqrt{SA\iota/K}.
        &\text{CCE-gap}(\pi) = \max\limits_{m\in[M]} \Big( V_{m,1}^{\dag, \pi_{-m}} - V_{m,1}^{\pi} \Big)(s_1) \lesssim H^3 \sqrt{S\mysT{K}/K^2}\iota^2 + H^3\sqrt{SA\iota/K}.
    \end{split}\end{equation} 
\end{theorem}
Under Assumption~\ref{ass:dmax}, it can be shown (with Lemma~\ref{lem:comp1_base2} in Appendix~\ref{sec:proof_comp_step1} in \cite{fullreport}):
\begin{equation*}\label{eq:holdmax}\begin{split}
    \mysT{K} =& \max_{m,h} \sum_{s\in\mathcal{S}}\myT{m}{h}{s}{\myn{m}{h}{s}{K}} = \max_{m,h} \sum_{s\in\mathcal{S}} \sum_{n=1}^{\myn{m}{h}{s}{K}} \left(n -e_{m,h}^n(s)\right) \leq d_{max}\max_{m,h}\sum_{s\in\mathcal{S}}\mynpr{m}{h}{s}{K} \leq Kd_{max}.
\end{split}\end{equation*} 
Substituting it into Theorem~\ref{thm:comp} gives the CCE-gap of order $\tilde{\mathcal{O}}(\frac{H^3\sqrt{Sd_{max}}+H^3\sqrt{SA}}{\sqrt{K}})$. In other words, in the worst case where the delays are always $d_{max}$ and every $(h,s)$ is visited for $K$ times, at most $K = \tilde{\mathcal{O}}\big(\frac{H^6S(d_{max}+A)}{\epsilon^2}\big)$ episodes are needed for an $\epsilon$-CCE. The influence of the reward delays is linearly bounded by term $\tilde{\mathcal{O}}(\frac{H^6Sd_{max}}{\epsilon^2})$ and tends to $0$ when $d_{max}$ goes to $0$.
Note that our result bears an extra factor $H$ compared with V-learning \citep{alg_ma_VL, alg_ma_VL2}, even when all delays are zero. This is because we have to choose the parameters generously so that our algorithm is adaptive to potential delays. 
 %We also note that in general the upper bound will be much smaller, because the delays are not always $d_{max}$, and count of usable visits of $(m,h,s)$ at episode $K$, i.e. $\myn{m}{h}{s}{K}$, will be much smaller than $K$. The influence of the reward delays are by $\mathcal{T}_K$, which is the maximum number of unreceived and unusable visits across all possible $(m,h,s)$ during the $K$ episodes.
 %\lina{I delete this because you revised your theorem to make it be $\sum_{s}...$ The previous discussion is not accurate. Moreover, I think it is straigtforward to notice that it is just an upper bound}

% \lina{Need to add discussion of the results here, even it is simple: effect of delays}
% In addition, the term $\myT{m}{h}{s}{k}$ can be calculated a posteriori, hence the bound is further refined after execution of the algorithm.  %\cathy{TODO: double check if this is a correct statement} \yy{Is it OK to leave the bound randomized?}

% \cathy{Cathy TBD: comparison with the original V-learning alg}

% \lina{I think the discussion with naive alg might be good to move to here. But I am flexible on this as I am not sure yet.}

\subsection{Proof Sketch of Theorem~\ref{thm:comp}}\label{sec:sketch1}
The proof can be broken down into the following three steps. %\textbf{In step one}, we establish policy optimization regret bound for every pair $(m,h,s,k)$ (ref. Lemma~\ref{lem:reg}). \textbf{In step two}, we show that bonuses $\overline{\beta}_{n}$ and $\underline{\beta}_{n}$ (previously mentioned in the `VALUE\_UPDATE$_{m,h,s}$' subroutine) upper bounds the performance degradation due to the policy optimization regret (ref. Lemma~\ref{lem:opt} and Lemma~\ref{lem:pes}). This fact ensures that the two value estimates $\overline{V}$ and $\underline{V}$ upper and lower bound performance of the output policy of Algorithm~\ref{alg:damavl}. \textbf{In step three}, we bound the gap between the two estimates, which directly translates to the convergence of the output policy.

\vspace{5pt}\noindent\textbf{STEP 1: Bound the `Policy Optimization Regret'.} For every pair $(m,h,s,n)$, we first define the policy optimization regret $\myR{m}{h}{s}{n}$. For notational simplicity, we let $k_n$ denote $\myk{h}{s}{n}$.
% It serves as the weighted measure of the how good the first $n$ Markov policies $\hat{\pi}_{m,h}^{k_i}$ are given the policies of other agents $\hat{\pi}_{-m,h}^{k_i}$. 
\begin{equation}
    R_{m,h}^{n}(s) \!=\!\! \max\limits_{a_m\in \mathcal{A}_m} \sum\limits_{i=1}^{n} \alpha_{n}^i \Big[ \mathbb{E}_{} \Big(r_{m,h}(s,\bm{a}) + \myVOver{m}{h+1}{s'}{k_i}\Big)-\Big(r_{m,h}^{k_i} + \myVOver{m}{h+1}{s_{h+1}^{k_i}}{k_i} \Big)\Big],
\end{equation}
where $\bm{a} = (a_m, \bm{a}_{-m})$, $\alpha_{n}^i$ is the weight which we define in Equation~\eqref{eq:alpha} in Appendix~\ref{sec:notation} in \cite{fullreport}, and the expectation is taken over $\bm{a}_{-m}\sim\hat\pi_{-m,h}^{k_i}(\cdot|s)$ and $s'\sim \mathbb{P}_h(\cdot|s,\bm{a})$. Intuitively, it measures the performance of the first $n$ outputs of subroutine `POLICY\_OPT$_{m,h,s}$' in Algorithm~\ref{alg:damavl}, i.e. Markov policies $\{\hat{\pi}_{m,h}^{k_i}(s)\}_{i\in[n]}$. Under Assumption~\ref{ass:dmax}, we give the following upper bound:
\begin{mylemma}\label{lem:reg}
    Let Assumption \ref{ass:dmax} holds. For $\forall (m,h,s,k)\in[M]\times[H]\times \mathcal{S}\times [K]$, the following inequality holds with probability at least $1-\delta/2$
    \begin{equation}\label{eq:lem_reg}
        R_{m,h}^{n}(s) \leq 12H^2 \sqrt{\dfrac{nA+\myT{m}{h}{s}{n}}{n^2}\iota} + 2H^2\dfrac{d_{max}}{n} \iota.
    \end{equation}
\end{mylemma}
In this lemma, the key difference from V-learning, and main technical difficulty, is that the subroutine needs to learn the $n$-th output, i.e. $\hat{\pi}_{m,h}^{k_n}(s)$, without access to all reward information of the first $n-1$ visits of $(h,s)$ due to the reward delays. We have to measure the influence of the delays on outputs. By comparing it with the no-delay versions% their cheating versions, which are the outputs of the subroutine when the delays are zero
, we can show that the influence of the delays can be reflected by term $\sqrt{\myT{m}{h}{s}{n}/n^2}$ and $d_{max}/n$ in Equation~\eqref{eq:lem_reg}.

\vspace{5pt}
\noindent\textbf{STEP 2: Optimism and Pessimism.} Utilizing the regret defined above, we carefully design bonuses $\mybetaOver{m}{h}{s}{n}$ and $\mysbetaUnd{n}$ in subroutine `VALUE\_UPDATE' as follows:
\vspace{-6pt}
\begin{equation}
    \mybetaOver{m}{h}{s}{n} = \myR{m}{h}{s}{n} + 2H^2 \dfrac{d_{max}}{n} \iota, \quad 
    \mysbetaUnd{n} = 2\sqrt{\dfrac{H^3}{n}\iota} + 2H^2\dfrac{d_{max}}{n}\iota.
        % 0, &n=0\\
    % & \mybetaUnd{m}{h}{s}{n} = \left\{\begin{array}{lr}
    %     ,\ \  &n\geq 1\\
    %     0, &n=0\\
    % \end{array}\right.\\
\end{equation}
With the bonuses, we can show that the value estimates $\myVOver{m}{h}{s}{k}$ and $\myVUnd{m}{h}{s}{k}$ in Algorithm \ref{alg:damavl} upper and lower bound the performance of policy $\pi_{m,h}^k$. 
\vspace{-6pt}
\begin{mylemma}\label{lem:opt}
    Let Assumption~\ref{ass:dmax} holds. For $\forall (m,h,s,k)\in [M]\times[H]\times \mathcal{S}\times[K]$, the following inequality holds with probability at least $1-\delta$
    \begin{equation}\label{eq:bonuses}
 \overline{V}_{m,h}^k(s) \geq V^{\dagger, \pi_{-m,h}^k}_{m,h}(s),\quad  \underline{V}_{m,h}^k(s) \leq V^{\pi_h^k}_{m,h}(s).
 \end{equation}
\end{mylemma}
In this lemma, policy $\pi_h^k(s)$ can be seen as part of the output policy $\pi(s)$ in Algorithm~\ref{alg:damavl_output}, that is used from step $h$ to $H$.
% When $h=1$ and $k$ is sampled uniformly from $[K]$, $\pi_h^k(s)$ is just the output policy. 
It is formally defined in Algorithm~\ref{alg:damavl_outputMed} in Appendix~\ref{sec:notation} in \cite{fullreport}. 

% Note that $\myVOver{m}{h}{s}{k}$ and $\myVUnd{m}{h}{s}{k}$ are calculated with reward information from policies $\{\hat{\pi}_{m,h}^{k_i}(s)\}_{i\in[\myn{m}{h}{s}{k}]}$, but they can bound the values related to part of output policy $\pi$, thanks to the design of the output policy (Algorithm~\ref{alg:damavl_output}). We refer the readers to Appendix~\ref{sec:proof_s2} for more details.

% Although Lemma \ref{lem:opt} looks similar to optimism and pessimism results (Lemma 13 and Lemma 14 in \cite{alg_ma_VL}), it is a non-trivial generalization
We note that it is technically difficult to ensure optimism and pessimism under the influence of heterogeneous reward delays among agents. 
% , because some agent will have more information than others. 
Notice that $\myVOver{m}{h}{s}{k}$ and $\myVUnd{m}{h}{s}{k}$ are calculated only with information of agent $m$. 
However, the output policy $\pi^k_h$, as in $V^{\dagger, \pi_{-m,h}^k}_{m,h}(s)$ and $V^{\pi_h^k}_{m,h}(s)$, is a \emph{correlated} policy that takes information of all agents into consideration. This information mismatch makes it technically challenging for $\myVOver{m}{h}{s}{k}$ and $\myVUnd{m}{h}{s}{k}$ to upper or lower bound $V^{\dagger, \pi_{-m,h}^k}_{m,h}(s)$ and $V^{\pi_h^k}_{m,h}(s)$, and breaks the original optimism and pessimism results in V-learning \citep{alg_ma_VL,alg_ma_VL2}. Here we carefully design Algorithm~\ref{alg:damavl_output} (especially line 3) to ensure that every agent is aware of the extra information of the most informed agent. Then with the carefully designed bonuses as in Equation~\ref{eq:bonuses}, we are able tackle this difficulty and ensure optimism and pessimism.%\lina{this paragraph is confusing. First of all, "on the one hand, on the other hand" is often used to describe two conflicting issues. I don't think it is what you mean here. What is the message you want to say here?}
% We refer readers to Appendix~\ref{sec:proof_damavl} in \cite{fullreport} for technical details.

\vspace{5pt}
\noindent\textbf{STEP 3: Bound the CCE-gap.} Finally, given Lemma \ref{lem:opt}, it suffices to bound the gap between the optimistic and pessimistic value estimates $
\sum\limits_{k=1}^{K} (\overline{V}_{m,1}^k - \underline{V}_{m,1}^k)(s_h^k)$.

As is mentioned in Step 2, the value estimates $\myVOver{m}{h}{s}{k}$ and $\myVUnd{m}{h}{s}{k}$ are calculated without access to all information due to the reward delays. This fact increases the variance of the value estimates. In the proof of this theorem, we carefully analyze the number of unreceived and unusable visits for every episode and analyze its cumulative influence across all episodes. 
% We refer the readers to Appendix~\ref{sec:proof_damavl} for more details.

\section{Extension to Infinite Delays}

\subsection{The Skipping Scheme}\label{sec:damavl_skip}
The performance of the DA-MAVL algorithm in Section \ref{sec:damavl} heavily relies on the assumption that delays are finite. One single infinite delay could prevent the algorithm from convergence because all visits that happen later are unusable. 
%However, when there exist infinite delays, $\mysT{K}$ could be very large, even growing quadratically with $K$, making the bound in Theorem~\ref{thm:comp} invalid. 
In this case, it is worth skipping some of the rewards for better performance. Following the intuitions of \cite{rw_bandit2}, we extend DA-MAVL and design a new skipping metric to deal with infinite delays in MARL. Details for the extended algorithm (DA-MAVL with Reward Skipping) are presented in Appendix~\ref{sec:sub_damavl_skip} in \cite{fullreport}. 
%In the training algorithm (Algorithm~\ref{alg:damavl_skip}) for DA-MAVL with Reward Skipping, a `Skipping' process is performed before the `Learning' process. 

%If some visit of $(h,s)$ has remained unreceived for large enough episodes, the algorithm skips it and runs default subroutines `VALUE\_UPDATE' and `POLICY\_OPT' for it.% $\myVOver{m}{h}{s}{}$ in subroutine `VALUE\_UPDATE' (Algorithm~\ref{alg:damavlskip_sub1} in \ref{sec:sub_damavl_skip} in \cite{fullreport}), where the reward and the optimistic value estimate of the next state are rendered as $1$ and $H-1$. Subroutine `POLICY\_OPT' (Algorithm~\ref{alg:damavl_sub2} in Appendix~\ref{sec:sub_damavl} in \cite{fullreport}) treats bandit loss of this visit as $0$ and the cumulative loss, as well as the policy, remains unchanged.

%\subsubsection{Intuitions for `SKIPPING'}

The critical part of the `Skipping' process is to determine when to skip a visit. When the $n$-th visit of $(h,s)$ happens, we maintain the skipping metric $\myphi{m}{h}{s}{i}{n} = \sum_{j=i+1}^n (j-i)$ if the $i$-th visit of $(h,s)$ is unreceived. Intuitively speaking, $\myphi{m}{h}{s}{i}{n}$ upper bounds the contribution of the $i$-th visit to $\myT{m}{h}{s}{n}$. 
% If visit $i$ is skipped, term $\myphi{m}{h}{s}{i}{n}$ can be removed from term $\myT{m}{h}{s}{n}$ in the regret. 
It is beneficial to skip the $i$-th visit if $\myphi{m}{h}{s}{i}{n}$ becomes large enough. Following the intuition of previous reward skipping method in \citet{rw_bandit2} in the adversarial bandit setting, we skip the $i$-th visit if $\myphi{m}{h}{s}{i}{n}$ exceeds threshold $\sqrt{\myT{m}{h}{s}{n}}$. 

However, we would like to point out that our design of the skipping metric $\myphi{m}{h}{s}{i}{n}$ is not a direct generalization of previous skipping method. 
Unlike the multi-agent setting considered in this paper, the adversarial bandit setting does not need to consider the heterogeneity of reward delays among agents, thus their algorithm update does not need to wait for visits to become usable. Correspondingly, the skipping metric $n-i$ in previous method would fail in our setting, because it no longer upper-bounds the contribution of the $i$-th visit to $\myT{m}{h}{s}{n}$.

% Moreover, we expect the reduction to be at most the same order as the remaining term $\myT{m}{h}{s}{n}$. Consequently, we need to set the threshold at least $\sqrt{\myT{m}{h}{s}{n}}$. This is the simple intuition behind the truncation threshold.

%Intuition for the new skipping metric $\{\myphi{m}{h}{s}{i}{n}\}_{i/n \in [K]}$ is the focus of this paper. On one hand, we abandon the metric $n-i$ in \citet{rw_bandit2}, because it fails to capture all contribution of the $i$-th visit to $\myT{m}{h}{s}{n}$. If the $i$-th visit remains unreceived, all visits after it can neither be used by the algorithm. %On the other hand, we do not carefully count the exact contribution of the $i$-th visit to $\myT{m}{h}{s}{n}$, because the rate at which the true count grows is unstable. It can grow as slow as 0 and as fast as $\sqrt{\myT{m}{h}{s}{n}}$. Consequently, it may inflate too much above threshold and damage the performance. Contrarily, $\myphi{m}{h}{s}{i}{n}$ grows steadily with $n$, with the rate being $\sqrt[4]{\myT{m}{h}{s}{n}}$.

\subsection{Performance Guarantee for DA-MAVL with Reward Skipping}
Recall the notation $\myk{h}{s}{n}$ stands for the episode when $n$-th visit of $(h,s)$ happens. With the skipping scheme, we can also relax Assumption~\ref{ass:dmax} to the following: 
\begin{assumption}\label{ass:c}
    For $\forall (m,h,s,n) \in [M]\times [H]\times \mathcal{S}\times [K]$, there exists a constant $C$ satisfying:
    \vspace{-6pt}\begin{equation}\begin{split}
        |\{i \leq n: \myd{m}{h}{s}{i} + \myk{h}{s}{i} \geq \myk{h}{s}{n}\}| \leq C.
    \end{split}\end{equation}
\end{assumption}
Intuitively,  Assumption \ref{ass:c} requires that for every pair $(m,h,s,n)$, there are at most $C$ unreceived visits before the $n$-th visit of $(h,s)$for agent $m$. This implies that either large delays do not appear too many times or delays are not large enough to influence performance. It is worth noting that the finite delay Assumption~\ref{ass:dmax} implies Assumption \ref{ass:c} with $C=d_{max}$. But Assumption~\ref{ass:c} is more general than Assumption~\ref{ass:dmax} because Assumption~\ref{ass:c} holds even if there are less than $C$ infinite delays. 

% \cathy{TBD: (?) Discussion on the trade off for the set L. Larger L will result in smaller $T^L$, but $|L|$ is larger}
% With the skipping scheme, the performance of our algorithm only depends on the number of unusable or unreceived visits if the large delays do not exist. So we define $\{\myT{m}{h}{s}{n,\mathcal{L}}\}_{m\in[M], h\in[H], s\in \mathcal{S}, n\in[K]}$ to give a count. \cathy{Needs revision} Given any $\mathcal{L} \subset [K]$,
Given a subset of visit indices $\mathcal{L}\subset[K]$, at episode $\myk{h}{s}{n}$ when the $n$-th visit of $(h,s)$ happens, we define variable $\mye{m}{h}{s}{n,\mathcal{L}}$ as the earliest unreceived visit outside of $\mathcal{L}$:
\vspace{-6pt}\begin{equation}
    \mye{m}{h}{s}{n,\mathcal{L}} := \min \Big\{j: \myd{m}{h}{s}{j} + \myk{h}{s}{j} > \myk{h}{s}{n}-1, j\in[n-1]\backslash \mathcal{L}\Big\}.\vspace{-6pt}
\end{equation}
If all of the first $n-1$ visits are received, we let $\mye{m}{h}{s}{n,\mathcal{L}} = n$. Now we define $\myT{m}{h}{s}{n,\mathcal{L}}$ as follows:
% $\myT{m}{h}{s}{n,\mathcal{L}}$ as follows:
\vspace{-6pt}\begin{equation}
    \myT{m}{h}{s}{n,\mathcal{L}} := \sum_{i=1}^{n} i - \mye{m}{h}{s}{i,\mathcal{L}}.
    % \vspace{-12pt}
    \vspace{-6pt}
\end{equation}
It counts the accumulated number of unusable and unreceived visits outside of $\mathcal{L}$ for the first $n$ visits of $(h,s)$.
%
%that When the delays are zero for all visits of $(h,s)$ that happen in episode in $\mathcal{L}$, i.e. $\{i:\myk{h}{s}{i}\in\mathcal{L}\}$, the count of unusable and unreceived visits during the first $n$ visits of $(h,s)$ equals $\myT{m}{h}{s}{n,\mathcal{L}}$. 
% In general cases where the delays are not zero for visits of $(h,s)$ in $\mathcal{L}$, $\myT{m}{h}{s}{n,\mathcal{L}}$ still takes the same value.
% In other words, it counts the number of unusable and unreceived visits during the first $n$ visits of $(h,s)$, 
Finally, we also define 
$\mysTtil{m,h}^{K,\mathcal{L}} := \sum_{s\in\mathcal{S}}\myT{m}{h}{s}{\mynpr{m}{h}{s}{K},\mathcal{L}}$. Intuitively, it counts the accumulated number of unusable and unreceived visits outside of $\mathcal{L}$ during the $K$ episodes. 

% \begin{myremark}[Assumption~\ref{ass:c} Compared with Assumption~\ref{ass:dmax}]
%     Consider any pair $(m,h,s)$. If Assumption~\ref{ass:dmax} holds, then for any $i < n-d_{max}$, we have $i + \myd{m}{h}{s}{i} < n-d_{max}+d_{max} = n$.     That is, Assumption \ref{ass:c} holds with $C=d_{max}$. On the other hand, Assumption \ref{ass:c} can hold even when there exist less than $C$ infinite delays. This implies Assumption~\ref{ass:c} is more general than Assumption~\ref{ass:dmax}. 
% \end{myremark}

Now we are ready to present the performance guarantee for DA-MAVL with Reward Skipping:
\begin{theorem}\label{thm:compskip}
    Under Assumption~\ref{ass:c}, for any $ \delta\in (0, 1)$, $K\geq C^6S^3\iota^3$ where $\iota = \log ( 4MHSAK /\delta)$, suppose Algorithm~\ref{alg:damavl_skip} is run for $K$ episodes, then the following equation holds for the output policy $\pi$ of Algorithm~\ref{alg:damavl_output} with probability at least $1-\delta$
    \begin{equation}\begin{split}
        &\text{CCE-gap}(\pi) \lesssim CH^3 \max_{m,h}\min_{\mathcal{L}} \Bigg\{\frac{S|\mathcal{L}|}{K} + \sqrt{\frac{S\mysTtil{m,h}^{K,\mathcal{L}}}{K^2}}\Bigg\}\iota^2 + H^3\sqrt{\frac{SA}{K}\iota}.\\
    \end{split}\end{equation} 
\end{theorem}
% In general, the CCE-gap converges in the same order compared to Theorem \ref{thm:comp}. 
Theorem~\ref{thm:compskip} implies that DA-MAVL with Reward Skipping can still obtain convergence to CCE when there are infinite delays. Consider the case where all delays are upper bounded by constant $d_{max}$, except for $C$ infinite delays for every $(h,s)$. 
Let $\mathcal{L}_{m,h}=\{n: \exists s, \myd{m}{h}{s}{n}=\infty\}$ denote all visit indices where the delay is infinite for some state $s$ and fixed pair $(m,h)$. We then have $|\mathcal{L}_{m,h}| \leq CS$ and $\mysT{m,h}^{K,\mathcal{L}_{m,h}}\leq Kd_{max}$. Substituting into Theorem~\ref{thm:compskip} gives CCE-gap of order $\tilde{\mathcal{O}}(\frac{H^3\sqrt{Sd_{max}}+H^3\sqrt{SA}}{\sqrt{K}})$, which is exactly the same as the result of Theorem~\ref{thm:comp}. %In other words, DA-MAVL with Reward Skipping is barely influenced by infinite delays in this case.
% and When faced with one infinite delay, CCE-gap bound of Theorem \ref{thm:comp} diverges because all later visits are held back by the infinite delay. However, this infinite delay will be put into set $\mathcal{L}$ thanks to the skipping scheme. Consequently, the algorithm performance is barely influenced. It is worth mentioning that the constant $C$ can be large constants, e.g. $C = \log K$, without significantly influencing performance of the algorithm. 

% % The proof follows similar three-step routine as the proof for Theorem~\ref{thm:comp}. \textbf{In step one}, we establish policy optimization regret bound for every pair $(m,h,s,k)$. \textbf{In step two}, we show that bonuses $\overline{\beta}_{n}$ and $\underline{\beta}_{n}$ (previously mentioned in the `VALUE\_UPDATE$_{m,h,s}$' subroutine) upper bounds the performance degradation due to the policy optimization regret (ref. Lemma~\ref{lem:opt} and Lemma~\ref{lem:pes}). This fact ensures that the two value estimates $\overline{V}$ and $\underline{V}$ upper and lower bound performance of the output policy of Algorithm~\ref{alg:damavl}. \textbf{In step three}, we bound the gap between the two estimates, which directly translates to the regret bound of the output policy.

\vspace{-6pt}\section{Simulations}\label{sec:simulation}
\begin{figure}[ht]
\centering
\vspace{-6pt}\begin{minipage}[b]{.32\linewidth}
    \centering
    \includegraphics[width=1\linewidth]{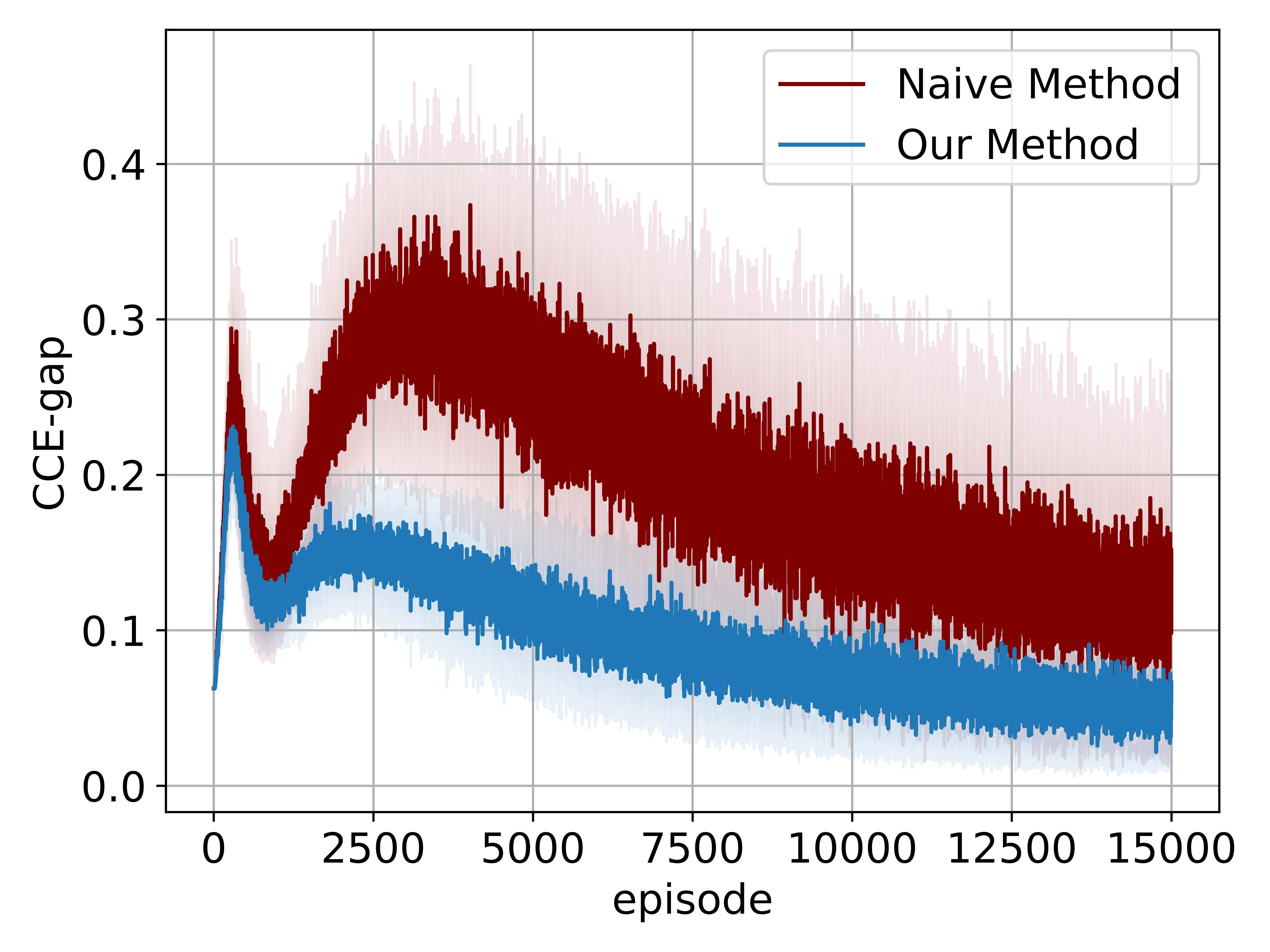}
\end{minipage}
\begin{minipage}[b]{.32\linewidth}
    \centering
    \includegraphics[width=1\linewidth]{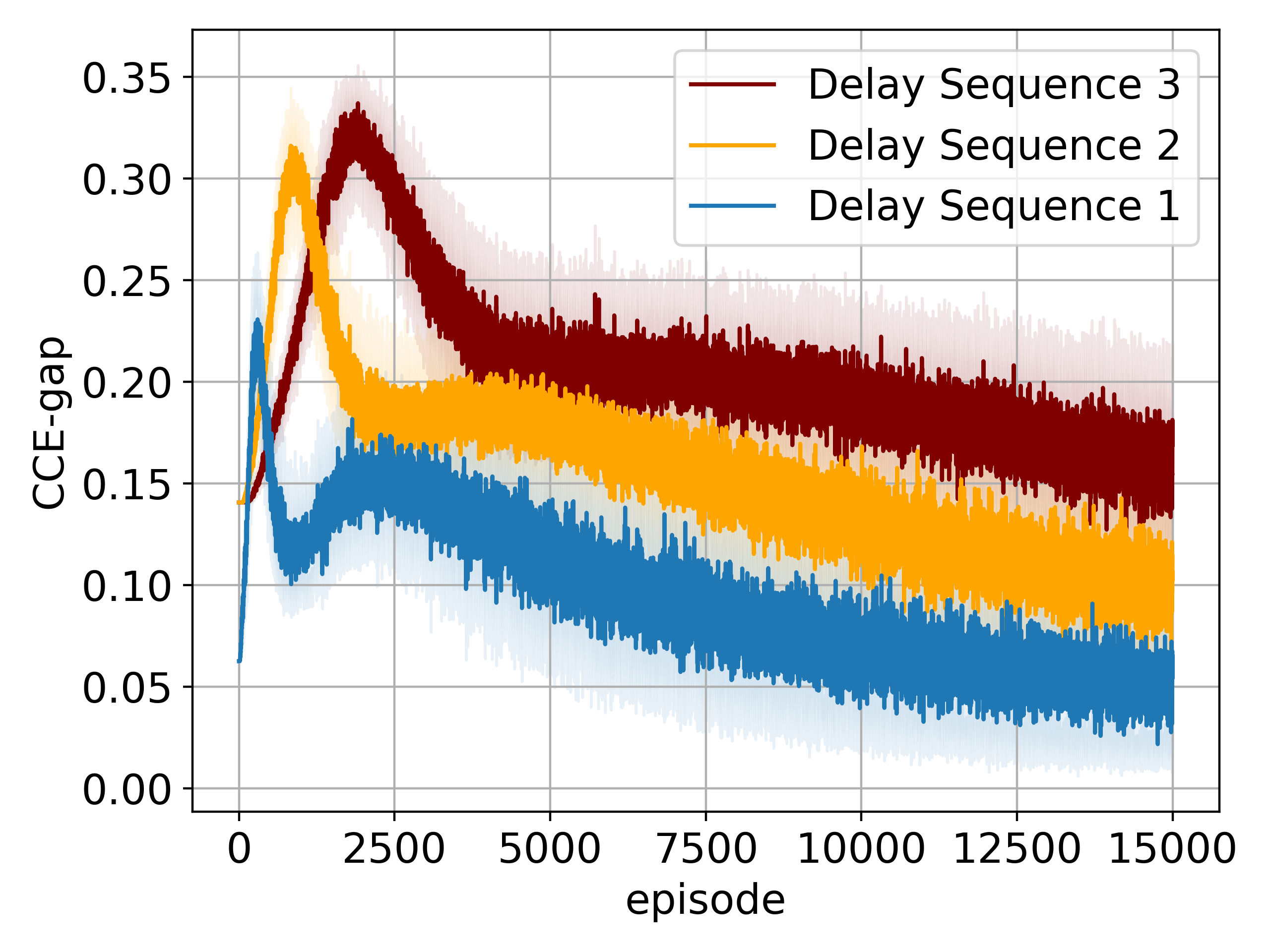}
\end{minipage}
\begin{minipage}[b]{.32\linewidth}
    \centering
    \includegraphics[width=1\linewidth]{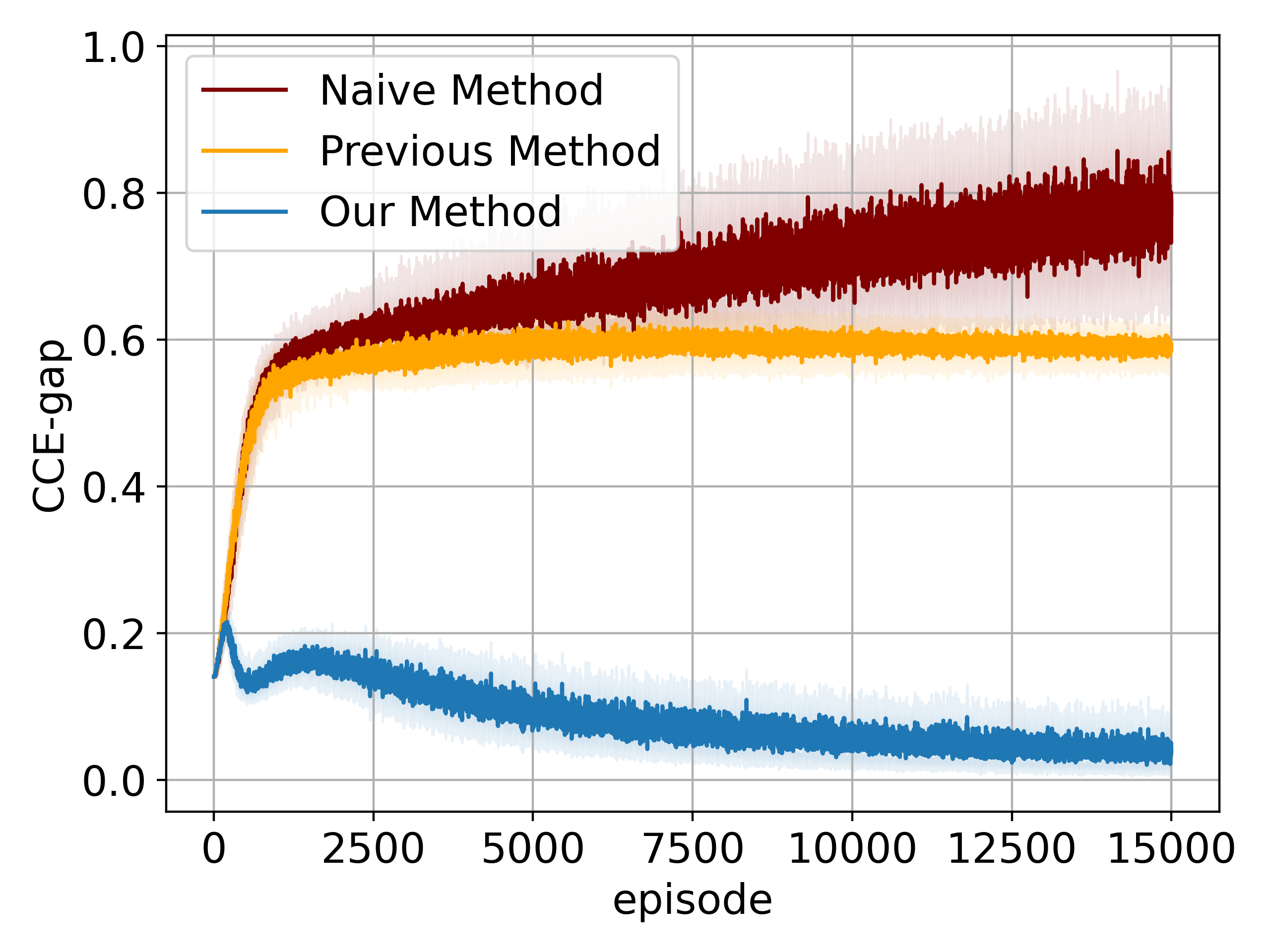}
\end{minipage}
\vspace{-12pt}\caption{\small \textbf{Left}: CCE-gap for Output Policy of DA-MAVL (Our Method) and the Naive Algorithm (Naive Method); \textbf{Center}: CCE-gap for Output Policy of DA-MAVL under Different Delay Sequences; \textbf{Right}: CCE-gap for Skipping Metrics in DA-MAVL with Reward Skipping (Our Method), in Previous Work~\citep{rw_bandit2} (Previous Method) and No Skipping (Naive Method).
}
\label{fig:sim}
\end{figure}

We simulate our algorithms in a simple MG with $M = 3, S = 3, A = 2, H = 2$. Due to the space limit, the simulation settings are deferred to Appendix~\ref{sec:simulation-appendix} in \cite{fullreport}. We only present the simulation results in  Figure~\ref{fig:sim}. We can see that our algorithm outperforms the naive algorithm (mentioned in Section~\ref{sec:damavl}) when delays are finite. Moreover, our novel skipping metric outperforms previous skipping method (mentioned in Section~\ref{sec:damavl_skip}) when delays are infinitely large.

\vspace{-28pt}\section{Conclusion}
% In this paper, we develop MARL algorithms to handle both finite and infinite reward delays. We provided high probability bounds on the CCE-gap. There are many future directions including providing lower bounds on the CCE-gap for MARL with reward delays, relaxing Assumption~\ref{ass:c} for infinite delays, extending current results to MGs with function approximation, etc.

This paper studies MARL with reward delays. For finite delays, we propose MARL algorithms with a novel mechanism to choose proper visits for learning, so that agents can reach a CCE even when facing heterogeneous delays. We also adapt our algorithm to cases with infinite delays using a novel reward skipping metric. High probability bounds are given on the CCE-gap of our algorithms. There are many interesting future directions, such as proving or disproving the convergence of the naive algorithm (Appendix~\ref{sec:sub_naive} in \cite{fullreport}),  providing lower bounds on the CCE-gap for MARL with reward delays, relaxing Assumption~\ref{ass:c} for infinite delays, extending current results to MGs with function approximation, etc.% We also leave it as an interesting open question to prove or disprove the convergence of the naive algorithm (Appendix~\ref{sec:sub_naive} in \cite{fullreport}).

%bound studies MARL with reward delays. We propose MARL algorithms with novel mechanism to choose proper visits for learning, so that the agents' behaviour are aligned. We also adapt our algorithm to cases with infinite delays using a novel reward skipping metric. High probability bounds are given on the CCE-gap of our algorithms.

% We hope this paper can provide intuitions for further studies including but not restricted to Linear Quadratic Games \citep{4_further}, Stochastic Potential Games \citep{4_further2}, etc. that might involve reward delays. Moreover, in the presence of delays, the lower bounds on the sample complexity for $\epsilon$-CCEs is still unknown. The exact form of the bound and refined algorithms to reach that bound are left for future work.

% Acknowledgments---Will not appear in anonymized version
\acks{This work is supported by the NSF grants CNS 2003111 and AI institute 2112085 and by the ONR YIP award N00014-19-1-2217.}

%\nocite{*}
% \bibliographystyle{ieeetr}
\bibliography{ref}

% \newpage
\newpage\appendix

\section{Related Work}
\textbf{MARL algorithms with theoretical guarantees.} In MARL, it is a standard objective to find CCEs of the underlying MG \citep{alg_ma_VL}. There is a recent line of work providing non-asymptotic guarantees for learning CCEs of general-sum MGs. Generally speaking, existing model-based algorithms \citep{alg_ma1, alg_2a3} suffer from the curse of dimensionality in finding CCEs \citep{alg_ma_VL, alg_ma_VL2}. They require samples exponentially related to the number of agents to achieve the learning objective. Recent model-free algorithms \citep{alg_ma_VL, alg_ma_VL2,alg_ma2, alg_ma3} have successfully broken the curse, providing sample complexity not directly related to the number of agents. V-Learning in \cite{alg_ma_VL,alg_ma_VL2} is one of the algorithms that achieve this breakthrough. 

Another common objective in this setting is the Nash Equilibrium. However, it is proven PPAD-hard by previous work \citep{intro_NE_PPAD}.

\vspace{5pt}\noindent\textbf{Delays in MARL.} Different kinds of delays may occur in MARL, including but not restricted to state delays (or observation delays), action delays and reward delays (or feedback delays) \citep{rw_sdelay3}. We acknowledge that there exist lines of work on state and action delays \citep{rw_sdelay3, rw_sdelay4, rw_sdelay5}, but they are beyond the scope of this paper. In MARL with state delays, the major challenge is how to predict the current state from previous information. In MARL with action delays, the challenge is to predict when the actions will take effect. However, the focus of our paper is to better evaluate the current state and action with available information instead of predicting what they are. Most paper concerning reward delays in MARL is empirical \citep{rw_rdelay, rw_rdelay2, rw_rdelay3}. As discussed previously, they resort to alternative mechanisms to guide the agents instead of using rewards directly. Currently, the mechanisms are formed using deep neural networks that are hard to explain theoretically.

\vspace{5pt}\noindent\textbf{Reward delays in MDP.} Recently, empirical single-agent RL algorithms have achieved great progress in handling reward delays \citep{rw_mdp, rw_mdp2, rw_mdp3}. In contrast, theoretical aspect of the problem is relatively unexplored. The two available previous work \citep{rw_mdp4,rw_mdp5} studies adversarial reward delays. Take the latter as an example. For an adversarial MDP with state space size $S$, action space size $A$, planning horizon $H$ and total delay $D$, their algorithm achieves the optimal gap $\mathcal{\tilde{O}}(H^2S\sqrt{A/K} + H^{5/4}(SA)^{1/4}\sqrt{D/K^2})$ with high probability after $K$ episodes. This result matches ours (Theorem~\ref{thm:comp}) in terms of episode number $K$ and total delay $D$. However, as mentioned before, these model-based methods maintain exponentially many parameters that prohibit them from being tractable in the multi-agent setting. Our algorithm, based on V-Learning, is a completely different model-free algorithm that successfully breaks the curse.

\vspace{5pt}\noindent\textbf{Reward delays in multi-arm bandit (MAB).} There exist extensive theoretic work in MAB that deals with delays. \citet{rw_bandit, rw_bandit2, rw_bandit3, rw_bandit4} tackle delays chosen by adversarial while \citet{rw_bandit5, rw_bandit6} concern constant delays. Typically, their algorithms gives optimal gap $\tilde{\mathcal{O}}\sqrt{A/K+D/K^2}$, where $A$ is the number of actions, $D$ is the total delay, and $K$ is the number of episodes. All the above literature provides precious insights for this paper. But the MAB setting is completely different from MARL, since there are no state transitions and cooperation between agents. 

\section{Simulation Settings} \label{sec:simulation-appendix}

We simulate a simple MG with $M=3, S=3, A=2, H=2$. In the fixed initial state $s_1$, reward $r=1$ is given if all agents choose action one, $r=0.5$ is given if all agents choose action two, and reward $r=0$ otherwise. With $r=0$ at $h=0$, the agents transit to $h=1, s=s_2$, where no reward is given. With $r>0$ at $h=0$, the agents transit to $h=1, s=s_3$, where the reward follows the same distribution as in $s_1$.

\vspace{5pt}\noindent\textbf{Finite Delays.}
We first simulate DA-MAVL (Section~\ref{sec:damavl}) with respect to the naive algorithm (Appendix~\ref{sec:sub_naive}) under delay sequence 1 as follows:
\begin{equation}
    \myd{1}{1}{s_1}{n} = 20 - 2\cdot(i\ \text{mod}\ 10),\quad \myd{2}{1}{s_1}{n} = 5,\quad \myd{3}{1}{s_1}{n} = 5.
\end{equation}
We plot the CCE-gap $\max_{m\in[M]} (V^{\dagger,\pi^k_{-m,1}}_{m,1} - V^{\pi^k_1}_{m,1})(s_1)$ of every episode $k$ in Figure~\ref{fig:sim} (left). DA-MAVL (Our Method) achieves satisfying convergence results, which aligns with our intuition in section~\ref{sec:alg}. Contrarily, the naive algorithm (Naive Method) fails to converge in limited episodes.

We then show the influence of the delays on DA-MAVL in Figure~\ref{fig:sim} (center). Delay sequence 2 and delay sequence 3 are four and nine times the value of delay sequence 1. 
The numeric result matches Theorem~\ref{thm:comp} in two ways: i). The CCE-gap of our output policy converges to $0$; ii). The CCE-gap of our output policy is positively related to $d_{max}$.

\vspace{5pt}\noindent\textbf{Infinite Delays.}
For the simulations of infinite delays, we set the delays as follows:
\begin{equation}
    \myd{1}{1}{s_1}{n} = \left\{\begin{array}{ll}
        \infty &, n\ \text{mod}\ 10 \leq 5 \\
        0 &, else\\
    \end{array}\right.,\quad \myd{2}{1}{s_1}{n} = 5, \quad \myd{3}{1}{s_1}{n} = 5.
\end{equation}
Namely, for agent $1$, there will be five infinite delays every ten visits of $s_1$. The numerical results for skipping metric in 
DA-MAVL with Reward Skipping (Algorithm~\ref{alg:damavl_skip}) (Our Method), skipping metric in previous work~\citep{rw_bandit2} (Previous Method) and no skipping scheme in DA-MAVL (Algorithm~\ref{alg:damavl})(Naive Method) is shown in Figure~\ref{fig:sim} (right). As is suggested in the figure, the algorithm without delay skipping will not converge because of the infinite delays. However, our algorithm with method skips all infinite delays and behaves as if they do not exist, and enjoys superiority over previous skipping methods in the setting of this paper.

\section{Algorithms and Subroutines}

\subsection{Subroutines for DA-MAVL Training (Algorithm \ref{alg:damavl})}\label{sec:sub_damavl}
\begin{algorithm2e}[H]\label{alg:damavl_sub1}
    \caption{Subroutine VALUE\_UPDATE$_{m,h,s}$ for agent $m$ for Algorithm~\ref{alg:damavl}}
    \KwInit{$n \gets 0$, $\myVOvertil{m}{h}{s}{} \gets H+1-h$, $\myVUndtil{m}{h}{s}{} \gets 0$; $\myVOver{m}{h}{s}{} \gets H+1-h$; $\myVUnd{m}{h}{s}{} \gets 0$\;}
    Receive $\mathcal{F}$, $\mysn{}$\;
        
    $\myVOvertil{m}{h}{s}{} \gets \myVOvertil{m}{h}{s}{} - \mybetaOver{m}{h}{s}{n}$;
    $\myVUndtil{m}{h}{s}{} \gets \myVUndtil{m}{h}{s}{} + \mysbetaUnd{n}$; 
    $n\gets \mysn{}$\;
    \For{$(i, a, \hat{\pi}, \overline{V}', \underline{V}', r) \in \mathcal{F}$}{
        
        % \ \ $\myaT{h}{s}{i} \gets \myaT{h}{s}{i-1} + \myatau{h}{s}{i}$\;
        $\myVOvertil{m}{h}{s}{} \gets (1-\alpha_i)\myVOvertil{m}{h}{s}{} + \alpha_i (r + \overline{V}')$\;
        % + \mybetaOver{m}{h}{s}{i}-\mybetaOver{m}{h}{s}{i-1}\cdot(1-\alpha_i)$\;
        $\myVUndtil{m}{h}{s}{} \gets (1-\alpha_i)\myVUndtil{m}{h}{s}{} + \alpha_i (r + \underline{V}')$\;
    }
    $\myVOvertil{m}{h}{s}{} \gets \myVOvertil{m}{h}{s}{} + \mybetaOver{m}{h}{s}{n}$; $\myVUndtil{m}{h}{s}{} \gets \myVUndtil{m}{h}{s}{} - \mysbetaUnd{n}$\;
    $\myVOver{m}{h}{s}{} \gets \min\big\{ H+1-h, \myVOvertil{m}{h}{s}{}, \myVOver{m}{h}{s}{} \big\}$;$\myVUnd{m}{h}{s}{} \gets \max\big\{ 0, \myVUndtil{m}{h}{s}{}, \myVUnd{m}{h}{s}{}\big\}$\;
\end{algorithm2e}

\begin{algorithm2e}[H]
    \caption{Subroutine POLICY\_OPT$_{m,h,s}$ for agent $m$ for Algorithm~\ref{alg:damavl}}
    \label{alg:damavl_sub2}
    \KwInit{$\forall a\in \mathcal{A}, \myLhat{m}{h}{s}{}{a}\gets 0$\;}
    Receive $\mathcal{F}, \mysnpr{}$\;  
    \For{$(i, a, \hat{\pi}, \overline{V}', \underline{V}', r) \in \mathcal{F}$}{
        \For(){$a'\in \mathcal{A}$}{
            $\mylhat{m}{h}{s}{i}{a'} = \mathbb{I}(a'=a) \Big[\frac{H - r - \overline{V}'}{H}\Big] \Big/ \Big[\hat{\pi} + \mygamma{m}{h}{s}{i}\Big]$\;
            $\myLhat{m}{h}{s}{}{a'} = \myLhat{m}{h}{s}{}{a'} + w_i \mylhat{m}{h}{s}{i}{a'}$\;    
        }
        $\hat{\pi}_h(\cdot|s) \propto \exp\big(-(\myeta{m}{h}{s}{\mysnpr{}}/w_{\mysnpr{}}) \myLhat{m}{h}{s}{}{\cdot}\big)$\;
    }
    
    Empty $\mathcal{F}$\;
\end{algorithm2e}

\noindent Parameters for the above subroutines are defined as follows:
\begin{equation}\label{eq:subParams}\begin{split}
    &\alpha_n = \dfrac{H+1}{H+n}, \ n\geq 1, \quad  \mygamma{m}{h}{s}{n} = \myeta{m}{h}{s}{n} = \sqrt{\dfrac{\iota}{nA+\myT{m}{h}{s}{n}} },\ \ n\geq 1,\\
    % &\overline{\beta}_{h,s,n} = \left\{\begin{array}{ll}
    %     0, & n=0\\
    %     12H^2 \sqrt{\dfrac{nA+\mysT{n}}{n^2}\iota} +  \dfrac{6d_{max}H^2}{n} \iota, & n \geq 1\\
    % \end{array}
    % \right.\\.
    & \mybetaOver{m}{h}{s}{n} = \left\{\begin{array}{lr}
        12H^2 \sqrt{\dfrac{nA+\myT{m}{h}{s}{n}}{n^2}\iota} +  4H^2\dfrac{d_{max}}{n} \iota, &n\geq 1\\
        0, &n=0\\
    \end{array}\right.,\\
    & \mysbetaUnd{n} = \left\{\begin{array}{lr}
        2\sqrt{\dfrac{H^3}{n}\iota} + 2H^2\dfrac{d_{max}}{n}\iota,\ \  &n\geq 1\\
        0, &n=0\\
    \end{array}\right.,\quad w_n = \left\{\begin{array}{lr}
        \alpha_n \prod_{i=2}^n (1-\alpha_i)^{-1}, &n \geq 2 \\
        1, &n = 1\\
    \end{array}\right.,
    % &\mygamma{m}{h}{s}{0} = \myeta{m}{h}{s}{0} = 1\\ 
\end{split}\end{equation}
where $\log ( 4MHSAK /\delta)$, and $\myT{m}{h}{s}{n}$ is a parameter maintained by Algorithm~\ref{alg:damavl}. Note that here $\mysbetaUnd{n}$ does not depend on $(m,h,s)$.

\subsection{Naive Multi-Agent V-Learning (Naive-MAVL)}\label{sec:sub_naive}

\vspace{-9pt}\begin{algorithm2e}[h]
    \caption{Naive-MAVL Training for Agent $m$}
    \label{alg:damavl_naive}
    \KwInit{$\forall (s,h)$, $\mynpr{m}{h}{s}{0} \gets 0$; $\myn{m}{h}{s}{0} \gets 0$; $\myT{m}{h}{s}{0} \gets 0$; $\myaF{m}{h}{s}\gets \emptyset$; $\myaMm{m}{h}{s}\gets \emptyset$\;}

    \For(){Episode\ \  $k = 1, \dots, K$}{
        Receive initial state $s^k_1$\;
        \For(){Step\ \  $h = 1, \dots, H$}{
            \textcolor{cyan}{//Preparation}\\
            $s \gets s^k_h$;
            $\mysnpr{} \gets \mynpr{m}{h}{s}{k} = \mynpr{m}{h}{s}{k-1} + 1$; $\mysn{}\gets \myn{m}{h}{s}{k} = \myn{m}{h}{s}{k-1} + |\myaF{m}{h}{s}|$\;

            $\myT{m}{h}{s}{\mysnpr{}} \gets \myT{m}{h}{s}{\mysnpr{}-1} + |\myM{m}{h}{s}{}|$\;
            % \textcolor{cyan}{// Preparation}\\
            % $n \gets n_{h,s,k} = \mathbb{I}\big\{\myaM{h}{s}\neq \emptyset\big\}\cdot \big(\arg\min \{\myaM{h}{s}\}-1\big) + \mathbb{I}\big\{\myaM{h}{s}= \emptyset\big\}\cdot n'$\;
            % $n_{f} = \mathbb{I}\big\{\myaF{h}{s}\neq \emptyset\big\}\cdot \arg\min \{\myaF{h}{s}\} + \mathbb{I}\big\{\myaF{h}{s}= \emptyset\big\}\cdot \big(n+1\big)$\;
            % $\mathcal{T}_{h,s,n'} = \mathcal{T}_{h,s,n'-1} + (n'-n)$\;

            \textcolor{cyan}{// Learning}\\
            $\myVOver{m}{h}{s}{k}, \myVUnd{m}{h}{s}{k} \gets \text{VALUE\_UPDATE}_{m,h,s}\big(\myaF{m}{h}{s}, \mysn{}\big)$\;
            $\hat{\pi}_h^k(\cdot|s) \gets \text{POLICY\_OPT}_{m,h,s}\big(\myaF{m}{h}{s}, \mysn{}, \mysnpr{}\big)$\;   
            \textcolor{cyan}{// Sampling}\\
            Take action $a_h^k \sim \hat{\pi}_h^k(\cdot | s)$;\ \ Observe next state $s_{h+1}^k$\;
            \For(){$s'\in \mathcal{S}\backslash s$}{
                $\mynpr{m}{h}{s'}{k} \gets \mynpr{m}{h}{s'}{k-1}$;\ \ $\myn{m}{h}{s'}{k} \gets \myn{m}{h}{s'}{k-1}$\;
                $\myVOver{m}{h}{s}{k} \gets \myVOver{m}{h}{s}{k-1}$;\ \  $\myVUnd{m}{h}{s}{k} \gets \myVUnd{m}{h}{s}{k-1}$;\ \  $\hat{\pi}^k_{m,h}(\cdot|s)\gets \hat{\pi}^{k-1}_{m,h}(\cdot|s)$\;
            }}
        % \For(){Step $h = 1, \dots, H$}{
        %     Save $\big(\mynpr{m}{h}{s^k_h}{k}, a_h^k, \hat{\pi}_h^k(a_h^k|s^k_h), 
        %     \myVOver{m}{h+1}{s^k_{h+1}}{k}, \myVUnd{m}{h+1}{s^k_{h+1}}{k}\big)$ to $\myaMm{m}{h}{s^k_h}$\;
        %     % Save $\mytau{m}{h}{s}{\mysnpr{}}$ to $\mathcal{M}em_{m,h}(s)$\;
        % }
        % Receive delayed reward for all states $s$\;
        % \For(){Delayed Reward $(m, h, s, i, r)$}{
        %     Extract and remove $\big(i, a, \hat{\pi}, \overline{V}', \underline{V}'\big)$ from $\myaMm{m}{h}{s}$\;
        %     Save $\big(i, a, \hat{\pi}, \overline{V}', \underline{V}', r\big)$ to $\myaF{m}{h}{s}$\;
        % }
    \For(){Step $h = 1, \dots, H$}{
            Save $\big(\mynpr{m}{h}{s^k_h}{k}, a_h^k, \hat{\pi}_h^k(a_h^k|s^k_h), 
            \myVOver{m}{h+1}{s^k_{h+1}}{k}, \myVUnd{m}{h+1}{s^k_{h+1}}{k}\big)$ to $\myaMm{m}{h}{s^k_h}$\;
            % Save $\mytau{m}{h}{s}{\mysnpr{}}$ to $\mathcal{M}em_{m,h}(s)$\;
        }
        Receive delayed reward for all states $s$\;
        \For(){Delayed Reward $(m, h, s, i, r)$}{
            Extract and remove $\big(i, a, \hat{\pi}, \overline{V}', \underline{V}'\big)$ from $\myaMm{m}{h}{s}$\;
            Save $\big(i, a, \hat{\pi}, \overline{V}', \underline{V}', r\big)$ to $\myaF{m}{h}{s}$\;
        }
    }
\end{algorithm2e}

\begin{algorithm2e}[H]
    \caption{Execution of Output Policy $\pi_{m}$ for Naive-MAVL}
    Sample $k \sim \text{Uniform}([K])$\;
    \For(){step\ \ $h=1, \dots, H$}{
        Observe current state $s_h$; $n \gets \myn{m}{h}{s_h}{k}$\;
        Sample $i$ from $[n]$ with probability $\alpha_n^i$; $k \gets \myk{m,h}{s_h}{i}$\;
        Take action $a_{m,h} \sim \hat{\pi}_{m,h}^k(\cdot|s_h)$\;
    }
\end{algorithm2e}
\noindent Notice that the random samples can not be shared across all agents, because the visits are used in different orders for different agents in the subroutines. 

\begin{algorithm2e}[H]
    \caption{Subroutine VALUE\_UPDATE$_{m,h,s}$ for agent $m$ for Algorithm~\ref{alg:damavl_naive}}
    \label{alg:damavlnaive_sub1}
    \KwInit{$n \gets 0$; $\myVOvertil{m}{h}{s}{} \gets H+1-h$; $\myVUndtil{m}{h}{s}{} \gets 0$; $\myVOver{m}{h}{s}{} \gets H+1-h$; $\myVUnd{m}{h}{s}{} \gets 0$\;}
    Receive $\mathcal{F}, \mysn{}, \mysnpr{}$\;
    $\myVOvertil{m}{h}{s}{} \gets \myVOvertil{m}{h}{s}{} - \mybetaOver{m}{h}{s}{n}$; $\myVUndtil{m}{h}{s}{} \gets \myVUndtil{m}{h}{s}{} + \mysbetaUnd{n}$; $n \gets \mysn{} - |\mathcal{F}|$\;
    \For{$(j, a, \hat{\pi}, \overline{V}', \underline{V}', r) \in \mathcal{F}$}{
        $n\gets n+1$\;
        % \ \ $\myaT{h}{s}{i} \gets \myaT{h}{s}{i-1} + \myatau{h}{s}{i}$\;
        $\myVOvertil{m}{h}{s}{} \gets (1-\alpha_n)\myVOvertil{m}{h}{s}{} + \alpha_n \Big(r + \overline{V}'\Big)$\;
        $\myVUndtil{m}{h}{s}{} \gets (1-\alpha_n)\myVUndtil{m}{h}{s}{} + \alpha_n \Big(r + \underline{V}'\Big)$\;}
    $\myVOvertil{m}{h}{s}{} \gets \myVOvertil{m}{h}{s}{} + \mybetaOver{m}{h}{s}{n}$; $\myVUndtil{m}{h}{s}{} \gets \myVUndtil{m}{h}{s}{} - \mysbetaUnd{n}$\;
    $\myVOver{m}{h}{s}{} \gets \min\big\{ H+1-h, \myVOvertil{m}{h}{s}{}, \myVOver{m}{h}{s}{} \big\}$;$\myVUnd{m}{h}{s}{} \gets \max\big\{ 0, \myVUndtil{m}{h}{s}{}, \myVUnd{m}{h}{s}{}\big\}$\;
\end{algorithm2e}

\begin{algorithm2e}[H]
    \caption{Subroutine POLICY\_OPT$_{m,h,s}$ for agent $m$ for Algorithm~\ref{alg:damavl_naive}}
    \label{alg:damavlnaive_sub2}
    \KwInit{$\forall a\in \mathcal{A}, \myLhat{m}{h}{s}{}{a}\gets 0$\;}
    Receive $\mathcal{F}, \mysn{}, \mysnpr{}$; $n \gets \mysn{} - |\mathcal{F}|$\;
    \For{$(j, a, \hat{\pi}, \overline{V}', \underline{V}', r) \in \mathcal{F}$}{
        $n \gets n+1$\;
        \For(){$a'\in \mathcal{A}$}{
            $\mylhat{m}{h}{s}{n}{a'} = \mathbb{I}(a'=a) \Big[\frac{H - r - \overline{V}'}{H}\Big] \Big/ \Big[\hat{\pi} + \mygamma{m}{h}{s}{j}\Big]$\;
            $\myLhat{m}{h}{s}{}{a'} = \myLhat{m}{h}{s}{}{a'} + w_{n} \mylhat{m}{h}{s}{n}{a'}$\;    
        }
    }
    $\hat{\pi}_h(\cdot|s) \propto \exp\big(-(\myeta{m}{h}{s}{\mysnpr{}}/w_{\mysnpr{}}) \myLhat{m}{h}{s}{}{\cdot}\big)$\;        
    Empty $\mathcal{F}$\;
\end{algorithm2e}

All parameters share the same definition with those in previous subsection.

\subsection{DA-MAVL with Reward Skipping}\label{sec:sub_damavl_skip}
\begin{algorithm2e}
    \caption{DA-MAVL Training with Reward Skipping for Agent $m$}
    \label{alg:damavl_skip}
    \KwInit{$\forall (s,h)$, $\mynpr{m}{h}{s}{0} \gets 0$, $\myn{m}{h}{s}{0} \gets 0$, $\myT{m}{h}{s}{0} \gets 0$, $\myaF{m}{h}{s}\gets \emptyset$, $\myaM{m}{h}{s}\gets \emptyset$\;}

    \For(){Episode\ \  $k = 1, \dots, K$}{
        Receive initial state $s^k_1$\;
        \For(){Step\ \  $h = 1, \dots, H$}{
            $s \gets s^k_h$\;
            % \For(){$(n,a,\hat{\pi},\overline{V}', \underline{V}',r) \in \myaMp{m}{h}{s}$}{
            %     \If{$\forall i<n, i \notin \arg \{\myaMm{m}{h}{s}\}$}{
            %         Save $(n,a,\hat{\pi},\overline{V}', \underline{V}',r)$ to $\myaF{m}{h}{s}$;
            %         Remove $(n,a,\hat{\pi},\overline{V}', \underline{V}')$ from $\myaMp{m}{h}{s}$\;
            %     }
            % }
            \textcolor{cyan}{// Skipping}\\
            $\mysnpr{} \gets \mynpr{m}{h}{s}{k} = \mynpr{m}{h}{s}{k-1} + 1$;
            $\mysn{} \gets \myn{m}{h}{s}{k} = \myn{m}{h}{s}{k-1}$\;
            \textcolor{purple}{
            % $i_{0}\gets \min_i (i, \dots)\in \{\myaMm{m}{h}{s}\}$\;
            % $\myphi{m}{h}{s}{i_{0}}{\mysnpr{}} \gets |\{j:(j, \dots)\in \myaMp{m}{h}{s}\wedge j> i_0\}|$\;
            \For{$(i, a, \hat{\pi}, \overline{V}', \underline{V}', r) \in \myaMm{m}{h}{s}$}{
                $\myphi{m}{h}{s}{i}{\mysnpr{}} \gets \myphi{m}{h}{s}{i}{\mysnpr{}} + \mysnpr{}-i$\;
                \If{$ \myphi{m}{h}{s}{i}{\mysnpr{}} >  \sqrt{\myTtil{m}{h}{s}{\mysnpr{}}}$}{
                    Save $(i,a,\hat{\pi},H,0,0)$ to $\myaF{m}{h}{s}$; Remove $(i, a,\hat{\pi},\overline{V}',\underline{V}')$ from $\myaMm{m}{h}{s}$\;
                    $\mysn{} \gets \mysn{} + 1$\;
                }
            }}
            \textcolor{cyan}{// Preparation}\\
            \textcolor{purple}{
            \For(){$(i,a,\hat{\pi},\overline{V}', \underline{V}',r) \in \myaMp{m}{h}{s}$}{
                \If{$\forall j<i, j \notin \arg \{\myaMm{m}{h}{s}\}$}{
                    Save $(i,a,\hat{\pi},\overline{V}', \underline{V}',r)$ to $\myaF{m}{h}{s}$;
                    Remove $(i,a,\hat{\pi},\overline{V}', \underline{V}')$ from $\myaMp{m}{h}{s}$\;
                }
            }}

            % $\mytau{m}{h}{s}{\mysnpr{}} \gets |\myM{m}{h}{s}{}|$\;
            $\myT{m}{h}{s}{\mysnpr{}} \gets \myT{m}{h}{s}{\mysnpr{}-1} + |\myM{m}{h}{s}{}|$\;
            \textcolor{cyan}{// Learning}\\
            $\myVOver{m}{h}{s}{k}, \myVUnd{m}{h}{s}{k} \gets \text{VALUE\_UPDATE}_{m,h,s}\big(\myaF{m}{h}{s}, \mysn{}, \mysnpr{}\big)$\;
            $\hat{\pi}_h^k(\cdot|s) \gets \text{POLICY\_OPT}_{m,h,s}\big(\myaF{m}{h}{s}, \mysn{}, \mysnpr{}\big)$\;   
            \textcolor{cyan}{// Execution}\\
            Take action $a_h^k \sim \hat{\pi}_h^k(\cdot | s)$;\ \ Observe next state $s_{h+1}^k$\;
            \For(){$s'\in \mathcal{S}\backslash s$}{
                $\mynpr{m}{h}{s'}{k} \gets \mynpr{m}{h}{s'}{k-1}$; $\myn{m}{h}{s'}{k} \gets \myn{m}{h}{s'}{k-1}$\;
                $\myVOver{m}{h}{s'}{k} \gets \myVOver{m}{h}{s'}{k-1}$; $\myVUnd{m}{h}{s'}{k} \gets \myVUnd{m}{h}{s'}{k-1}$; $\hat{\pi}^k_{m,h}(\cdot|s')\gets \hat{\pi}^{k-1}_{m,h}(\cdot|s')$\;
            }
            }
        \For(){Step $h = 1, \dots, H$}{
            Save $\big(\mynpr{m}{h}{s^k_h}{k}, a_h^k, \hat{\pi}_h^k(a_h^k|s^k_h), 
            \myVOver{m}{h+1}{s^k_{h+1}}{k}, \myVUnd{m}{h+1}{s^k_{h+1}}{k}\big)$ to $\myaMm{m}{h}{s^k_h}$\;
            % Save $\mytau{m}{h}{s}{\mysnpr{}}$ to $\mathcal{M}em_{m,h}(s)$\;
        }
        Receive delayed reward for all states $s$\;
        \For(){Delayed Reward $(m, h, s, i, r)$}{
            \textcolor{purple}{\If{$\big(i, a, \hat{\pi}, \overline{V}', \underline{V}'\big)\notin\myaMm{m}{h}{s}$}{
                continue\;
            }}
            Extract and remove $\big(i, a, \hat{\pi}, \overline{V}', \underline{V}'\big)$ from $\myaMm{m}{h}{s}$\;
            Save $\big(i, a, \hat{\pi}, \overline{V}', \underline{V}', r\big)$ to $\myaMp{m}{h}{s}$\;
        }
    }
\end{algorithm2e}

\begin{algorithm2e}
    \caption{Subroutine VALUE\_UPDATE$_{m,h,s}$ for agent $m$ for Algorithm~\ref{alg:damavl_skip}}
    \label{alg:damavlskip_sub1}
    \KwInit{$n \gets 0$, $n' \gets 0$, $\myVOvertil{m}{h}{s}{} \gets 0$, $\myVUndtil{m}{h}{s}{} \gets 0$; $\myVOver{m}{h}{s}{} \gets 0$; $\myVUnd{m}{h}{s}{} \gets 0$\;}
    \For{Episode $k = 1, \dots, K$}{
        Receive $\mathcal{F}$, $\mysn{}$, $\mysnpr{}$\;
        
        $\myVOvertil{m}{h}{s}{} \gets \myVOvertil{m}{h}{s}{} - \mybetaOver{m}{h}{s}{n,n'}$; $\myVUndtil{m}{h}{s}{} \gets \myVUndtil{m}{h}{s}{} + \mybetaUnd{m}{h}{s}{n,n'}$; $n\gets \mysn{}$; $n'\gets \mysnpr{}$\;
        \For{$(i, a, \hat{\pi}, \overline{V}', \underline{V}', r) \in \mathcal{F}$}{
            
            % \ \ $\myaT{h}{s}{i} \gets \myaT{h}{s}{i-1} + \myatau{h}{s}{i}$\;
            $\myVOvertil{m}{h}{s}{} \gets (1-\alpha_i)\myVOvertil{m}{h}{s}{} + \alpha_i \Big(r + \overline{V}'\Big)$\;
            % + \mybetaOver{m}{h}{s}{i}-\mybetaOver{m}{h}{s}{i-1}\cdot(1-\alpha_i)$\;
            $\myVUndtil{m}{h}{s}{} \gets (1-\alpha_i)\myVUndtil{m}{h}{s}{} + \alpha_i \Big(r + \underline{V}'\Big)$\;
        }
        $\myVOvertil{m}{h}{s}{} \gets \myVOvertil{m}{h}{s}{} + \mybetaOver{m}{h}{s}{n,n'}$;$\myVUndtil{m}{h}{s}{} \gets \myVUndtil{m}{h}{s}{} - \mybetaUnd{m}{h}{s}{n,n'}$\; 
        % $\myVUndtil{m}{h}{s}{} \gets \myVOvertil{m}{h}{s}{} + \mybetaOver{m}{h}{s}{n} + \mybetaTil{m}{h}{s}{n'}$\;
        $\myVOver{m}{h}{s}{} \gets \min\Big\{ H+1-h, \myVOvertil{m}{h}{s}{}, \myVOver{m}{h}{s}{} \Big\}$;$\myVUnd{m}{h}{s}{} \gets \max\Big\{ 0, \myVUndtil{m}{h}{s}{}, \myVUnd{m}{h}{s}{}\Big\}$\;
    }
\end{algorithm2e}
\noindent Algorithm~\ref{alg:damavl_skip} shares the same subroutine `POLICY\_OPT$_{m,h,s}$' with Algorithm~\ref{alg:damavl}. The definition of `VALUE\_UPDATE$_{m,h,s}$' is presented in Subroutine~\ref{alg:damavlskip_sub1}. Parameters for Subroutine~\ref{alg:damavlskip_sub1} share the same definitions with those in  Subroutine~\ref{alg:damavl_sub1} except:
\begin{equation}\label{eq:subParam_skip}\begin{split}
    &\mybetaOver{m}{h}{s}{n,n'} = \left\{\begin{array}{lr}
        24H^2C \sqrt{\dfrac{\myT{m}{h}{s}{n'}}{n^2}}\iota +  18H^2\sqrt{\dfrac{A}{n}}\iota,\ \ &n\geq 1\\
        0, &n=0\\
    \end{array}\right.,\\
    &\mybetaUnd{m}{h}{s}{n,n'} = \left\{\begin{array}{lr}
        2H^2 \dfrac{\sqrt[4]{4\myT{m}{h}{s}{n'}}+2}{n} +  2\sqrt{\dfrac{H^3}{n}\iota}, &n\geq 1\\
        0, &n=0\\
    \end{array}\right.,\\
    &\mygamma{m}{h}{s}{n} = \myeta{m}{h}{s}{n} = \sqrt{\dfrac{\iota}{nA+\myT{m}{h}{s}{n}} },\ \ n\geq 1.\\
\end{split}\end{equation}
where $\myT{m}{h}{s}{n}$ is a parameter maintained by Algorithm~\ref{alg:damavl_skip}.

\section{Notations}\label{sec:notation}

In this section, we summarize and introduce the important notations. 

Recall $\log ( 4MHSAK /\delta)$. Recall $\{\alpha_n\}_{n\in\mathbb{N}}$ and $\{w_n\}_{n\in\mathbb{N}}$ as defined in Equation~\eqref{eq:subParams}. We also define an auxiliary sequence $\{\alpha_n^i\}_{i\in[n], n\in\mathbb{N}}$ as follows:
\begin{equation}\label{eq:alpha}
    \alpha^i_n = \left\{\begin{array}{lr}
        \alpha_i \prod_{j=i+1}^n (1-\alpha_j), &n > i \geq 1\\
        \alpha_i, & n = i \geq 1\\ 
    \end{array}\right., \ \ 
    \alpha^0_n = \left\{\begin{array}{lr}
        0, & n > 0\\
        1, & n = 0\\ 
    \end{array}\right..
\end{equation}
We summarize the important properties of this sequence in Lemma~\ref{lem:comp1_base1}.

% Recall that for any pair $(h,s)$, the visits can be sorted in the order that they happened or arrived. 
% For the $i$-th arrived visit, $\phi_{h,s,i}$ denotes it is the $\phi_{h,s,i}$-th happened visit. 

Consider agent $m$ and $(h,s)$. $\myn{m}{h}{s}{k}$ denotes the count of usable visits, $\mynpr{m}{h}{s}{k}$ denotes the count of happened visits at episode $k$. 
$\pi_{m,h}^k(\cdot|s)$, $a_{m,h}^k$ and $r_{m,h}^k$ denote the policy, action and reward at episode $k$. $\myMm{m}{h}{s}{k}$, $\myMp{m}{h}{s}{k}$ and $\myF{m}{h}{s}{k}$ denote the set of \textbf{unreceived visits}, \textbf{unusable visits} and \textbf{visits to be used} at the beginning of episode $k$. In other words, they refer to sets $\myaMm{m}{h}{s}$, $\myaMp{m}{h}{s}$ and $\myaF{m}{h}{s}$ before the `Learning' process of episode $k$. We also let $\myM{m}{h}{s}{k} = \myMm{m}{h}{s}{k} \cup \myMp{m}{h}{s}{k}$. When $(m,h,s)$ is fixed in the context, the above notations will be abbreviated as $\mysn{k}$, $\mysnpr{k}$, $\pi_k(\cdot)$, $a_k$, $r_k$, $\mysM{k}=\mysMm{k}\cup\mysMp{k}$ and $\mysF{k}$. 

Consider agent $m$ and the $n$-th visit of $(h,s)$. $\myk{h}{s}{n}$ denotes the episode when it happens. $\mye{m}{h}{s}{n}$ denotes the earliest unreceived visit when it happens. 
For this visit, parameters $\myeta{m}{h}{s}{n}$, $\mygamma{m}{h}{s}{n}$, $\mybetaOver{m}{h}{s}{n}$, $\mysbetaUnd{n}$ are maintained by Algorithm~\ref{alg:damavl} and its subroutines (Algorithm~\ref{alg:damavl_sub1} and \ref{alg:damavl_sub2}). Parameters $\myeta{m}{h}{s}{n}$, $\mygamma{m}{h}{s}{n}$, $\{\mybetaOver{m}{h}{s}{i,n}\}_{i\in[n]}$, $\{\mybetaUnd{m}{h}{s}{i,n}\}_{i\in[n]}$ and $\{\myphi{m}{h}{s}{i}{n}\}_{i\in[n]}$ are maintained by the Algorithm~\ref{alg:damavl_skip} and its subroutines (Algorithm~\ref{alg:damavlskip_sub1} and \ref{alg:damavl_sub2}). It is worth notice that the skipping metric $\myphi{m}{h}{s}{i}{n}$ in Algorithm~\ref{alg:damavl_skip} can be written as:
\begin{equation}\label{eq:skip_phidef}\begin{split}
    \myphi{m}{h}{s}{i}{n} = \sum_{j=i+1}^{n} (j-i) \cdot \mathbb{I}\{i\in \arg\myM{m}{h}{s}{k_j}\}
\end{split}\end{equation}
When $(m,h,s)$ is fixed in the context, the above notations are abbreviated as $k_n$, $\myse{n}$ $\eta_n$, $\gamma_n$, $\{\mysbetaOver{i,n}\}_{i\in[n]}$, $\{\mysbetaUnd{i,n}\}_{i\in[n]}$ and $\{\mysphi{i}{n}\}_{i\in[n]}$.

For pair $(m,h,s,n)$, we also define three bandit losses $\mylhat{m}{h}{s}{n}{a}$, $\myl{m}{h}{s}{n}{a}$ and $\mylbar{m}{h}{s}{n}{a}$:
\begin{equation}\begin{split}
    \mylhat{m}{h}{s}{n}{a} &= \dfrac{1}{H}\mathbb{I}(a=a^{k_n}_{m,h}) \Big[H - r_{m,h}^{k_n} - \overline{V}^{k_n}_{m,h+1}(s_{h+1}^{k_n})\Big] \Big/ \Big[\hat{\pi}^{k_n}_{m,h}(a|s) + \mygamma{m}{h}{s}{n}\Big],\\
    \myl{m}{h}{s}{n}{a} & = \frac{1}{H} \Big[H - r_{m,h}(s,\bm{a}) - \overline{V}^{k_n}_{m,h+1}(s_{h+1}^{k_n}) \Big], \text{where}\quad \bm{a} = (a,a_{-m,h}^{k_n}),\\
    \mylbar{m}{h}{s}{n}{a} & = \frac{1}{H} \mathop{\mathbb{E}}_{\substack{\bm{a}=(a,a_{-m,h})\\ a_{-m,h} \sim \hat{\pi}_{-m,h}^{k_n},\ s'\sim \mathbb{P}_h(s,\bm{a})}} \Big[H - r_{m,h}(s,\bm{a}) - \overline{V}^{k_n}_{m,h+1}(s')\Big].\\
\end{split}\end{equation}
Here the first loss $\mylhat{m}{h}{s}{n}{a}$ is the bandit loss used in Subroutine~\ref{alg:damavl_sub2}, while the other two are its variants. When $(m,h,s)$ is fixed in the context, they can be abbreviated $\myslhat{n}{a}, \mysl{n}{a}$ and $\myslbar{n}{a}$. 

Consider agent $m$ and the first $n$ visits of $(h,s)$. Recall $\myR{m}{h}{s}{n}$ denotes the policy optimization regret. Recall $\myT{m}{h}{s}{n}$ denotes the count of unusable and unreceived visits. In Algorithm~\ref{alg:damavl}, it can be written as:
\begin{equation}
    \myT{m}{h}{s}{n} = \sum\limits_{i=1}^n i - \mathop{\arg\min}\limits_{j} \Big[ \myd{m}{h}{s}{j} + k_j \geq k_i\Big] = \sum\limits_{i=1}^n i - \mye{m}{h}{s}{n} = \sum_{i=1}^n |\myM{m}{h}{s}{k_i}|,
\end{equation}
where $\mye{m}{h}{s}{n}$ and $\myM{m}{h}{s}{k_i}$ refer to the variable and set related Algorithm~\ref{alg:damavl}.
In Algorithm~\ref{alg:damavl_skip} with reward skipping, we define the set of skipped visits of $(h,s)$ during the first $n$ visits of $(h,s)$ as $\myO{m}{h}{s}{n}$. Then $\myT{m}{h}{s}{n}$ can be written as:
\begin{equation}
    \myT{m}{h}{s}{n} = \sum\limits_{i=1}^n i - \mathop{\arg\min}\limits_{j\notin \myO{m}{h}{s}{i}} \Big[ \myd{m}{h}{s}{j} + k_j \geq k_i\Big] = \sum\limits_{i=1}^n i - \mye{m}{h}{s}{n} = \sum_{i=1}^n |\myM{m}{h}{s}{k_i}|,
\end{equation}
where $\mye{m}{h}{s}{n}$ and $\myM{m}{h}{s}{k_i}$ refer to the variable and set related Algorithm~\ref{alg:damavl_skip}. We also define $\myT{m}{h}{s}{n,\mathcal{L}}$ to denote the count of unreceived and unusable visits during the first $n$ visits of $(h,s)$, if the delays of visits in $\mathcal{L}$ are set to $0$:
\begin{equation}
    \myT{m}{h}{s}{n} = \sum\limits_{i=1}^{n} i - \mathop{\arg\min}\limits_{j \notin \mathcal{L}} \Big[ \myd{m}{h}{s}{j} + k_j \geq k_i\Big].
\end{equation}
When $(m,h,s)$ is fixed in the context, the above notations are abbreviated as $\mysR{n}$, $\mysO{n}$, $\mysT{n}$, $\mysT{n,\mathcal{L}}$.

Finally, we introduce policies $\{\pi_{m,h}^k\}_{m\in[M], h\in[H], k\in[K]}$ defined by their execution procedures:

\begin{algorithm2e}[H]\label{alg:damavl_outputMed}
    \caption{Policy Certification for $\pi^k_{m, h}$}
    \For(){Episode\ \ $h'=h, \dots, H$}{
        Observe current state $s_{h'}$;\ \ $n \gets \max_m \myn{m}{h'}{s_{h'}}{k}$\;
        Sample $i$ from $[n]$ with probability $\alpha_n^i$;\ \ $k \gets \myk{h'}{s_{h'}}{i}$\;
        Take action $a_{m,h'} \sim \hat{\pi}_{m,h'}^k(\cdot|s_{h'})$\;
    }
\end{algorithm2e}
\noindent While definitions of $\{\pi_{m,h}^k\}_{m\in[M], h\in[H], k\in[K]}$ and $\{\pi_{m}\}_{m\in[M]}$ are similar, their differences are two-fold: (1) $\pi_{m,h}^k$ begins from a given $k$ while $\pi_m$ begins by sampling a $k$ from $[K]$; (2) $\pi_{m,h}^k$ is for steps from $h$ to $H$ while $\pi_m$ is for steps from $1$ to $H$. This definition mainly follows the certification process in \cite{alg_ma_VL, alg_ma_VL2}. We refer the readers to their work for further details. 

\section{Performance Guarantee for DA-MAVL}\label{sec:proof_damavl}

As is stated in Section~\ref{sec:sketch1}, proof of Theorem~\ref{thm:comp} is under assumption~\ref{ass:dmax} and can be broken down into three steps.
\textbf{In step one}, we bound the policy optimization regret for Subroutine~\ref{alg:damavl_sub2} (Lemma~\ref{lem:reg}). 
\textbf{In step two}, we establish the optimism and pessimism of our value estimates in Subroutine~\ref{alg:damavl_sub1} (Lemma~\ref{lem:opt}).
\textbf{In step three}, we bound the gap between the optimistic and pessimistic value estimates and prove the main theorem (Theorem~\ref{thm:comp}).

\subsection{\textbf{Step One:} Proof of Lemma~\ref{lem:reg}}\label{sec:proof_comp_step1}

Consider any fixed pair $(m,h,s,n)$. Let $R_{n}$ denote $\myR{m}{h}{s}{n}$, and let $a^*$ denote the optimal action for the first $n$ visit of $(h,s)$. Utilizing notations $\mysl{i}{a}, \myslbar{i}{a}$, regret $R_n$ can be rewritten as:
\begin{equation}\begin{split}
    R_n =& \max\limits_{a\in \mathcal{A}_m} \sum\limits_{i=1}^{n} \alpha_{n}^i \Big[ \mathop{\mathbb{E}}_{\substack{\bm{a}=(a,a_{-m,h})\\ a_{-m,h} \sim \hat{\pi}_{-m,h}^{k_i},\ s'\sim \mathbb{P}_h(s,\bm{a})}} \Big(r_{m,h}(s,\bm{a}) + \myVOver{m}{h+1}{s'}{k_i}\Big)-\Big(r_{m,h}^{k_i} + \myVOver{m}{h+1}{s_{h+1}^{k_i}}{k_i} \Big)\Big]\\
    =& H \sum\limits_{i=1}^n \alpha_n^i \Big[ \mysl{i}{a_{k_i}} - \myslbar{i}{a^*}\Big].
\end{split}\end{equation}
We then decompose the regret as follows:
\begin{equation}\begin{split}
    R_n &= H \sum\limits_{i=1}^n \alpha_n^i \Big[ \mysl{i}{a_{k_i}} - \myslbar{i}{a^*}\Big]\\
    &= H \prod_{i=2}^n (1-\alpha_i) \cdot \sum\limits_{i=1}^n w_i \Big[ \mysl{i}{a_{k_i}} - \myslbar{i}{a^*}\Big]\\
    &= H \prod_{i=2}^n (1-\alpha_i) \cdot R'_n.\\
    % &= H^2d_{max}\cdot \alpha_n + H\cdot \sum\limits_{i=Hd_{max}}^n \alpha_n^i \Big[ \mysl{i}{a_{k_i}} - \myslbar{i}{a^*}\Big]\\
    % &\leq \frac{2H^3d_{max}}{n} + H \cdot \prod_{i=2}^n (1-\alpha_i) \cdot \sum\limits_{i=Hd_{max}}^n w_i \Big[ \mysl{i}{a_{k_i}} - \myslbar{i}{a^*}\Big]\\
    % &= \frac{2H^3d_{max}}{n} + H \cdot \prod_{i=2}^n (1-\alpha_i) \cdot R'_n\\
\end{split}\end{equation}

Recall that $\mysM{k_i}$ denotes the set of unusable or unreceived visits of $(h,s)$ for agent $m$ at the beginning of the $i$-th visit. We then define two cumulative losses for $i$:
\begin{equation}\begin{split}
    &\mysLhat{i}{a} = \sum\limits_{j\in [i-1]\backslash \mysM{k_i}} w_j \myslhat{j}{a},\\
    &\mysLtil{i+1}{a} = \sum\limits_{j\in [i]} w_j \myslhat{j}{a} = \mysLhat{i}{a} + w_i\myslhat{i}{a} + \sum\limits_{j\in \mysM{k_i}} w_j \myslhat{j}{a}.
\end{split}\end{equation}
Cumulative loss $\mysLhat{i}{a}$ generates policy $\hat{\pi}_{k_i}$ in algorithm \ref{alg:damavl}. Cumulative loss $\mysLtil{i+1}{a}$ is the cheating version of $\mysLhat{i}{a}$, which includes all information of the first $i$ visits. Then we have two corresponding policies: 
\begin{equation}\begin{split}
    \hat{\pi}_{k_i}(\cdot) &\propto \exp\Big[-(\eta_i/w_i) \mysLhat{i}{\cdot}\Big],\\
    \tilde{\pi}_{k_{i+1}}(\cdot) &\propto \exp\Big[-(\eta_i/w_i) \tilde{L}_{i+1}(\cdot)\Big].
\end{split}\end{equation}
Then the weighted regret $R'_n$ can be further decomposed as follows:
\begin{equation}\begin{split}
    R'_n = & \sum\limits_{i=1}^n w_i \Big[ \mysl{i}{a_{k_i}} - \myslbar{i}{a^*}\Big]\\
    = & \underbrace{\sum\limits_{i=1}^n w_i\Big[\mysl{i}{a_{k_i}} - \myslhat{i}{a_{k_i}} \hat{\pi}_{k_i}(a_{k_i}) \Big]}_{I_{1,1}} + \underbrace{\sum\limits_{i=1}^n w_i\Big[\myslhat{i}{a_{k_i}} \hat{\pi}_{k_i}(a_{k_i}) - \myslhat{i}{a_{k_i}} \tilde{\pi}_{k_{i+1}}(a_{k_i}) \Big]}_{I_{1,2}}\\
    & \hspace{2em} + \underbrace{\sum\limits_{i=1}^n w_i\Big[\myslhat{i}{a_{k_i}} \tilde{\pi}_{k_{i+1}}(a_{k_i}) - \myslhat{i}{a^*}\Big]}_{I_{1,3}} + \underbrace{\sum\limits_{i=1}^n w_i\Big[\myslhat{i}{a^*} - \myslbar{i}{a^*}\Big]}_{I_{1,4}}.\\
\end{split}\end{equation}
We then give the upper bounds for $I_{1,1}$, $I_{1,2}$, $I_{1,3}$, $I_{1,4}$ in the following lemmas.

\begin{mylemma}[The upper bound of $I_{1,1}+I_{1,2}$]\label{reg_12}
    For $\forall (m,h,s,n) \in [M]\times [H]\times \mathcal{S}\times [K]$, the following equation holds with probability at least $1-\delta/4$:
    \begin{equation}\begin{split}
        I_{1,1} + I_{1,2} \leq w_n d_{max} \iota + \sum\limits_{i=1}^n w_i \mygamma{m}{h}{s}{i} \Big(2A+|\myM{m}{h}{s}{\myk{h}{s}{i}}|\Big). \\
    \end{split}\end{equation}
\end{mylemma}

\begin{mylemma}[The upper bound of $I_{1,3}$]\label{reg_3}
    For $\forall (m,h,s,n) \in [M]\times [H]\times \mathcal{S}\times [K]$, the following equation holds:
    \begin{equation}\begin{split}
        I_{1,3} &\leq \frac{w_n}{\myeta{m}{h}{s}{n}} \iota.
    \end{split}\end{equation}        
\end{mylemma}

\begin{mylemma}[The upper bound of $I_{1,4}$]\label{reg_4}
    For $\forall (m,h,s,n) \in [M]\times [H]\times \mathcal{S}\times [K]$, the following equation holds with probability at least $1-\delta/4$:
    \begin{equation}\begin{split}
        I_{1,4} &\leq \frac{w_n}{\myeta{m}{h}{s}{n}} \iota.
    \end{split}\end{equation}        
\end{mylemma}

Combining the three lemmas and by union bound, the following equations hold for $\forall (m,h,s,n) \in [M]\times [H]\times \mathcal{S}\times [K]$ with probability at least $1-\delta/2$:
\begin{equation}\begin{split}
    R_n =& H \cdot \prod_{i=2}^n (1-\alpha_i) \cdot \Big(I_{1,1} + I_{1,2} + I_{1,3} + I_{1,4}\Big)\\
    \leq & H \alpha_n d_{max} \iota + H\sum\limits_{i=1}^n\alpha_n^i \gamma_i \Big(2A+|\mysM{k_i}|\Big) + 2H\dfrac{\alpha_n}{\eta_n}\iota\\
    % \leq & H \alpha_n d_{max} \iota + 2H\sum\limits_{i=1}^n\alpha_n^i \gamma_i \Big(A+|\mysM{k_i}|\Big) + H\dfrac{\alpha_n}{\eta_n}\iota\\
    \leq & \frac{4H^2}{n} \sum\limits_{i=1}^n \Big(A+|\mysM{k_i}|\Big)\sqrt{\dfrac{\iota}{iA+\mysT{i}}} + 4H^2 \sqrt{\dfrac{nA+\mysT{n}}{n^2}\iota} + \dfrac{2d_{max}H^2}{n}\iota\\
    \leq & 12H^2 \sqrt{\dfrac{nA+\mysT{n}}{n^2}\iota} +  \dfrac{2d_{max}H^2}{n} \iota.
\end{split}\end{equation}
Here the third line comes from Lemma~\ref{lem:comp1_base1}, the last line utilizes Lemma 1 in \cite{3_Lemma1}, and the fact that $\mysT{i} = \sum_{j=1}^i |\mysM{k_j}|$.

\subsubsection{Supporting Details}\label{sec:comp1_reg_supporting}
\begin{mylemma}\label{lem:comp1_base1}
    The following properties hold for $\forall n \geq i \geq 1$:
    \begin{itemize}
    \setlength\itemsep{0.5em}
        \item $\frac{1}{\sqrt{n}} \leq \sum_{i=1}^n \frac{\alpha^i_n}{\sqrt{n}} \leq  \frac{2}{\sqrt{n}}$\quad and\quad $\frac{1}{n} \leq \sum_{i=1}^n \frac{\alpha^i_n}{n} \leq  \frac{2}{n}$,
        \item $\max_{i\in[n]} \alpha_n^i \leq \frac{2H}{n}$,
        \item $\sum_{n=1}^{\infty} \alpha^i_n = 1+\frac{1}{H}$,
        \item $\alpha^i_n = w_i\prod_{j=2}^n (1-\alpha_j)  $.
    \end{itemize}
\end{mylemma}

\begin{proof}
    Here the first three lines is from Lemma 4.1 in \cite{alg_ma_VL}, while the last line is from Proof of Corollary 19 in \cite{alg_ma_VL}.
\end{proof}

\begin{mylemma}\label{lem:comp1_base2}
    The following property hold for $\forall (m,h,s,n)\in [M]\times [H]\times \mathcal{S}\times [K]$:
    \begin{equation}\label{eq:comp1_base2_1}
        |\myM{m}{h}{s}{\myk{h}{s}{n}}|  = n - \mye{m}{h}{s}{n} \leq d_{max}.
    \end{equation}
    Similar property hold for $\forall (m,h,s,k)\in [M]\times [H]\times \mathcal{S}\times [K]$:
    \begin{equation}\label{eq:comp1_base2_2}
        |\myM{m}{h}{s}{k}|  = \mynpr{m}{h}{s}{k} - \myn{m}{h}{s}{k} - 1 \leq d_{max}.
    \end{equation}
\end{mylemma}

\begin{proof}
    We first proof Equation~\eqref{eq:comp1_base2_1}. Consider any fixed pair $(m,h,s,n)$. Recall that $\mysk{n}$ denotes the episode when the $n$-th visit of $(h,s)$ happens, $\myse{n}$ is from Equation~\eqref{eq:mye} and \eqref{eq:mye_2}. 
    
    If $|\mysM{\mysk{n}}| = 0$, all first $n-1$ visits are received, which gives $\myse{n} = n$. The lemma clearly holds:
    \begin{equation}
         |\mysM{\mysk{n}}|  = n - \myse{n} = 0.
    \end{equation}
    
    If $|\mysM{\mysk{n}}| > 0$, $\myse{n}$ is the first unreceived visit. According to the definitions of unusable and unrecieved visits, all visits with happening order $\myse{n}, \dots, n-1$ are unusable and unreceived visits. Consequently the count of unusable and unreceived visits are $n-\myse{n}$. On the other hand, Algorithm~\ref{alg:damavl} ensures all unusable and unreceived visits are included in set $\mysM{\mysk{n}}$. Therefore,
    \begin{equation}
        |\mysM{\mysk{n}}|  = n - \myse{n} - 1.
    \end{equation}
    
    We prove the other half of the equation by contradiction. Suppose $\myse{n} \leq n - d_{max} -1$, then the $\myse{n}$-th visit has been delayed for at least $d_{max}+1$ episodes. This contradict with Assumption~\ref{ass:dmax}. Therefore $\myse{n} \geq n - d_{max}$, which completes the proof. 
    
    We now prove Equation~\eqref{eq:comp1_base2_2}. At episode $k$, visits with happening order $1, \dots, \mysn{k}$ are usuable visits, while visits with happening order $\mysn{k}+1, \dots, \mysnpr{k}-1$ are unusable or unreceived visits. This directly implies:
    \begin{equation}
        |\mysM{k}| = \mysnpr{k} - \mysn{k}.
    \end{equation}
    The other half of the equation follows directly from Equation~\eqref{eq:comp1_base2_1}.
\end{proof}

\begin{mylemma}\label{lem:comp1_reg1}    For $\forall (m,h,s,n) \in [M]\times [H]\times \mathcal{S}\times [K]$, the following inequality holds:
    \begin{equation}\begin{split}
        I_{1,1} \leq \sum\limits_{i=1}^n \sum\limits_{a\in \mathcal{A}_m} w_i \mygamma{m}{h}{s}{i} \mylhat{m}{h}{s}{i}{a}.
    \end{split}\end{equation}        
\end{mylemma}

\begin{proof}[Proof of Lemma \ref{lem:comp1_reg1}]
    Consider any fixed pair of $(m,h,s,n)$.
    \begin{equation}\begin{split}
        I_{1,1} &= \sum\limits_{i=1}^n w_i \Big[\mysl{i}{a_{k_i}} - \myslhat{i}{a_{k_i}} \hat{\pi}_{k_i}(a_{k_i})\Big]\\
        &= \sum\limits_{i=1}^n w_i \Big[\hat{\pi}_{k_i}(a_{k_i})+\gamma_i-\hat{\pi}_{k_i}(a_{k_i})\Big]\myslhat{i}{a_{k_i}}\\
        &= \sum\limits_{i=1}^n w_i \gamma_i \myslhat{i}{a_{k_i}}\\ 
        &= \sum\limits_{i=1}^n \sum\limits_{a\in \mathcal{A}_m} w_i \gamma_{i} \myslhat{i}{a}.\\
        % &= \sum\limits_{i=1}^n \sum\limits_{a\in \mathcal{A}_m} w_i \gamma_i \myslhat{i}{a}.\\
    \end{split}\end{equation}
    Here the second and the last line follows directly from the definitions of $\mysl{i}{a}$ and $\myslhat{i}{a}$. 
\end{proof}

\begin{mylemma}\label{lem:comp1_reg2}    For $\forall (m,h,s,n) \in [M]\times [H]\times \mathcal{S}\times [K]$, the following equation holds:
    \begin{equation}\begin{split}
        I_{1,2} \leq \sum\limits_{j=1}^n \sum\limits_{a\in \mathcal{A}_m} \mylhat{m}{h}{s}{j}{a} \Big(\myeta{m}{h}{s}{j}w_j+\sum\limits_{i: j\in \myM{m}{h}{s}{k_i}} \myeta{m}{h}{s}{i} w_i\mathbb{I}\Big\{a_{k_i}=a\Big\}\Big).
    \end{split}\end{equation}        
\end{mylemma}

\begin{proof}[Proof of Lemma \ref{lem:comp1_reg2}]
    Consider any fixed pari $(m,h,s,n)$.
    \begin{equation}\label{eq:comp1_reg2_1}\begin{split}
        I_{1,2} &= \sum\limits_{i=1}^n w_i \Big[\myslhat{i}{a_{k_i}} \hat{\pi}_{k_i}(a_{k_i}) - \myslhat{i}{a_{k_i}} \tilde{\pi}_{k_{i+1}}(a_{k_i}) \Big]\\
        &= \sum\limits_{i=1}^n w_i \myslhat{i}{a_{k_i}} \hat{\pi}_{k_i}(a_{k_i}) \cdot \Big[1-\dfrac{\tilde{\pi}_{k_{i+1}}(a_{k_i})}{\hat{\pi}_{k_i}(a_{k_i})}\Big].\\
    \end{split}\end{equation}
    By the definition of $\tilde{\pi}_{k_{i+1}}(a_{k_i})$ and $\hat{\pi}_{k_i}(a_{k_i})$, we have: 
    \begin{equation}\begin{split}
        \frac{\tilde{\pi}_{k_{i+1}}(a_{k_i})}{\hat{\pi}_{k_i}(a_{k_i})} &= \frac{\exp \Big\{-(\eta_{i}/w_i) \mysLtil{i+1}{a_{k_i}}\Big\}}{\exp \Big\{-(\eta_i/w_i) \mysLhat{i}{a_{k_i}}\Big\}} \cdot \frac{\sum_a \exp \Big\{-(\eta_i/w_i) \mysLhat{i}{a}\Big\}}{\sum_a \exp \Big\{-(\eta_i/w_i) \mysLtil{i+1}{a}\Big\}}\\
        & \geq \frac{\exp \Big\{-(\eta_i/w_i) \hat{L}_i(a_{k_i}) - (\eta_i/w_i) \sum_{j\in \mysM{k_i}} w_j\myslhat{j}{a_{k_i}}-(\eta_iw_i/w_i) \myslhat{i}{a_{k_i}}\Big\}}{\exp \Big\{-(\eta_i/w_i) \hat{L}_i(a_{k_i})\Big\}}\\
        & = \exp \Big\{- \frac{\eta_i}{w_i} \sum_{j\in \mysM{k_i}} w_j\myslhat{j}{a_{k_i}} - \eta_i \myslhat{i}{a_{k}}\Big\}\\
        & \geq 1 - \frac{\eta_i}{w_i} \sum_{j\in \mysM{k_i}} w_j\myslhat{j}{a_{k_i}} - \eta_i\myslhat{i}{a_{k_i}}.
    \end{split}\end{equation}
    Here the second line is due to the fact $\tilde{L}_{i+1}(a) \geq \hat{L}_i(a)$. Substituting into Equation~\eqref{eq:comp1_reg2_1}, we get:
    \begin{equation}\begin{split}
        I_{1,2} &\leq \sum\limits_{i=1}^n w_i \myslhat{i}{a_{k_i}}\hat{\pi}_{k_i}(a_{k_i}) \cdot \Big(\dfrac{\eta_i}{w_i}  \sum\limits_{j\in \mysM{k_i}} w_j \myslhat{j}{a_{k_i}} + \eta_i\myslhat{i}{a_{k_i}}\Big)\\
        &= \sum\limits_{i=1}^n \eta_{i} \sum\limits_{j\in \mysM{k_i}} w_{j}\myslhat{j}{a_{k_i}} + \sum_{i=1}^n \eta_iw_i \myslhat{i}{a_{k_i}}\\
        &= \sum\limits_{i=1}^n \eta_{i} \sum\limits_{j\in \mysM{k_i}}\sum_{a\in\mathcal{A}_m} w_{j}\myslhat{j}{a}\mathbb{I}\Big\{a_{k_i}=a\Big\} + \sum_{i=1}^n \sum_{a\in\mathcal{A}_m} \eta_iw_i \myslhat{i}{a}\\
        &= \sum\limits_{j=1}^n \sum_{a\in\mathcal{A}_m} \myslhat{j}{a} \sum\limits_{i: j\in \mysM{k_i}} \eta_i w_{j} \mathbb{I}\Big\{a_{k_i}=a\Big\} + \sum_{i=1}^n \sum_{a\in\mathcal{A}_m} \eta_iw_i \myslhat{i}{a}\\
        &= \sum\limits_{j=1}^n \sum_{a\in\mathcal{A}_m} \myslhat{j}{a} \sum\limits_{i: j\in \mysM{k_i}} \eta_i w_i \mathbb{I}\Big\{a_{k_i}=a\Big\} + \sum_{i=1}^n \sum_{a\in\mathcal{A}_m} \eta_iw_i \myslhat{i}{a}\\
        &\leq \sum\limits_{j=1}^n \sum\limits_{a\in \mathcal{A}_m} \myslhat{j}{a} \Big(\eta_jw_j+\sum\limits_{i: j\in \mysM{k_i}} \eta_i w_i\mathbb{I}\Big\{a_{k_i}=a\Big\}\Big).\\
    \end{split}\end{equation}
    Here the first line is due to the following fact:
    \begin{equation}\begin{split}
        \myslhat{i}{a_{k_i}} \hat{\pi}_{k_i}(a_{k_i}) \leq  \frac{\hat{\pi}_{k_i}(a_{k_i})}{\hat{\pi}_{k_i}(a_{k_i})+\gamma_i} \leq 1.
    \end{split}\end{equation}
    and the last inequality is because $j\leq i$ and because $w_i$ monotonically increases with $i$. 
\end{proof}

\begin{proof}[Proof of Lemma \ref{reg_12}]
    Consider any fixed pair of $(m,h,s,n)$. With results of Lemma~\ref{lem:comp1_reg1}, Lemma~\ref{lem:comp1_reg2}, we directly have:
    \begin{equation}\label{eq:comp_reg12_1}\begin{split}
        I_{1,1}+I_{1,2}\leq &\sum\limits_{i=1}^{n} \sum\limits_{a\in \mathcal{A}_m} \myslhat{i}{a} \Big[2w_i\gamma_i + \sum\limits_{j: i\in \mysM{k_j}} \eta_{j} w_j \mathbb{I}\Big\{a_{k_j}=a\Big\} \Big]\\
        \leq &\sum\limits_{i=1}^{n} \sum\limits_{a\in \mathcal{A}_m} \myslbar{i}{a} \Big[2w_i\gamma_i + \sum\limits_{j: i\in \mysM{k_j}} \eta_{j} w_j \mathbb{I}\Big\{a_{k_j}=a\Big\} \Big]\\
        &\hspace{2em} + \sum\limits_{i=1}^{n} \sum\limits_{a\in \mathcal{A}_m} \Big[\myslhat{i}{a}-\myslbar{i}{a}\Big] \cdot \Big[2w_i\gamma_i + \sum\limits_{j: i\in \mysM{k_j}} \eta_{j} w_j \mathbb{I}\Big\{a_{k_j}=a\Big\} \Big]\\
        \leq & \sum\limits_{i=1}^n 2w_i \gamma_i A + \sum\limits_{i=1}^{n} \sum\limits_{j:i\in \mysM{k_j}} \eta_jw_j\\
        &\hspace{2em} + \sum\limits_{i=1}^{n} \sum\limits_{a\in \mathcal{A}_m} \Big[\myslhat{i}{a}-\myslbar{i}{a}\Big] \cdot \Big[2w_i\gamma_i + \sum\limits_{j: i\in \mysM{k_j}} \eta_{j} w_j \mathbb{I}\Big\{a_{k_j}=a\Big\} \Big].\\
    \end{split}\end{equation}
    Here the last line is because $\myslbar{i}{a} \leq 1$. 
    
    For the second term of the last line, switching the summation gives:
    \begin{equation}\begin{split}
        &\sum\limits_{i=1}^n \sum\limits_{j:i\in \mysM{k_j}} \eta_jw_j =  \sum\limits_{j=1}^n \eta_jw_j \sum\limits_{i:i\in \mysM{k_j}} 1 = \sum\limits_{j=1}^n \eta_jw_j |\mysM{k_j}|.\\
    \end{split}\end{equation}
    Therefore
    \begin{equation}\begin{split}
        I_{1,1}+I_{1,2} \leq & \sum\limits_{i=1}^n w_i \gamma_i (2A+|\mysM{k_i}|)\\
        &\hspace{2em} + \sum\limits_{i=1}^n \sum\limits_{a\in \mathcal{A}_m} \Big[\myslhat{i}{a}-\myslbar{i}{a}\Big] \cdot \Big[2w_i\gamma_i + \sum\limits_{j: i\in \mysM{k_j}} \eta_{j} w_j \mathbb{I}\Big\{a_{k_i}=a\Big\} \Big].\\
    \end{split}\end{equation}
    Notice that 
    \begin{equation}\begin{split}
        & 2w_i\gamma_i + \sum\limits_{j: i\in \mysM{k_j}} \eta_{j} w_j \mathbb{I}\Big\{a_{k_i}=a\Big\} \leq w_n \Big(2\gamma_i + \eta_i d_{max} \Big) \leq 2w_n d_{max} \gamma_i.\\
    \end{split}\end{equation}
    where the first inequality is because $i \geq j$ and $i$ can be delayed for at most $d_{max}$ episodes. Then by Lemma 4.3 in \cite{rw_bandit}, the following equation holds for any fixed pair $(m, h, s, n)\in [M]\times [H]\times \mathcal{S}\times [K]$ with probability at least $1-\delta/(4MHSK)$:
    \begin{equation}\begin{split}
        \sum\limits_{i=1}^n \sum\limits_{a\in \mathcal{A}_m} \Big[\myslhat{i}{a}-\myslbar{i}{a}\Big] \cdot \Big[2w_i\gamma_i + \sum\limits_{j: i\in \mysM{k_j}} \eta_j w_j \mathbb{I}\Big\{a_j=a\Big\} \Big] \leq w_n d_{max} \iota.
    \end{split}\end{equation}
    By union bound, the above equation holds for all $(m,h,s,n)\in [M]\times[H]\times \mathcal{S}\times [K]$ with probability at least $1- \delta/4$. Substituting the above results into Equation~\eqref{eq:comp_reg12_1}, we have the following equation holds with probability at least $1-\delta/4$:
    \begin{equation}\begin{split}
        & I_{1,1} + I_{1,2} \leq \sum\limits_{i=1}^n w_i \gamma_i \Big(2A+|\mysM{k_i}|\Big) + w_n d_{max} \iota.\\
    \end{split}\end{equation}
\end{proof}

\begin{proof}[Proof of Lemma \ref{reg_3}]
    Consider any fixed pair of $(m,h,s,n)$.
    \begin{equation}\begin{split}
        I_{1,3} &= \sum\limits_{i=1}^n w_i \Big[ \myslhat{i}{a_{k_i}} \tilde{\pi}_{k_{i+1}}(a_{k_i}) - \myslhat{i}{a^*} \Big].\\
    \end{split}\end{equation}
    To bound the term, we apply Theorem 3 in \citet{rw_bandit3} with:
    \begin{equation}\begin{aligned}
        p_i(\pi) &= 0, & \forall i\in [n],\\
        q_0(\pi) &= \frac{w_0}{\eta_0}\sum_{a\in \mathcal{A}_m} \pi(a)\log\pi(a), & \\
        q_i(\pi) &= (\frac{w_i}{\eta_i}-\frac{w_{i-1}}{\eta_{i-1}})\sum_{a\in \mathcal{A}_m} \pi(a)\log\pi(a), & \forall i\in [n].\\
    \end{aligned}\end{equation}
    Here we define $w_0 = \eta_0 = 1$. Finally, we have:
    \begin{equation}\begin{split}
        I_{1,3} \leq -\sum_{i=1}^n q_i(\tilde{\pi}_{k_{i+1}}) \leq \log A\sum_{i=1}^n (\frac{w_i}{\eta_i}-\frac{w_{i-1}}{\eta_{i-1}}) \leq \frac{w_n}{\eta_n} \iota.
    \end{split}\end{equation}
\end{proof}

\begin{proof}[Proof of Lemma \ref{reg_4}]
    The lemma follows directly from Lemma 4.3 in \cite{rw_bandit}.
\end{proof}

\subsection{\textbf{Step Two:} Proof of Lemma \ref{lem:opt}}\label{sec:proof_s2}
% Proofs in this section mainly follow Lemma 13 and Lemma 14 in \cite{alg_ma_VL}.

\begin{proof}[Proof of lemma \ref{lem:opt}]
We first prove the optimism part of the lemma for any fixed pair $(m,h,s,k)$. Conditioned on the successful event of Lemma~\ref{lem:reg}, which holds for probability at least $1-\delta/2$, we prove the lemma by induction. For $k=0$, it is clear that:
    \begin{equation}
        \overline{V}^0_{m,h}(s) = H+1-h \geq V_{m,h}^{\dagger, \pi_{-m,h}^1}(s).
    \end{equation}
    
    Recall that $\mysbetaOver{\mysn{k}} = 12 \sqrt{\dfrac{\mysn{k}A+\mysT{\mysn{k}}}{\mysn{k}^2}\iota} + \dfrac{4H^2d_{max}}{\mysn{k}}\iota$. Let $N_k = \max_m \myn{m}{h}{s}{k}$. Suppose the optimism part holds for all $k' < k$. Then for episode $k$ and any $(m,h,s)$:
    \begin{equation}\label{eq:opt_1}\begin{aligned}
        \myVOver{m}{h}{s}{k} &= \alpha^0_{\mysn{k}} \cdot H + \sum\limits_{i=1}^{\mysn{k}} \alpha^i_{\mysn{k}} \Big( r_{m,h}^{k_i} + \myVOver{m}{h+1}{s^{k_i}_{h+1}}{k_i}\Big) + \mysbetaOver{\mysn{k}}\\
        & = \alpha^0_{\mysn{k}} \cdot H + H  \sum\limits_{i=1}^{\mysn{k}} \alpha^i_{\mysn{k}} \Big[ 1 - \mysl{i}{a_{m,h}^{k_i}} \Big] + \mysbetaOver{\mysn{k}}\\
        & \geq \alpha^0_{\mysn{k}} \cdot H + H \sum\limits_{i=1}^{\mysn{k}} \alpha^i_{\mysn{k}} \Big[1 - \myslbar{i}{a^*}\Big] + \Big(\mysbetaOver{\mysn{k}} - \mysR{\mysn{k}}\Big)\\
        & = \alpha^0_{\mysn{k}} \cdot H + \frac{2d_{max}H^2}{\mysn{k}}\iota+ \sum\limits_{i=1}^{\mysn{k}} \alpha^i_{\mysn{k}}  \mathop{\mathbb{E}}_{\substack{\bm{a}=(a^*,a_{-m,h})\\ a_{-m,h} \sim \hat{\pi}_{-m,h}^{k_i},\ s'\sim \mathbb{P}_h(s,\bm{a})}} \Big[r_{m,h}(s,\bm{a}) + \overline{V}^{k_i}_{m,h+1}(s')\Big].
    \end{aligned}\end{equation}
    Here the third line is becuase of Lemma~\ref{lem:reg}.
    
    On the other hand, 
    \begin{equation}\label{eq:opt_2}\begin{aligned}
        V^{\dagger, \pi^k_{-m,h}}_{m,h}(s) =& \max_{\mu_h}\max_{\mu_{h+1:H}} \sum_{i=1}^{N_k} \alpha_{N_k}^i \mathop{\mathbb{E}}_{\substack{\bm{a}=(a, a_{-m,h})\\a\sim\mu_h, a_{-m,h}\sim\hat{\pi}_{-m,h}^{k_i},s'\sim \mathbb{P}_h(s,\bm{a})\\}}\Big( r_{m,h}(s,\bm{a}) + V^{\mu_{h+1:H}, \pi_{-m,h+1}^{k_i}}_{m,h+1}(s') \Big)\\
        \leq& \max_{\mu_h} \sum_{i=1}^{N_k} \alpha_{N_k}^i \mathop{\mathbb{E}}_{\substack{\bm{a}=(a, a_{-m,h})\\a\sim\mu_h, a_{-m,h}\sim\hat{\pi}_{-m,h}^{k_i},s'\sim \mathbb{P}_h(s,\bm{a})}}\Big( r_{m,h}(s,\bm{a}) + V^{\dagger, \pi_{-m,h+1}^{k_i}}_{m,h+1}(s') \Big)\\
        \leq& \max_{\mu_h} \sum_{i=1}^{\mysn{k}} \alpha_{N_k}^i \mathop{\mathbb{E}}_{\substack{\bm{a}=(a, a_{-m,h})\\a\sim\mu_h, a_{-m,h}\sim\hat{\pi}_{-m,h}^{k_i},s'\sim \mathbb{P}_h(s,\bm{a})}}\Big( r_{m,h}(s,\bm{a}) + V^{\dagger, \pi_{-m,h+1}^{k_i}}_{m,h+1}(s') \Big)\\
        &\hspace{2em} + \max_{\mu_h} \sum_{i=\mysn{k}+1}^{N_k} \alpha_{N_k}^i \mathop{\mathbb{E}}_{\substack{\bm{a}=(a, a_{-m,h})\\a\sim\mu_h, a_{-m,h}\sim\hat{\pi}_{-m,h}^{k_i},s'\sim \mathbb{P}_h(s,\bm{a})}}\Big( r_{m,h}(s,\bm{a}) + V^{\dagger, \pi_{-m,h+1}^{k_i}}_{m,h+1}(s') \Big)\\
        \leq& \max_{\mu_h} \sum_{i=1}^{\mysn{k}} \alpha_{\mysn{k}}^i \mathop{\mathbb{E}}_{\substack{\bm{a}=(a, a_{-m,h})\\a\sim\mu_h, a_{-m,h}\sim\hat{\pi}_{-m,h}^{k_i},s'\sim \mathbb{P}_h(s,\bm{a})}}\Big( r_{m,h}(s,\bm{a}) + V^{\dagger, \pi_{-m,h+1}^{k_i}}_{m,h+1}(s') \Big) + \sum_{i=\mysn{k}+1}^{N_k} \alpha_{N_k} H\\
        \leq& \sum_{i=1}^{\mysn{k}} \alpha_{\mysn{k}}^i \mathop{\mathbb{E}}_{\substack{\bm{a}=(a^*, a_{-m,h})\\a_{-m,h}\sim\hat{\pi}_{-m,h}^{k_i},s'\sim \mathbb{P}_h(s,\bm{a})}}\Big( r_{m,h}(s,\bm{a}) + \myVOver{m}{h+1}{s'}{k_i} \Big) + 2H^2\frac{N_k-\mysn{k}}{N_k}\\
        \leq& \sum_{i=1}^{\mysn{k}} \alpha_{\mysn{k}}^i \mathop{\mathbb{E}}_{\substack{\bm{a}=(a^*, a_{-m,h})\\a_{-m,h}\sim\hat{\pi}_{-m,h}^{k_i},s'\sim \mathbb{P}_h(s,\bm{a})}}\Big( r_{m,h}(s,\bm{a}) + \myVOver{m}{h+1}{s'}{k_i} \Big) + 2H^2\frac{d_{max}}{N_k}.\\
    \end{aligned}\end{equation}
    Here the second line is because of the convexity of the maximum, the last line is due to the fact that $N_k-\mysn{k} \leq (\mysnpr{k}-1)-\mysn{k} = |\mathcal{M}_k| \leq d_{max}$, where the first equation is because $\mysn{k} $ which is the direct result of Lemma~\ref{lem:comp1_base2}.
    
    Finally, combining equations~\ref{eq:opt_1} and \ref{eq:opt_2}, we finish the proof of the induction and prove the optimism part of the lemma.
    % When $\mysn{k}\leq d_{max}$, we have $(\mysbetaOver{\mysn{k}} - R_{m,h}^{\mysn{k}}(s)) \geq 2H^2 \geq \frac{2(N_k-\mysn{k})H^2}{N_k}$. 
    % When $\mysn{k} > d_{max}$, we have $(\mysbetaOver{\mysn{k}} - R_{m,h}^{\mysn{k}}(s)) \geq \frac{2d_{max}H^2}{\mysn{k}} \geq \frac{2(N_k-\mysn{k})H^2}{N_k}$.
    % Both cases lead to $\myVOver{m}{h}{s}{k} > V^{\dagger, \pi^k_{-m,h}}_{m,h}(s)$ and complete the proof.

    Now we prove the pessimism part. For a fixed pair $(m,h,s,k)$, the following hold with probability at least $1-\delta/(2MHSK)$:
    
    \begin{equation}\begin{split}
        \underline{V}^k_{m,h}(s) &= \sum\limits_{i=1}^{\mysn{k}} \alpha^i_{\mysn{k}} \Big( r_{m,h}^{k_i} + \myVUnd{m}{h+1}{s^{k_i}_{h+1}}{k_i}\Big) - \mysbetaUnd{\mysn{k}}\\
        & \leq \sum\limits_{i=1}^{\mysn{k}} \alpha^i_{\mysn{k}} \mathop{\mathbb{E}}_{\substack{\bm{a}\sim \hat{\pi}^{k_i}_h, s'\sim \mathbb{P}_h(\cdot|s,\bm{a})}}\Big( r_{m,h}(s,\bm{a}) + \underline{V}^{k_i}_{m,h+1}(s')\Big) + 2\sqrt{\dfrac{H^3}{\mysn{k}}\iota} - \mysbetaUnd{\mysn{k}}\\
        & \leq \sum\limits_{i=1}^{\mysn{k}} \alpha^i_{\mysn{k}} \mathop{\mathbb{E}}_{\substack{\bm{a}\sim \hat{\pi}^{k_i}_h, s'\sim \mathbb{P}_h(\cdot|s,\bm{a})}} \Big( r_{m,h}(s,\bm{a}) + \underline{V}^{k_i}_{m,h+1}(s')\Big) - 2H^2 \frac{d_{max}}{\mysn{k}}\iota\\
        & = \sum\limits_{i=1}^{\mysn{k}} \alpha^i_{N_k} \mathop{\mathbb{E}}_{\substack{\bm{a}\sim \hat{\pi}^{k_i}_h, s'\sim \mathbb{P}_h(\cdot|s,\bm{a})}} \Big( r_{m,h}(s,\bm{a}) + \underline{V}^{k_i}_{m,h+1}(s')\Big) - 2H^2 \frac{d_{max}}{\mysn{k}}\iota\\
        &\hspace{2em} + \Big(1-\prod_{j=\mysn{k}+1}^{N_k}(1-\alpha_j)\Big)\sum\limits_{i=1}^{\mysn{k}} \alpha^i_{\mysn{k}} \mathop{\mathbb{E}}_{\substack{\bm{a}\sim \hat{\pi}^{k_i}_h, s'\sim \mathbb{P}_h(\cdot|s,\bm{a})}} \Big( r_{m,h}(s,\bm{a}) + \underline{V}^{k_i}_{m,h+1}(s')\Big)\\
        & \leq \sum\limits_{i=1}^{\mysn{k}} \alpha^i_{N_k} \mathop{\mathbb{E}}_{\substack{\bm{a}\sim \hat{\pi}^{k_i}_h, s'\sim \mathbb{P}_h(\cdot|s,\bm{a})}} \Big( r_{m,h}(s,\bm{a}) + \underline{V}^{k_i}_{m,h+1}(s')\Big) - 2H^2 \frac{d_{max}}{\mysn{k}}\iota + (N_k-\mysn{k})\alpha_{\mysn{k}} \sum\limits_{i=1}^{\mysn{k}} \alpha^i_{\mysn{k}} H\\
        & \leq \sum\limits_{i=1}^{\mysn{k}} \alpha^i_{N_k} \mathop{\mathbb{E}}_{\substack{\bm{a}\sim \hat{\pi}^{k_i}_h, s'\sim \mathbb{P}_h(\cdot|s,\bm{a})}} \Big( r_{m,h}(s,\bm{a}) + \underline{V}^{k_i}_{m,h+1}(s')\Big) - 2H^2 \frac{d_{max}}{\mysn{k}}\iota + 2H^2\frac{d_{max}}{\mysn{k}}\\
        & \leq \sum\limits_{i=1}^{N_k} \alpha^i_{N_k} \mathop{\mathbb{E}}_{\substack{\bm{a}\sim \hat{\pi}^{k_i}_h, s'\sim \mathbb{P}_h(\cdot|s,\bm{a})}} \Big( r_{m,h}(s,\bm{a}) + \underline{V}^{k_i}_{m,h+1}(s')\Big)\\
        & = V^{\pi^k_h}_{m,h}(s).
    \end{split}\end{equation}
    Here the second line follows Azuma's inequality with propbability at least $1-\delta/(2MHSK)$. While the last line is the definition of $V_{m,h}^{\pi^k_h}(s)$. Finally, by union bound over all $(m,h,s,k) \in [M]\times[H]\times \mathcal{S}\times [K]$, we finish the proof.
\end{proof}

\subsection{Proof of Theorem \ref{thm:comp}}\label{sec:proof_s3}

\begin{proof}[Proof of theorem \ref{thm:comp}]
    Consider any fixed pair $(m,h)$. We start by upper bounding the term $\sum\limits_{k=1}^K (\overline{V}_{m,h}^k - \underline{V}_{m,h}^k)(s_h^k)$. For any episode $k$, we slightly overload notations $\mysn{k} = \myn{m}{h}{s^k_h}{k}$, $k_i = \myk{h}{s^k_h}{i}$, $\mysbetaOver{n} = \mybetaOver{m}{h}{s^k_h}{n}$. Then we have:
    
    \begin{equation}\begin{split}
        & (\overline{V}_{m,h}^k - \underline{V}_{m,h}^k)(s_h^k) \\
        \leq & \alpha^0_{\mysn{k}} \cdot H + \sum\limits_{i=1}^{\mysn{k}} \alpha^i_{\mysn{k}} \Big[ r_{m,h}^{k_i} + \overline{V}_{m,h+1}^{k_i}(s_{h+1}^{k_i})\Big] + \mysbetaOver{\mysn{k}}\\
        &\hspace{2em}- \sum\limits_{i=1}^{\mysn{k}} \alpha^i_{\mysn{k}} \Big[ r_{m,h}^{k_i} + \underline{V}_{m,h+1}^{k_i}(s_{h+1}^{k_i})\Big] - \mysbetaUnd{\mysn{k}}\\
        \leq & \alpha^0_{\mysn{k}} \cdot H + \sum\limits_{i=1}^{\mysn{k}} \alpha^i_{\mysn{k}} (\overline{V}_{m,h+1}^{k_i} - \underline{V}_{m,h+1}^{k_i})(s_{h+1}^{k_i})  + \mysbetaOver{\mysn{k}} - \mysbetaUnd{\mysn{k}}.\\
    \end{split}\end{equation}
    Taking the summation over episode $k$ gives:
    \begin{equation}\begin{split}
        & \sum\limits_{k=1}^{K} (\overline{V}_{m,h}^k - \underline{V}_{m,h}^k)(s_h^k) \\
        =& \sum\limits_{k=1}^{K} (\overline{V}_{m,h}^k - \underline{V}_{m,h}^k)(s_h^k) \cdot \mathbb{I}\Big\{\mysn{k} < d_{max}\Big\} + \sum\limits_{k=1}^{K} (\overline{V}_{m,h}^k - \underline{V}_{m,h}^k)(s_h^k) \cdot \mathbb{I}\Big\{\mysn{k} \geq d_{max}\Big\} \\
        \leq & H\sum\limits_{k=1}^K \mathbb{I}\Big\{\mysn{k} < d_{max}\Big\} + \underbrace{\sum\limits_{k=1}^{K} \sum\limits_{i=1}^{\mysn{k}} \alpha^i_{\mysn{k}}(\overline{V}_{m,h+1}^{k_i} - \underline{V}_{m,h+1}^{k_i})(s_{h+1}^{k_i})}_{I_{2,1}}\\
        &\hspace{2em}+ \underbrace{\sum\limits_{k=1}^{K} (\mysbetaOver{\mysn{k}}+\mysbetaUnd{\mysn{k}})\cdot \mathbb{I}\Big\{\mysn{k} \geq d_{max}\Big\}}_{I_{2,2}}.\\
    \end{split}\end{equation}
    We then give the upper bounds for $I_{2,1}$ and $I_{2,2}$ in the following lemmas.

    \begin{mylemma}[The upper bound of $I_{2,1}$]\label{lem:comp_1}
        For $\forall (m, h)\in [M]\times[H]$,
        \begin{equation}\begin{split}
            I_{2,1} \leq (1+\dfrac{1}{H}) \sum\limits_{k=1}^{K} \Big( \overline{V}_{m,h+1}^k-\underline{V}_{m,h+1}^k \Big)(s^k_{h+1}) + 4d_{max}H^2 S\iota.
        \end{split}\end{equation}
    \end{mylemma} 

    \begin{mylemma}[The upper bound of $I_{2,2}$]\label{lem:comp_2}
        For $\forall (m, h)\in [M]\times[H]$,
        \begin{equation}\begin{split}
            I_{2,2} \leq 12d_{max}H^2S \iota^2 + 56H^2\sqrt{SAK\iota} + 24H^2\sqrt{S\mysT{K}}\iota^2.
        \end{split}\end{equation}
    \end{mylemma} 
    Substituting into previous equations, we get:
    \begin{equation}\begin{split}
        & \sum\limits_{k=1}^{K} (\overline{V}_{m,h}^k - \underline{V}_{m,h}^k)(s_h^k) \\
        \leq & H\sum\limits_{k=1}^K \mathbb{I}\Big\{\mysn{k} < d_{max}\Big\} + (1+\dfrac{1}{H}) \sum\limits_{k=1}^{K} \Big( \overline{V}_{m,h+1}^k-\underline{V}_{m,h+1}^k \Big)(s^k_{h+1})\\
        & \hspace{4em} + 4d_{max}H^2S \iota + 12d_{max}H^2S \iota^2 + 56H^2\sqrt{SAK\iota} + 24H^2\sqrt{S\mysT{K}}\iota^2\\
        \leq & 2d_{max}HS + (1+\dfrac{1}{H}) \sum\limits_{k=1}^{K} \Big( \overline{V}_{m,h+1}^k-\underline{V}_{m,h+1}^k \Big)(s^k_{h+1})\\
        & \hspace{4em} + 16d_{max}H^2S \iota^2 + 56H^2\sqrt{SAK\iota} + 24H^2\sqrt{S\mysT{K}}\iota^2\\.
    \end{split}\end{equation}
    To see why the last line holds, we notice that the $(d_{max}-1)$-th visit of any $(h,s)$ will be received when the $2d_{max}$-th visit of $(h,s)$ happens. This leads to 
    \begin{equation}
        H\sum\limits_{k=1}^K \mathbb{I}\Big\{\mysn{k} < d_{max}\Big\} \leq H\sum_{s\in\mathcal{S}} 2d_{max} \leq 2d_{max}HS.
    \end{equation}
    
    Iterating over h, we get:
    \begin{equation}\begin{split}
        & \sum\limits_{k=1}^{K} (\overline{V}^k_{m,1} - \underline{V}^k_{m,1})(s_1) \\
        \leq & 18d_{max}H^3S \iota^2 + 56H^3\sqrt{SAK\iota} + 24H^3\sqrt{S\mysT{K}}\iota^2\\
        & \hspace{4em} + (1+\dfrac{1}{H})^H \sum\limits_{k=1}^{K} \Big( \overline{V}_{m,H+1}^k-\underline{V}_{m,H+1}^k \Big)(s^k_{H+1})\\
        \lesssim & d_{max}H^3 S\iota^2 + H^3 \sqrt{SAK\iota} + H^3 \sqrt{S\mysT{K}}\iota^2.\\
    \end{split}\end{equation}
    Following from Lemma~\ref{lem:opt}, the following inequality holds with probability at least $1-\delta$:
    \begin{equation}\begin{split}
        \max\limits_{m\in[M]} \sum\limits_{k=1}^K \Big( V_{m,1}^{\dag, \pi_{-m,k}} - V_{m,1}^{\pi_k} \Big)(s_1) \lesssim d_{max}H^3 S\iota^2 + H^3 \sqrt{SAK\iota} + H^3 \sqrt{S\mysT{K}}\iota^2.\\
    \end{split}\end{equation}
    By the definition of policy $\pi$, we have the following equation when $K \geq d_{max}^2S\iota^3$:
    \begin{equation}\begin{split}
        \max\limits_{m\in[M]} \Big( V_{m,1}^{\dag, \pi_{-m}} - V_{m,1}^{\pi} \Big)(s_1) &\lesssim H^3 \sqrt{S\mysT{K}/K^2}\iota^2 + H^3\sqrt{SA\iota/K}.
    \end{split}\end{equation}
\end{proof}

\subsubsection{Supporting Details}

% \begin{mylemma}\label{lem:comp_hold}
%     For $\forall (m,h,s,n) \in [M]\times [H]\times \mathcal{S}\times [K]$, $\myTover{m}{h}{s}{n} \geq \myT{m}{h}{s}{n}$.
% \end{mylemma}

% \begin{proof}[Proof of Lemma \ref{lem:comp_hold}.]
%     Consider any two visits $n$ and $i$ for fixed pair $(h,s)$. Notice that the following inequalities always hold:
%     \begin{equation}\begin{split}
%         k_n - k_i \geq n - i
%     \end{split}\end{equation}
%     This is because $n-i$ is the interval between the two visits, when only state $s$ is visited at step $h$. However, $k_n - k_i$ is the interval between the two visits, when all other states may be visited. 
    
%     Consequently, for any pair $(m,h,s,n)$
%     \begin{equation}\begin{split}
%         \mathop{\arg\min}\limits_{i} \Big[ d_i \geq n-i \Big] \leq \mathop{\arg\min}\limits_{i} \Big[ d_i \geq k_n - k_i  \Big] 
%     \end{split}\end{equation}
%     This directly gives $\myTover{m}{h}{s}{n} \geq \myT{m}{h}{s}{n}$ according to their definitions.
% \end{proof}

\begin{proof}[Proof of Lemma \ref{lem:comp_1}.]
    Consider any fixed pair $(m,h)$, we define the following set $X_n(s)$:
    \begin{equation}
        X_n(s) = \Big\{x: s^x_h = s,  \myn{m}{h}{s}{x} \geq n \Big\}.
    \end{equation}
    Intuitively speaking, it collects episodes where $(h,s)$ is visited and the $n$-th visit of $(h,s)$ is usuable.
    
    Rearranging the summation gives:
    \begin{equation}\label{eq:comp_1_1}\begin{split}
        & \sum\limits_{k=1}^{K} \sum\limits_{i=1}^{\mysn{k}} \alpha^i_{\mysn{k}} \Big(\overline{V}_{m,h+1}^{k_i} - \underline{V}_{m,h+1}^{k_i}\Big)(s_{h+1}^{k_i})\\
        = & \sum\limits_{k=1}^{K} \Big( \overline{V}_{m,h+1}^k-\underline{V}_{m,h+1}^k \Big)(s_{h+1}^k) \sum\limits_{x\in X_{\mysnpr{k}}(s^k_h)} \alpha^{\mysnpr{k}}_{\mysn{x}}.\\
    \end{split}\end{equation} 
    For any episode $x\in X_{\mysnpr{k}}(s^k_h)$, there are at most $d_{max}$ unreceived and unusable visits of $(h,s^k_h)$ according to Lemma~\ref{lem:comp1_base2}. So we have $\mysn{x} \geq \mysnpr{x} - d_{max} + 1$. On the other hand, according to the definition of $X_{\mysnpr{k}}(s^k_h)$, $\mysn{x} \geq \mysnpr{k}$. This gives:
    \begin{equation}
        \mysnpr{x} \geq \mysn{x} \geq \max \{\mysnpr{k}, \mysnpr{x}-d_{max}+1\}.
    \end{equation}
    Notice here $\mysnpr{x}$ strictly increases with $x$. 
    
    Based on the above observations, we conclude that for the $i$-th episode ($i \leq d_{max}$) in $X_{\mysnpr{k}}(s^k_h)$, if denoted as $x$, we have:
    \begin{equation}
        \mysn{x} \geq \mysnpr{k}.
    \end{equation}
    For the $i$-th element ($i > d_{max}$) in $X_{\mysnpr{k}}(s^k_h)$, if denoted as $x$, we have:
    \begin{equation}\begin{split}
        \mysn{x} & \geq \mysnpr{x} - d_{max} + 1 \geq \mysnpr{k}+(i-1) - d_{max} + 1\\
        & = \mysnpr{k} + (i-d_{max}).\\
    \end{split}\end{equation}

    Substituting the above two inequalities on $\mysn{x}$ into Equation~\eqref{eq:comp_1_1}, then:
    \begin{equation}\begin{split}
        & \sum\limits_{k=1}^{K} \Big( \overline{V}_{m,h+1}^k-\underline{V}_{m,h+1}^k \Big)(s_{h+1}^k) \sum\limits_{x\in X_{\mysnpr{k}}(s^k_h)} \alpha^{\mysnpr{k}}_{\mysn{x}}\\
        \leq & \sum\limits_{k=1}^{K} \Big( \overline{V}_{m,h+1}^k-\underline{V}_{m,h+1}^k \Big)(s^k_{h+1}) 
            \Big[d_{max} \alpha_{\mysnpr{k}} + \sum\limits_{i=d_{max}+1}^{\infty} \alpha^{\mysnpr{k}}_{\mysnpr{k}+i-d_{max}}\Big]\\
        \leq & \sum\limits_{k=1}^{K} \Big( \overline{V}_{m,h+1}^k-\underline{V}_{m,h+1}^k \Big)(s^k_{h+1})
            \Big[d_{max} \alpha_{\mysnpr{k}} + (1+\dfrac{1}{H})\Big]\\
        \leq & (1+\dfrac{1}{H}) \sum\limits_{k=1}^{K} \Big( \overline{V}_{m,h+1}^k-\underline{V}_{m,h+1}^k \Big)(s^k_{h+1}) + d_{max}H \sum\limits_{k=1}^{K} \alpha_{\mysnpr{k}}\\
        \leq & (1+\dfrac{1}{H}) \sum\limits_{k=1}^{K} \Big( \overline{V}_{m,h+1}^k-\underline{V}_{m,h+1}^k \Big)(s^k_{h+1}) + 2d_{max}H^2 \sum\limits_{s\in \mathcal{S}} \sum\limits_{i=1}^{\mynpr{m}{h}{s}{K}} \dfrac{1}{i} \\
        \leq & (1+\dfrac{1}{H}) \sum\limits_{k=1}^{K} \Big( \overline{V}_{m,h+1}^k-\underline{V}_{m,h+1}^k \Big)(s^k_{h+1}) + 2d_{max}H^2 S(\ln K+1)\\
        \leq & (1+\dfrac{1}{H}) \sum\limits_{k=1}^{K} \Big( \overline{V}_{m,h+1}^k-\underline{V}_{m,h+1}^k \Big)(s^k_{h+1}) + 4d_{max}H^2 S\iota.
    \end{split}\end{equation}
    Here the second line follows the monotonicity of $\{\alpha^i_n\}_{n\in[K]}$, the third and fifth line follows Lemma~\ref{lem:comp1_base1}, and the last line is because $\mynpr{m}{h}{s}{K}$.
\end{proof}

\begin{proof}[Proof of Lemma \ref{lem:comp_2}.]
    Consider any fixed pair $(m,h)$. We inherit the definition of $X_n(s)$ from the proof of Lemma \ref{lem:comp_1}. We first bound term $\sum\limits_{k=1}^{K} \mysbetaOver{\mysn{k}}\cdot\mathbb{I}\Big\{\mysn{k}\geq d_{max}\Big\}$
    \begin{equation}\begin{split}
        & \sum\limits_{k=1}^{K} \mysbetaOver{\mysn{k}} \cdot\mathbb{I}\Big\{\mysn{k}\geq d_{max}\Big\}\\
        =& 4d_{max}H^2 \sum\limits_{k=1}^{K} \dfrac{\iota}{\mysn{k}}\cdot \mathbb{I}\Big\{\mysn{k}\geq d_{max}\Big\}+ 12H^2 \sum\limits_{k=1}^{K}  \sqrt{\dfrac{\mysn{k}A+\myT{m}{h}{s^k_h}{\mysn{k}}}{\mysn{k}^2}\iota}\cdot \mathbb{I}\Big\{\mysn{k}\geq d_{max}\Big\}\\
        \leq & 4d_{max}H^2 \sum\limits_{k=1}^{K} \dfrac{\iota}{\mysn{k}}\cdot \mathbb{I}\Big\{\mysn{k}\geq d_{max}\Big\} + 12H^2 \sum\limits_{k=1}^{K}  \sqrt{\dfrac{A}{\mysn{k}}\iota}\cdot \mathbb{I}\Big\{\mysn{k}\geq d_{max}\Big\}\\
        &\hspace{2em}+ 12H^2 \cdot \sum\limits_{k=1}^{K} \sqrt{\dfrac{\myT{m}{h}{s^k_h}{\mysn{k}}}{\mysn{k}^2}\iota}\cdot \mathbb{I}\Big\{\mysn{k}\geq d_{max}\Big\}.\\
    \end{split}\end{equation}
    
    For the first term:
    \begin{equation}\begin{split}
        & d_{max}H^2 \sum\limits_{k=1}^{K} \dfrac{\iota}{\mysn{k}} \cdot\mathbb{I}\Big\{\mysn{k}\geq d_{max}\Big\} = d_{max}H^2 \iota \sum_{s\in \mathcal{S}} \sum_{x\in X_{d_{max}}(s)} \dfrac{1}{\mysn{x}}.\\
    \end{split}\end{equation}
    Here set $X_{d_{max}}(s)$ collects all episodes when $(h,s)$ is visited and the $d_{max}$-th visit of $(h,s)$ is usable. Following the analysis for Lemma \ref{lem:comp_1}, for the $i$-th element $(i \leq d_{max})$ in set $X_{d_{max}}(s)$, if denoted as $x$, we have:
    \begin{equation}\begin{split}
        \mysn{x} \geq d_{max}.
    \end{split}\end{equation}
    For the $i$-th element $(i > d_{max})$ in $X_{d_{max}}(s)$, if denoted as $x$, we have:
    \begin{equation}\begin{split}
        \mysn{x} \geq (d_{max}+i) - d_{max} = i.
    \end{split}\end{equation}
    These inequalities lead to:
    \begin{equation}\begin{split}
        \sum_{x\in X_{d_{max}}(s)} \dfrac{1}{\mysn{x}} &\leq 1 + \sum_{i=d_{max}+1}^{\mynpr{m}{h}{s}{K}} \dfrac{1}{i} \leq \ln \mynpr{m}{h}{s}{K} + 2 \leq 3\iota,
    \end{split}\end{equation}
    where the first inequality is due to the monotinicity of $1/n$.
    So the first term is bounded as follows:
    \begin{equation}\label{eq:comp_1_3}\begin{split}
        & d_{max}H^2 \sum\limits_{k=1}^{K} \dfrac{\iota}{\mysn{k}} \cdot\mathbb{I}\Big\{\mysn{k}\geq d_{max}\Big\} \leq 3d_{max}H^2S \iota^2.
    \end{split}\end{equation}

    For the second term:
    \begin{equation}\label{eq:comp_1_2}\begin{split}
        H^2 \sum\limits_{k=1}^{K}  \sqrt{\dfrac{A}{\mysn{k}}\iota} \cdot \mathbb{I}\Big\{\mysn{k} \geq d_{max}\Big\} \leq & H^2\sqrt{A\iota}  \cdot \sum\limits_{s\in \mathcal{S}} \sum_{x\in X_{d_{max}}(s)} \sqrt{\dfrac{1}{\mysn{x}}}\\
        \leq & H^2\sqrt{A\iota} \cdot \sum\limits_{s\in \mathcal{S}} \Big[1 + \sum_{i=d_{max}+1}^{\mynpr{m}{h}{s}{K}} \dfrac{1}{\sqrt{i}}\Big]\\
        \leq & 2H^2\sqrt{A\iota} \sum\limits_{s\in \mathcal{S}} (\sqrt{\mynpr{m}{h}{s}{K}}+1)\\
        \leq & 2H^2\sqrt{A\iota}  \sqrt{S \sum\limits_{s\in \mathcal{S}} \mynpr{m}{h}{s}{K}} + 2H^2S\sqrt{A\iota}\\
        \leq & 2H^2\sqrt{SAK\iota} + 2H^2S\sqrt{A\iota}\\
        \leq & 4H^2\sqrt{SAK\iota}.\\
    \end{split}\end{equation}
    Here the second inequalities follows similar analysis of the first term, the fifth inequality holds because $\sum_{s\in\mathcal{S}} \mynpr{m}{h}{s}{K} = K$ and the last inequality holds because $K \geq d_{max}^2S\iota^3$.

    For the third term:
    \begin{equation}\begin{split}
        H^2\sum\limits_{k=1}^{K} \sqrt{\dfrac{\myT{m}{h}{s^k_h}{\mysn{k}}}{\mysn{k}^2}\iota}\cdot \mathbb{I}\Big\{\mysn{k}> d_{max}\Big\} \leq & H^2\sum\limits_{s\in \mathcal{S}} \sum_{x\in X_{d_{max}}(s)} \sqrt{\dfrac{\myT{m}{h}{s}{\mysn{x}}}{\mysn{x}^2}\iota}\\
        \leq & H^2\sum\limits_{s\in \mathcal{S}} \sqrt{\myT{m}{h}{s}{\myn{m}{h}{s}{K}}\iota}\sum_{x\in X_{d_{max}}(s)}  \frac{1}{\mysn{x}}\\
        \leq & H^2\sum\limits_{s\in \mathcal{S}} \sqrt{\myT{m}{h}{s}{\myn{m}{h}{s}{K}}\iota}\cdot \Big[ 1 + \sum_{i=d_{max}+1}^{\mynpr{m}{h}{s}{K}} \frac{1}{i} \Big]\\
        \leq & 2H^2\sum\limits_{s\in \mathcal{S}} \sqrt{\myT{m}{h}{s}{\myn{m}{h}{s}{K}}}\iota^2\\
        % \leq & H^2\sum\limits_{s\in \mathcal{S}} \sqrt{\myTover{m}{h}{s}{\mynpr{m}{h}{s}{K}}}\iota^2\\
        \leq & 2H^2\sqrt{S\mysT{K}}\iota^2.\\
        % \leq & H^2 \sum\limits_{s\in \mathcal{S}} \Big[\sum_{i=1}^{d_{max}} \sqrt{\dfrac{1}{2}} + \sum_{i=d_{max}+1}^{\mynpr{m}{h}{s}{K}} \dfrac{\sqrt{\overline{\mathcal{T}}_i}}{i-d_{max}}\Big]\iota\\
        % \leq & H^2 \sum\limits_{s\in \mathcal{S}} \Big[d_{max} + \sum_{i=1}^{\mynpr{m}{h}{s}{K}} \dfrac{\sqrt{\overline{\mathcal{T}}_{i+d_{max}}}}{i}\Big]\iota\\
        % \leq & d_{max}H^2 S\iota + H^2S\sqrt{\overline{\mathcal{T}}_K}\iota^2
    \end{split}\end{equation}
    Here the last line utilizes the definition of $\mysT{K}$.

    Finally, we bound the term $\sum\limits_{k=1}^{K} \mysbetaUnd{\mysn{k}} \cdot \mathbb{I}\Big\{\mysn{k} > d_{max}\Big\}$ as follows:
    \begin{equation}\label{eq:comp_1_4}\begin{split}
        \sum\limits_{k=1}^{K} \mysbetaUnd{\mysn{k}} \cdot \mathbb{I}\Big\{\mysn{k} > d_{max}\Big\}= & 2H^2 \sum\limits_{k=1}^{K} \sqrt{\dfrac{\iota}{\mysn{k}}}\ \mathbb{I}\Big\{\mysn{k} > d_{max}\Big\}\\
        \leq & 4H^2\sqrt{\iota} \sum\limits_{s\in \mathcal{S}} (\sqrt{\mynpr{m}{h}{s}{K}}+1)\\
        \leq & 8H^2 \sqrt{SK\iota} \\
    \end{split}\end{equation}
    Here the second inequality utilizes the analysis of Equation~\eqref{eq:comp_1_2}.

    Combining all above four terms, we have:
    \begin{equation}\begin{split}
        & \sum\limits_{k=1}^{K} (\mysbetaOver{\mysn{k}}+\mysbetaUnd{\mysn{k}})\cdot \mathbb{I}\Big\{\mysn{k} \geq d_{max}\Big\}\\
        \leq & 12d_{max}H^2S \iota^2 + 48H^2\sqrt{SAK\iota} + 8H^2\sqrt{SK\iota} + 24H^2\sqrt{S\mysT{K}}\iota^2\\
        \leq & 12d_{max}H^2S \iota^2 + 56H^2\sqrt{SAK\iota} + 24H^2\sqrt{S\mysT{K}}\iota^2\\
    \end{split}\end{equation}
\end{proof}

\section{Performance Guarantee for DA-MAVL with Reward Skipping}
% As in stated in Section~\ref{sec:sketch2}, 
% \textbf{The first three steps} follows similar routine as the proof of Theorem~\ref{thm:comp}. We consequently bound the policy optimization regret (ref. Lemma~\ref{lem:regskip}), establish the optimism and pessimism of our value estimates (ref. Lemma~\ref{lem:optskip}, Lemma~\ref{lem:pesskip}) and evaluate performance of our value estimates (ref. Lemma~\ref{lem:gap}).
% \textbf{In step four}, we show that delays of skipped visits can be neglected in the aforementioned results and utilize this fact to finish the proof of Theorem~\ref{thm:compskip}.
% In this section, we separately give proof for the Lemmas and Theorem mentioned above. 

Proof of Theorem~\ref{thm:compskip} is under assumption~\ref{ass:c} and consists of four steps. We inherit all notations from previous section, except that they refer to variables in DA-MAVL with reward skipping. 

\vspace{5pt}\noindent\textbf{STEP ONE: Bound the `Policy Optimization Regret'.} For any pair $(m,h,s,n)$, upper bound of $\myR{m}{h}{s}{n}$ is established:
\begin{mylemma}\label{lem:regskip}
    Let Assumption \ref{ass:c} holds. For $\forall (m,h,s,n)\in[M]\times[H]\times \mathcal{S}\times [K]$, the following inequality holds with probability at least $1-\delta/2$:
    \begin{equation}
      \myR{m}{h}{s}{n} \leq 20H^2C \sqrt{\dfrac{\mysTtil{n}}{n^2}}\iota + 14H^2 \sqrt{\dfrac{A}{n}}\iota.
    \end{equation}
\end{mylemma}
Lemma~\ref{lem:regskip} extends Lemma~\ref{lem:reg} to cases with infinite delays. In the proof of this lemma, we have to upper bound the regret by \textbf{the largest possible delay} and \textbf{the number of reward skips} instead of the maximum delay $d_{max}$, since the delays may be infinite. We then highlight their upper bounds, i.e. Lemma \ref{lem:regskip_dmax} and Lemma \ref{lem:regskip_omax}, which play significant roles in showing that the influence of the delays can be bounded by term $H^2C\sqrt{\myTtil{m}{h}{s}{n}/n^2}$.

\vspace{5pt}\noindent\textbf{STEP TWO: Optimism and Pessimism.} Utilizing regret $\myR{m}{h}{s}{n}$, we carefully design the bonuses (Equation~\eqref{eq:subParam_skip}) and show that value estimates in Algorithm~\ref{alg:damavl_skip} are optimistic and pessimistic:
\begin{mylemma}\label{lem:optskip}
    For  $\forall (m,h,s,k)\in [M]\times[H]\times \mathcal{S}\times[K]$, the following inequality holds with probability at least $1-\delta$:
    \begin{equation}
        \overline{V}_{m,h}^k(s) \geq V^{\dagger, \pi_{-m,h}^k}_{m,h}(s),\quad
        \underline{V}_{m,h}^k(s) \leq V^{\pi_h^k}_{m,h}(s).
    \end{equation}
\end{mylemma}
In the proof of this lemma, we separately consider the skipped and unskipped visits. Utilizing the upper bounds on the largest possible delay, we can show the bonuses can make up for the performance degradation of skipping visits and ensure optimism and pessimism.

\vspace{5pt}\noindent\textbf{STEP THREE:} Next, we bound the gap between the optimistic and pessimistic value estimates:
\begin{equation}
    \sum\limits_{k=1}^{K} (\overline{V}_{m,1}^k - \underline{V}_{m,1}^k)(s_h^k).
\end{equation}
The gap, together with Optimism and Pessimism, leads to the following bound:
\begin{mylemma}\label{lem:gap}
    Let Assumption \ref{ass:c} holds. For $\forall \delta\in (0, 1)$ and $\forall K\in \mathbb{N}$, let $\iota = \log \big( 4HSAK /\delta \big)$. Let policy $\pi$ be the output of Algorithm \ref{alg:damavl_output} after running Algorithm \ref{alg:damavl_skip} for $K$ episodes. The following equation holds with probability at least $1-\delta$:
    \begin{equation}\begin{split}
        & \sum_{k=1}^K \Big( V_{m,1}^{\dag, \pi_{-m,k}} - V_{m,1}^{\pi_k} \Big)(s_1) \lesssim CH^3\max_{h}\sum_{s\in\mathcal{S}}\sqrt{\myTtil{m}{h}{s}{\mynpr{m}{h}{s}{K}}}\iota^2 + H^3\sqrt{SAK\iota}.\\
    \end{split}\end{equation}    
\end{mylemma}

\vspace{5pt}\noindent\textbf{STEP FOUR: Bound the CCE-gap.} Finally, we give an upper bound on $\sum_{s\in\mathcal{S}}\sqrt{\myTtil{m}{h}{s}{\mynpr{m}{h}{s}{K}}}$.
\begin{mylemma}\label{lem:ignoreset}
    For $\forall (m,h,s,n) \in [M]\times[H]\times \mathcal{S}\times [K]$, the following inequality hold:
    \begin{equation}\begin{split}
        % \max\limits_{m\in[M]} \Big( V_{m,1}^{\dag, \pi_{-m}} - V_{m,1}^{\pi} \Big)(s_1) \lesssim H^3 \sqrt{SA/K}\iota + C\cdot H^3 S\cdot 
        \sqrt{\myTtil{m}{h}{s}{n}} \leq 2\min_{\mathcal{L}\in[K]} \Bigg\{ |\mathcal{L}| + \sqrt{\myTtil{m}{h}{s}{n,\mathcal{L}}}\Bigg\} + 64C^2.
    \end{split}\end{equation}
\end{mylemma}
Consequently, we can show that term $\sum_{s\in\mathcal{S}}\sqrt{\myTtil{m}{h}{s}{\mynpr{m}{h}{s}{K}}}$ can be upper bounded :
\begin{equation}\begin{split}
    \sum_{s\in\mathcal{S}}\sqrt{\myTtil{m}{h}{s}{\mynpr{m}{h}{s}{K}}} \leq& 2 \min_{\mathcal{L}\in[K]} \Bigg\{S|\mathcal{L}| + \sqrt{S\mysTtil{m,h}^{K,\mathcal{L}}}\Bigg\} + 64C^2S.\\
\end{split}\end{equation}
Intuitively, this upper bound shows that the influence of the skipped large delays is only reflected by some constant $|\mathcal{L}|$. As a direct consequence, we complete the proof of Theorem~\ref{thm:compskip}.

\subsection{\textbf{Step One:} Proof of Lemma \ref{lem:regskip}}

Consider any fixed pair $(m,h,s,n)$. Recall that $\mysO{n}$ denote that set of skipped visits of $(h,s)$ during the first $n$ visits of $(h,s)$. We first decompose the policy optimization regret $R_n$ as follows:
\begin{equation}\begin{split}
    R_n &= H \sum\limits_{i=1}^n \alpha_n^i \Big[ \mysl{i}{a_{k_i}} - \myslbar{i}{a^*}\Big]\\
    &= H \sum\limits_{i\in [n]\backslash \mysO{n}} \alpha_n^i \Big[ \mysl{i}{a_{k_i}} - \myslbar{i}{a^*}\Big] + H \sum\limits_{i\in \mysO{n}} \alpha_n^i \Big[ \mysl{i}{a_{k_i}} - \myslbar{i}{a^*}\Big]\\
    &\leq \alpha_n H|\mysO{n}| + H \sum\limits_{i\in [n]} \alpha_n^i \Big[ \mysl{i}{a_{k_i}} - \myslbar{i}{a^*}\Big] \cdot \mathbb{I}\{i\notin \mysO{n}\}.
\end{split}\end{equation}
Here the last line holds because of the monotonicity of $\{\alpha^i_n\}_{i\in[n]}$.

Consider the delays and losses defined as follows:
\begin{equation}\begin{split}
    &\mydpr{m}{h}{s}{i} = \left\{\begin{array}{lr}
        \myk{h}{s}{\min_{j} \{i\in \mysO{j}\}}- \myk{h}{s}{i} - 1, &i\in \mysO{n}\\
        \myd{m}{h}{s}{i}, &i\notin \mysO{n}
    \end{array}\right.,\\
    &\mylpr{m}{h}{s}{i}{a} = \left\{\begin{array}{lr}
        0, &i\in \mysO{n}\\
        \myl{m}{h}{s}{i}{a}, &i\notin \mysO{n}
    \end{array}\right.,\\  
    &\mylbarpr{m}{h}{s}{i}{a} = \mathbb{E} [\mylpr{m}{h}{s}{i}{a}] = \left\{\begin{array}{lr}
        0, &i\in \mysO{n}\\
        \mylbar{m}{h}{s}{i}{a}, &i\notin \mysO{n}
    \end{array}\right..
\end{split}\end{equation}
When a fixed $(m,h,s)$ is considered in the context, the above notations are abbreviated as $\mysdpr{i}$, $\myslpr{i}{a}$ and $\myslbarpr{i}{a}$. 
% \begin{equation}\begin{split}
%     &{d'}_{m,h}^i(s) = \left\{\begin{array}{lr}
%         \mathop{\arg\min}_{k\in [K]} \Big\{i\in \mysFm{k}\Big\}- k_i, &i\in \mysO{n}\\
%         \myd{m}{h}{s}{i}, &i\notin \mysO{n}
%     \end{array}\right.\\
%     &{l'}_{m,h}^i(s,a) = \left\{\begin{array}{lr}
%         0, &i\in \mysO{n}\\
%         \myl{m}{h}{s}{i}{a}, &i\notin \mysO{n}
%     \end{array}\right.,\\  
%     &{\bar{l}'}_{m,h}^i(s,a) = \mathbb{E} [{l'}_{m,h}^i(s,a)] / \hat{\pi}^{k_i}_{m,h}(a|s) = \left\{\begin{array}{lr}
%         0, &i\in \mysO{n}\\
%         \mylbar{m}{h}{s}{i}{a}, &i\notin \mysO{n}
%     \end{array}\right.
% \end{split}\end{equation}

Then the second term of $R_n$ is exactly the policy optimization regret (without reward skipping) with delays $\{\mysdpr{i}\}_{i\in[n]}$ and losses $\{\myslpr{i}{a}\}_{i\in [n], a\in \mathcal{A}_m}$ and $\{\myslbarpr{i}{a}\}_{i\in [n], a\in \mathcal{A}_m}$.
Let $d_{max}' = \max_{i\in[n]} \min\big\{{d'}_i, n-i\big\}$ be the maximum delay during the first $n$ visits of $(h,s)$. Then we directly apply results of Lemma \ref{lem:reg} and get the following with probability at least $1-\delta/2$:
\begin{equation}\begin{split}
    R_n &= \alpha_nH|\mysO{n}| + H \sum\limits_{i\in [n]} \alpha_n^i \Big[ \mysl{i}{a_{k_i}} - \myslbar{i}{a^*}\Big] \cdot \mathbb{I}\{i\notin \mysO{n}\}\\
    &\leq 2H^2 \dfrac{|\mysO{n}|}{n} + 12H^2 \sqrt{\dfrac{nA+\mysTtil{n}}{n^2}\iota} + 2H^2\dfrac{d_{max}'}{n} \iota.\\
\end{split}\end{equation}
We then upper bound $\mathcal{O}_n$ and $d'_{max}$:
\begin{mylemma}\label{lem:regskip_dmax}
    For $\forall (m,h,s,n,i)\in [M]\times[H]\times \mathcal{S}\times [K]\times[n]$, let $\mydpr{max,m}{h}{s}{n}$ denote the maximal delay of $(h,s)$ during the first $n$ visits of $(h,s)$:
    \begin{equation}
        \mydpr{max,m}{h}{s}{n} = \max_{i\in[n]} \min\big\{{d'}_{m,h}^i(s), n-i\big\}.
    \end{equation}
     the following equation holds:
    \begin{equation}\begin{split}
        &{d'}^n_{max,m,h}(s) \leq \sqrt[4]{4\myTtil{m}{h}{s}{n}}+1,\\
        &\myphi{m}{h}{s}{i}{n} \leq \sqrt{\myTtil{m}{h}{s}{n}} + \sqrt[4]{4\myTtil{m}{h}{s}{n}}+1.\\
    \end{split}\end{equation}
\end{mylemma}
\begin{mylemma}\label{lem:regskip_omax}
    For $\forall (m,h,s,n)\in [M]\times[H]\times \mathcal{S}\times [K]$, the following equation holds:
    \begin{equation}
      |\mathcal{O}_{m,h}^{n}(s)| \leq 2C\sqrt{\myTtil{m}{h}{s}{n}}.  
    \end{equation}
\end{mylemma}

Finally, with Lemma~\ref{lem:regskip_dmax} and Lemma~\ref{lem:regskip_omax}, we have:
\begin{equation}\begin{split}
    R_n &\leq 4H^2C \sqrt{\dfrac{\mysTtil{n}}{n^2}} + 12H^2 \sqrt{\dfrac{A}{n}\iota} + 12H^2 \sqrt{\dfrac{\mysTtil{n}}{n^2}\iota} + 4H^2\sqrt[4]{\dfrac{\mysTtil{n}}{n^4}}\iota + \frac{2H^2}{n}\iota\\
    &\leq 20H^2C \sqrt{\dfrac{\mysTtil{n}}{n^2}}\iota + 12H^2 \sqrt{\dfrac{A}{n}\iota} + \frac{2H^2}{n}\iota.\\
    &\leq 20H^2C \sqrt{\dfrac{\mysTtil{n}}{n^2}}\iota + 14H^2 \sqrt{\dfrac{A}{n}}\iota\\
    % &\leq 4H^2C\cdot  \dfrac{\sqrt{\mysTtil{n}}}{n}\iota + 6H^2 \sqrt{\dfrac{nA+\mysTtil{n}}{n^2}}\iota + 2H^2\dfrac{\sqrt[4]{4\mysTtil{n}}}{n} \iota + \dfrac{2H^2}{n}\iota\\
    % &\leq (10+4C)\cdot H^2\dfrac{\sqrt{\mysTtil{n}}}{n}\iota + 6H^2 \sqrt{\dfrac{A}{n}}\iota\\
\end{split}\end{equation}

\subsubsection{Supporting Details}

\begin{proof}[Proof of Lemma \ref{lem:regskip_dmax}]
    % For , it is clear that: 
    % \begin{equation}\begin{split}
    %     \myphi{m}{h}{s}{i}{n} =& \sum_{i'=i+1}^n \mathbb{I}\Big\{(i,\dots)\in\myMm{m}{h}{s}{k_{i'}}\Big\} \cdot \Big( i'-i \Big)\\
    %     \geq& \sum_{i'=i+1}^n \mathbb{I}\Big\{(i,\dots)\in\myMm{m}{h}{s}{k_{i'}}\Big\}\\
    %     =& d'^i_{m,h}
    % \end{split}\end{equation}
    Consider any fixed pair $(m,h,s,n)$. Recall $\mysphi{i}{n}$ is the skipping metric of the $i$-th visit when the $n$-th visit happens. It can be written as in Equation~\eqref{eq:skip_phidef}. Let $d'_{max}$ denote ${d'}^n_{max,m,h}(s)$. Let $i_0 = \arg\max_{i\in[n]} \min\{{d'}_i,n-i\}$ denote the visit with the largest delay. 
    
    We first bound $\mysphi{i}{n}$. Since $\mysphi{i}{n}$ grows monotonically with $\min\{{d'}_i, n-i\}$, we only need to consider $\mysphi{i_o}{n}$. 
    If $i_0\notin \mysO{n}$, namely the $i_0$-th visit of $(h,s)$ has not been skipped, then:
    \begin{equation}\label{eq:skip1_1}
        \mysphi{i_0}{n} \leq \sqrt{\mysTtil{n}}.
    \end{equation}
    On the other hand, from the definition of $\mysphi{i_0}{n}$, 
    \begin{equation}
        \mysphi{i_0}{n} = \sum_{j=1}^{d'_{max}} j \geq \frac{{d'}_{max}^2}{2}.
    \end{equation}
    This gives $d'_{max} \leq \sqrt[4]{4\mysTtil{n}}$.
    
    If $i_0 \in \mysO{n}$, namely the $i_0$-th visit of $(h,s)$ has been skipped, suppose it is skipped during the $n_0$-th visit of $(h,s)$. Then $d'_{max} = n_0-i_0$. Then:
    \begin{equation}\label{eq:skip1_3}
        \mysphi{i_0}{n_0-1} \leq \sqrt{\mysTtil{n_0-1}} \leq \sqrt{\mysTtil{n}}.
    \end{equation}
    On the other hand, from the definition of $\mysphi{i_0}{n_0-1}$, 
    \begin{equation}\label{eq:skip1_4}
        \mysphi{i_0}{n_0-1} = \sum_{j=1}^{n_0-1-i_0} j \geq \frac{(n_0-1-i_0)^2}{2} = \frac{(d'_{max}-1)^2}{2}.
    \end{equation}
    This is equivalent to $d'_{max} \leq \sqrt[4]{4\mysTtil{n}}+1$. Combining Equation~\eqref{eq:skip1_3} and \eqref{eq:skip1_4} we have:
    \begin{equation}\begin{split}
        \mysphi{i_0}{n} =& \mysphi{i_0}{n_0} = \mysphi{i_0}{n_0-1} + d'_{max}\\
        \leq& \sqrt{\mysTtil{n}} + \sqrt[4]{4\mysTtil{n}}+1.\\
    \end{split}\end{equation}
    Here the equation is because $\mysphi{i_0}{n}$ stops increasing after the $i_0$-th visit is skipped.
    
    Combining the above two cases finishes the proof.
\end{proof}

\begin{proof}[Proof of Lemma \ref{lem:regskip_omax}]
    Proof of this Lemma largely follows proof of Lemma 4 in \cite{rw_bandit2}. Consider any fixed pair $(m,h,s,n)$. Denote all visits in $\mysO{n}$ in order as $i_1$, $i_2$, \dots, $i_{|\mysO{n}|}$. And suppose they are skipped during visits $i'_1$, $i'_2$, $\dots$, $i'_{|\mathcal{O}_n|}$. Then following the skipping rule, we have:
    \begin{equation}\begin{split}
        \mysphi{i_x}{n} \geq& \sqrt{\mathcal{T}_{i'_x}} \geq \sqrt{\sum_{y=1}^{i_x} \mysphi{y}{i'_x}/C} \geq \sqrt{\sum_{y=1}^{x} \mysphi{i_y}{i'_x}/C} = \sqrt{\sum_{y=1}^{x} \mysphi{i_y}{n}/C}\\
        \geq& \dfrac{\sqrt{\mysphi{i_x}{n} + \sum_{y=1}^{x-1} \mysphi{i_y}{n}}}{\sqrt{C}}.
    \end{split}\end{equation}
    Here the second inequality is because of Lemma~\ref{lem:regskip_phi}, the first equation is because $\mysphi{i}{n}$ stops increasing after the $i$-th visit is skipped.
    
    Solve the inequality on $\mysphi{i_x}{n}$ gives:
    \begin{equation}\begin{split}
        \mysphi{i_x}{n} \geq \dfrac{1 + \sqrt{1+4C \sum_{y=1}^{x-1} \mysphi{i_y}{n}}}{2C}.
    \end{split}\end{equation}
    By induction, we can easily prove:
    \begin{equation}\begin{split}
        \mysphi{i_x}{n} \geq \dfrac{x}{2C}.
    \end{split}\end{equation}
    This directly gives:
    \begin{equation}\begin{split}
        \sum_{x=1}^{|\mysO{n}|} \mysphi{i_x}{n} \geq \dfrac{|\mysO{n}|^2}{4C}.
    \end{split}\end{equation}

    On the other hand, we have $\sum_{x=1}^{|\mysO{n}|} \mysphi{i_x}{n} \leq \sum_{x=1}^n \mysphi{x}{n} \leq C\mysTtil{n}$ from Lemma~\ref{lem:regskip_phi}. Combining the two inequalities gives the desired result.
\end{proof}

\begin{mylemma}\label{lem:regskip_phi}
    For $\forall (m,h,s,n)\in [M]\times[H]\times \mathcal{S}\times [K]$, the following equation holds:
    \begin{equation}
      \sum\limits_{i=1}^{n} \myphi{m}{h}{s}{i}{n} \leq C\myTtil{m}{h}{s}{n}.
    \end{equation}
\end{mylemma}

\begin{proof}[Proof of Lemma \ref{lem:regskip_phi}]
    Consider any fixed pair $(m,h,s,n)$. Recall that sequence $\{\mysphi{i}{n}\}_{i, n \in [K]}$ in algorithm \ref{alg:damavl_skip} can be written:
    \begin{equation}\begin{split}
        \mysphi{i}{n} = \sum_{j=i+1}^{n} (j-i) \cdot \mathbb{I}\{i\in \arg\mysM{k_j}\}.
    \end{split}\end{equation}
    Then simple algebra gives:
    \begin{equation}\begin{split}
        \sum\limits_{i=1}^{n-1} \mysphi{i}{n} &= \sum\limits_{i=1}^{n-1} \sum_{j=i+1}^{n} (j-i) \cdot \mathbb{I}\{i\in \arg\mysM{k_j}\}\\
        &= \sum\limits_{i=1}^{n-1} \sum_{j=i+1}^{n} \mathbb{I}\{i\in \arg\mysM{k_j}\} \sum_{x=i+1}^j 1\\
        &= \sum\limits_{i=1}^{n-1} \sum_{x=i+1}^{n} \sum_{j=x}^{n} \mathbb{I}\{i\in \arg\mysM{k_j}\} \\
        &= \sum\limits_{x=2}^{n} \sum_{i=1}^{x-1} \sum_{j=x}^{n} \mathbb{I}\{i\in \arg\mysM{k_j}\}.\\
    \end{split}\end{equation}
    Notice that $i\in \arg\mysM{k_j}$ implies the $i$-th visit is not usuable when the $j$-th visit happens. Since $x \geq i$, we have:
    \begin{equation}
      x \in \arg\mysM{k_j}.
    \end{equation}
    % Here $\min \big\{\arg\mysMm{k_j}\big\} = \mathop{\arg\min}_{x'} \Big\{(x', \dots) \in \mysMm{k_j}\Big\}$. 
    This observation gives:
    \begin{equation}\label{eq:regskip_2}\begin{split}
        \sum\limits_{i=1}^{n-1} \mysphi{i}{n} &= \sum\limits_{x=2}^{n} \sum_{j=x}^{n} \mathbb{I}\Big\{x \in \arg\mysM{k_j}\Big\}\cdot \Big[\sum_{i=1}^{x-1}\mathbb{I}\big\{i\in \arg\mysM{k_j}\big\} \Big].\\
    \end{split}\end{equation}
    
    On the other hand, we have:
    \begin{equation}\begin{split}
        \sum_{i=1}^{x-1}\mathbb{I}\{i\in \arg\mysM{k_j}\} \leq& \sum_{i=1}^{j-1}\mathbb{I}\{i\in \arg\mysM{k_j}\} = \Big|\{i\leq j-1: d_{i}+k_i \geq k_j\}\Big|\\
        \leq& \Big|\{i\leq j: d_{i}+i \geq j\}\Big| \leq C.\\
    \end{split}\end{equation}
    where the second inequality follows Assumption \ref{ass:c} and the fact that $j-i\leq k_j-k_i$. Substituting into Equation~\eqref{eq:regskip_2}:
    \begin{equation}\begin{split}
        \sum\limits_{i=1}^{n} \mysphi{i}{n} &\leq C\sum\limits_{x=2}^{n} \sum_{j=x}^{n} \mathbb{I}\Big\{x \in \arg\mysM{k_j}\Big\}\\
        &= C\sum\limits_{j=2}^{n} \sum_{x=2}^{j} \mathbb{I}\Big\{x \in \arg\mysM{k_j}\Big\}\\
        &= C\sum\limits_{j=2}^{n} |\mysM{k_j}|\\
        &\leq C\mysTtil{n}.
    \end{split}\end{equation}
\end{proof}

\subsection{\textbf{Step Two:} Proof of Lemma \ref{lem:optskip}}
% Proofs in this section mainly follow Lemma 13 and Lemma 14 in \cite{alg_ma_VL}.

\begin{proof}[Proof of lemma \ref{lem:optskip}]
    Consider any fixed pari $(m,h,s)$. Conditioned on the successful event of Lemma~\ref{lem:regskip}, which holds for probability at least $1-\delta/2$, we first prove the optimism part by induction. For $k=0$, it clear that for any $(m,h,s)$ :
    \begin{equation}
        \overline{V}^0_{m,h}(s) = H+1-h \geq V_{m,h}^{\dagger, \pi_{-m,h}^1}(s).
    \end{equation}
    Let $N_k = \max_m \myn{m}{h}{s}{k}$. Suppose the lemma holds for all $k' < k$. Then for episode $k$:
    \begin{equation}\begin{aligned}
        \myVOver{m}{h}{s}{k} &= \alpha^0_{\mysn{k}} \cdot H+ \sum\limits_{i=1}^{\mysn{k}} \alpha^i_{\mysn{k}} \Big( r_{m,h}^{k_i} + \myVOver{m}{h+1}{s^{k_i}_{h+1}}{k_i} \Big)\cdot \mathbb{I}\{i\notin \mathcal{O}_{\mysnpr{k}}\}\\
        &\hspace{2em} +  \sum\limits_{i=1}^{\mysn{k}} \alpha^i_{\mysn{k}} H \cdot \mathbb{I}\{i\in\mathcal{O}_{\mysnpr{k}}\} + \mybetaOver{m}{h}{s}{\mysn{k},\mysnpr{k}}\\
        &\geq \alpha^0_{\mysn{k}} \cdot H + \sum\limits_{i=1}^{\mysn{k}} \alpha^i_{\mysn{k}} \Big( r_{m,h}^{k_i} + \myVOver{m}{h+1}{s^{k_i}_{h+1}}{k_i}\Big) + \mybetaOver{m}{h}{s}{\mysn{k},\mysnpr{k}}\\
        % & = \alpha^0_{\mysn{k}} \cdot H + H \cdot \sum\limits_{i=1}^{\mysn{k}} \alpha^i_{\mysn{k}} \Big[ 1 - \mysl{i}{a_{m,h}^{k_i}} \Big] + \sum\limits_{i=1}^{\mysn{k}} \alpha^i_{\mysn{k}} \Big[\dfrac{\mybetaOver{m}{h}{s}{i}}{\alpha_i} - \dfrac{\mybetaOver{m}{h}{s}{i-1}}{\alpha_i}(1-\alpha_i)\Big]\\
        % & \geq \alpha^0_{\mysn{k}} \cdot H + H\cdot \sum\limits_{i=1}^{\mysn{k}} \alpha^i_{\mysn{k}} \Big[1 - \myslbar{i}{a^*}\Big] + \Big(\mysbetaOver{\mysn{k}} - R_{m,h}^{\mysn{k}}(s)\Big)\\
        & = \alpha^0_{\mysn{k}} \cdot H + 4H^2C\frac{\sqrt{\myTtil{m}{h}{s}{\mysnpr{k}}}}{\mysn{k}}\iota + 4H^2\sqrt{\frac{A}{\mysn{k}}}\iota\\
        &\hspace{2em}+ \sum\limits_{i=1}^{\mysn{k}} \alpha^i_{\mysn{k}} \mathop{\mathbb{E}}_{\substack{\bm{a}=(a^*,a_{-m,h})\\ a_{-m,h} \sim \hat{\pi}_{-m,h}^{k_i},\ s'\sim \mathbb{P}_h(s,\bm{a})}} \Big[r_{m,h}(s,\bm{a}) + \overline{V}^{k_i}_{m,h+1}(s')\Big].
    \end{aligned}\end{equation}
    Here the last line follows directly from proof of Lemma~\ref{lem:regskip}.
    
    On the other hand, the following holds according to Lemma~\ref{lem:opt}:
    \begin{equation}\begin{aligned}
        V^{\dagger, \pi^k_{-m,h}}_{m,h}(s) \leq& \sum_{i=1}^{\mysn{k}} \alpha_{\mysn{k}}^i \mathop{\mathbb{E}}_{\substack{\bm{a}=(a^*, a_{-m,h})\\a\sim\mu_h, a_{-m,h}\sim\hat{\pi}_{-m,h}^{k_i}}}\Big( r_{m,h}(s,\bm{a}) + \myVOver{m}{h+1}{s'}{k_i} \Big) + \frac{2(N_k-\mysn{k})H^2}{N_k}\\
        \leq& \sum_{i=1}^{\mysn{k}} \alpha_{\mysn{k}}^i \mathop{\mathbb{E}}_{\substack{\bm{a}=(a^*, a_{-m,h})\\a\sim\mu_h, a_{-m,h}\sim\hat{\pi}_{-m,h}^{k_i}}}\Big( r_{m,h}(s,\bm{a}) + \myVOver{m}{h+1}{s'}{k_i} \Big) + 2H^2\frac{\mysnpr{k}-\mysn{k}}{\mysnpr{k}}.
        % \leq& \sum_{i=1}^{\mysn{k}} \alpha_{\mysn{k}}^i \mathop{\mathbb{E}}_{\substack{\bm{a}=(a^*, a_{-m,h})\\a\sim\mu_h, a_{-m,h}\sim\hat{\pi}_{-m,h}^{k_i}}}\Big( r_{m,h}(s,\bm{a}) + \myVOver{m}{h+1}{s'}{k_i} \Big) + \frac{2d_{max}H^2}{N_k}\\
    \end{aligned}\end{equation}
    Here the last line is because $\mysnpr{k} > N_k$. From Lemma~\ref{lem:comp1_base2} and Lemma~\ref{lem:regskip_dmax}, we have:
    \begin{equation}\label{eq:nspr-ns}
        \mysnpr{k} - \mysn{k} = |\mathcal{M}_k| +1 \leq {d'}_{max,m,h}^{\mysnpr{k}}(s) +1 \leq \sqrt[4]{4\myT{m}{h}{s}{\mysnpr{k}}} + 2. 
    \end{equation}
    This leads to:
    \begin{equation}\begin{split}
        V^{\dagger, \pi^k_{-m,h}}_{m,h}(s) \leq& \sum_{i=1}^{\mysn{k}} \alpha_{\mysn{k}}^i \mathop{\mathbb{E}}_{\substack{\bm{a}=(a^*, a_{-m,h})\\a\sim\mu_h, a_{-m,h}\sim\hat{\pi}_{-m,h}^{k_i}}}\Big( r_{m,h}(s,\bm{a}) + \myVOver{m}{h+1}{s'}{k_i} \Big) + 2H^2\frac{\sqrt[4]{4\myT{m}{h}{s}{\mysnpr{k}}} + 2}{\mysnpr{k}}\\
        \leq& \sum_{i=1}^{\mysn{k}} \alpha_{\mysn{k}}^i \mathop{\mathbb{E}}_{\substack{\bm{a}=(a^*, a_{-m,h})\\a\sim\mu_h, a_{-m,h}\sim\hat{\pi}_{-m,h}^{k_i}}}\Big( r_{m,h}(s,\bm{a}) + \myVOver{m}{h+1}{s'}{k_i} \Big) + 4H^2\frac{\sqrt{\myT{m}{h}{s}{\mysnpr{k}}}}{\mysnpr{k}} + 4H^2 \frac{1}{\mysnpr{k}}\\
        \leq& \sum_{i=1}^{\mysn{k}} \alpha_{\mysn{k}}^i \mathop{\mathbb{E}}_{\substack{\bm{a}=(a^*, a_{-m,h})\\a\sim\mu_h, a_{-m,h}\sim\hat{\pi}_{-m,h}^{k_i}}}\Big( r_{m,h}(s,\bm{a}) + \myVOver{m}{h+1}{s'}{k_i} \Big) + 4H^2\frac{\sqrt{\myT{m}{h}{s}{\mysnpr{k}}}}{\mysn{k}} + 4H^2 \sqrt{\frac{A}{\mysnpr{k}}}\iota.
    \end{split}\end{equation}
    which directly leads to $\myVOver{m}{h}{s}{k} \geq V^{\dagger, \pi^k_{-m,h}}_{m,h}(s)$ and finishes the induction.
    % When $\mysn{k}\leq d_{max}$, we have $(\mysbetaOver{\mysn{k}} - R_{m,h}^{\mysn{k}}(s)) \geq 2H^2 \geq \frac{2(N_k-\mysn{k})H^2}{N_k}$. 
    % When $\mysn{k} > d_{max}$, we have $(\mysbetaOver{\mysn{k}} - R_{m,h}^{\mysn{k}}(s)) \geq \frac{2d_{max}H^2}{\mysn{k}} \geq \frac{2(N_k-\mysn{k})H^2}{N_k}$.
    % Both cases lead to $\myVOver{m}{h}{s}{k} > V^{\dagger, \pi^k_{-m,h}}_{m,h}(s)$ and complete the proof.

    We now prove the pessimism part of the lemma. For $\forall (m,h,s,k) \in [M]\times[H]\times \mathcal{S}\times [K]$, the following hold with probability at least $1-\delta/(2MHSK)$:
    \begin{equation}\begin{split}
        \underline{V}^k_{m,h}(s) &= \sum\limits_{i=1}^{\mysn{k}} \alpha^i_{\mysn{k}} \Big( r_{m,h}^{k_i} + \myVUnd{m}{h+1}{s^{k_i}_{h+1}}{k_i}\Big) \cdot \mathbb{I}\{i\notin \mathcal{O}_{\mysnpr{k}}\} - \mysbetaUnd{\mysn{k}, \mysnpr{k}}\\
        &\leq \sum\limits_{i=1}^{\mysn{k}} \alpha^i_{\mysn{k}} \Big( r_{m,h}^{k_i} + \myVUnd{m}{h+1}{s^{k_i}_{h+1}}{k_i}\Big) - \mysbetaUnd{\mysn{k}, \mysnpr{k}}\\
        & \leq \sum\limits_{i=1}^{\mysn{k}} \alpha^i_{\mysn{k}} \mathop{\mathbb{E}}_{\substack{\bm{a}\sim \hat{\pi}^{k_i}_h, s'\sim \mathbb{P}_h(\cdot|s,\bm{a})}} \Big( r_{m,h}(s,\bm{a}) + \underline{V}^{k_i}_{m,h+1}(s')\Big)\\
        &\hspace{2em}+ 2H^2 \frac{\mysnpr{k}-\mysn{k}}{\mysn{k}} + 2\sqrt{\frac{H^3}{\mysn{k}}\iota}  - \mysbetaUnd{\mysn{k}, \mysnpr{k}} \\
        & \leq \sum\limits_{i=1}^{\mysn{k}} \alpha^i_{\mysn{k}} \mathop{\mathbb{E}}_{\substack{\bm{a}\sim \hat{\pi}^{k_i}_h, s'\sim \mathbb{P}_h(\cdot|s,\bm{a})}} \Big( r_{m,h}(s,\bm{a}) + \underline{V}^{k_i}_{m,h+1}(s')\Big)\\
        &\hspace{2em}+ 2H^2 \frac{\sqrt[4]{4\myT{m}{h}{s}{\mysnpr{k}}}+2}{\mysn{k}} + 2\sqrt{\frac{H^3}{\mysn{k}}\iota}  - \mysbetaUnd{\mysn{k}, \mysnpr{k}} \\
        & = V^{\pi^k_h}_{m,h}(s).
    \end{split}\end{equation}
    Here the third line follows directly from the proof of Lemma~\ref{lem:opt}, while the last line is from the definition of $\mysbetaUnd{\mysn{k}, \mysnpr{k}}$ and Equation~\eqref{eq:nspr-ns}. Finally by taking the union bound over all $(m,h,s,k) \in [M]\times[H]\times \mathcal{S}\times [K]$, we finish the proof.
\end{proof}

\subsection{\textbf{Step Three:} Proof of Lemma \ref{lem:gap}}
Consider any fixed pair $(m,h)$. For episode $k$, we slightly overload the notations and let $\mysn{k} = \myn{m}{h}{s^k_h}{k}$, $\mysn{k} = \mynpr{m}{h}{s^k_h}{k}$, $k_i = \myk{h}{s^k_h}{i}$, $d'_{max}(s) = \mydpr{max,m}{h}{s}{\mynpr{m}{h}{s}{K}} \leq \sqrt[4]{4\myT{m}{h}{s}{\mynpr{m}{h}{s}{K}}}+1$. Then:
\begin{equation}\label{eq:gap_2}\begin{aligned}
    &\sum\limits_{k=1}^{K} (\overline{V}_{m,h}^k - \underline{V}_{m,h}^k)(s_h^k) \\
    =& \sum\limits_{k=1}^{K} (\overline{V}_{m,h}^k - \underline{V}_{m,h}^k)(s_h^k) \cdot \mathbb{I}\Big\{\mysn{k} < d'_{max}(s^k_h)\Big\} + \sum\limits_{k=1}^{K} (\overline{V}_{m,h}^k - \underline{V}_{m,h}^k)(s_h^k) \cdot \mathbb{I}\Big\{\mysn{k} \geq d'_{max}(s^k_h)\Big\} \\
    \leq & H\sum\limits_{k=1}^K \mathbb{I}\Big\{\mysn{k} < d'_{max}(s^k_h)\Big\} + \sum\limits_{k=1}^{K} (\overline{V}_{m,h}^k - \underline{V}_{m,h}^k)(s_h^k) \cdot \mathbb{I}\Big\{\mysn{k} \geq d'_{max}(s^k_h)\Big\}.\\
        % &\hspace{2em}+ \sum\limits_{k=1}^{K} (\mybetaOver{m}{h}{s^k_h}{\mysn{k}}+\sum\limits_{i=1}^{\mysn{k}}  \alpha^i_{\mysn{k}} \mysbetaUnd{i})\cdot \mathbb{I}\Big\{\mysn{k} \geq d'_{max}(s^k_h)\Big\}\\
\end{aligned}\end{equation}
To bound the second term, notice that:
\begin{equation}
    \begin{aligned}
    &(\overline{V}_{m,h}^k - \underline{V}_{m,h}^k)(s_h^k) \\
    \leq& \alpha^0_{\mysn{k}} H + \sum\limits_{i=1}^{\mysn{k}} \alpha^i_{\mysn{k}} H \cdot \mathbb{I}\{i\in\mathcal{O}_{\mysnpr{k}}\} + \sum\limits_{i=1}^{\mysn{k}} \alpha^i_{\mysn{k}} \Big( r_{m,h}^{k_i} + \myVOver{m}{h+1}{s^{k_i}_{h+1}}{k_i} \Big)\cdot \mathbb{I}\{i\notin \mathcal{O}_{\mysnpr{k}}\} + \mybetaOver{m}{h}{s^k_h}{\mysn{k},\mysnpr{k}}\\
    &\hspace{2em}- \sum\limits_{i=1}^{\mysn{k}} \alpha^i_{\mysn{k}} \Big( r_{m,h}^{k_i} + \myVUnd{m}{h+1}{s^{k_i}_{h+1}}{k_i}\Big) \cdot \mathbb{I}\{i\notin \mathcal{O}_{\mysnpr{k}}\} + \mybetaUnd{m}{h}{s^k_h}{\mysn{k}, \mysnpr{k}}\\
    \leq& \alpha^0_{\mysn{k}} H + \mybetaOver{m}{h}{s^k_h}{\mysn{k},\mysnpr{k}} + \mybetaUnd{m}{h}{s^k_h}{\mysn{k}, \mysnpr{k}} + \sum\limits_{i=1}^{\mysn{k}} \alpha^i_{\mysn{k}} H \cdot \mathbb{I}\{i\in\mathcal{O}_{\mysnpr{k}}\}\\
    &\hspace{2em} + \sum\limits_{i=1}^{\mysn{k}} \alpha^i_{\mysn{k}} \Big( \myVOver{m}{h+1}{s^{k_i}_{h+1}}{k_i} - \myVUnd{m}{h+1}{s^{k_i}_{h+1}}{k_i}\Big)\cdot \mathbb{I}\{i\notin \mathcal{O}_{\mysnpr{k}}\}\\
    \leq& \alpha^0_{\mysn{k}} H + \mybetaOver{m}{h}{s^k_h}{\mysn{k},\mysnpr{k}} + \mybetaUnd{m}{h}{s^k_h}{\mysn{k}, \mysnpr{k}} + 2\alpha_{\mysn{k}}CH\sqrt{\myT{m}{h}{s^k_h}{\mysnpr{k}}}\\
    &\hspace{2em}+ \sum\limits_{i=1}^{\mysn{k}} \alpha^i_{\mysn{k}}  \Big(\myVOver{m}{h+1}{s^{k_i}_{h+1}}{k_i} - \myVUnd{m}{h+1}{s^{k_i}_{h+1}}{k_i}\Big).\\
    \end{aligned}
\end{equation}
Here the last line follows Lemma~\ref{lem:regskip_omax}. Substituting into Equation~\eqref{eq:gap_2}, we get:
\begin{equation}\begin{aligned}
    & \sum\limits_{k=1}^{K} (\overline{V}_{m,h}^k - \underline{V}_{m,h}^k)(s_h^k) \\
    % =& \sum\limits_{k=1}^{K} (\overline{V}_{m,h}^k - \underline{V}_{m,h}^k)(s_h^k) \cdot \mathbb{I}\Big\{\mysn{k} < d'_{max}(s^k_h)\Big\} + \sum\limits_{k=1}^{K} (\overline{V}_{m,h}^k - \underline{V}_{m,h}^k)(s_h^k) \cdot \mathbb{I}\Big\{\mysn{k} \geq d'_{max}(s^k_h)\Big\} \\
    \leq & \underbrace{H\sum\limits_{k=1}^K \mathbb{I}\Big\{\mysn{k} < d'_{max}(s^k_h)\Big\}}_{I_{3,1}} + \underbrace{\sum\limits_{k=1}^{K} \sum\limits_{i=1}^{\mysn{k}} \alpha^i_{\mysn{k}}(\overline{V}_{m,h+1}^{k_i} - \underline{V}_{m,h+1}^{k_i})(s_{h+1}^{k_i})}_{I_{3,2}}\\
    &\hspace{2em}+ \underbrace{\sum\limits_{k=1}^{K} (\mybetaOver{m}{h}{s^k_h}{\mysn{k},\mysnpr{k}} + \mybetaUnd{m}{h}{s^k_h}{\mysn{k}, \mysnpr{k}} + 4CH^2\frac{\sqrt{\myT{m}{h}{s^k_h}{\mysnpr{k}}}}{\mysn{k}})\cdot \mathbb{I}\Big\{\mysn{k} \geq d'_{max}(s^k_h)\Big\}}_{I_{3,3}}.\\
\end{aligned}\end{equation}

The three terms are bounded by the following Lemmas:
\begin{mylemma}\label{lem:gap_1}
    For any fixed pair $(m,h)\in [M]\times [H]$, 
    \begin{equation}\begin{split}
    I_{3,1} \leq 4H \sum_{s\in\mathcal{S}} \sqrt[4]{\myTtil{m}{h}{s}{\mynpr{m}{h}{s}{K}}} + 2HS.
\end{split}\end{equation}
\end{mylemma}

\begin{mylemma}\label{lem:gap_2}
    For any fixed pair $(m,h)\in [M]\times [H]$, 
    \begin{equation}\begin{split}
        I_{3,2} \leq (1+\frac{1}{H}) \sum_{k=1}^K \Big(\overline{V}_{m,h+1}^{k}-\underline{V}_{m,h+1}^{k}\Big)(s^k_{h+1}) + 8H^2\sum_{s\in\mathcal{S}}\sqrt{\myT{m}{h}{s}{\mynpr{m}{h}{s}{K}}}\iota+4H^2S\iota.
    \end{split}\end{equation}
\end{mylemma}

\begin{mylemma}\label{lem:gap_3}
    For any fixed pair $(m,h)\in [M]\times [H]$, 
    \begin{equation}\begin{split}
        I_{3,3} \leq 96CH^2\sum\limits_{s\in \mathcal{S}} \sqrt{\myTtil{m}{h}{s}{\mynpr{m}{h}{s}{K}}}\iota^2 + 96H^2\sqrt{SAK\iota}.\\  
    \end{split}\end{equation}
\end{mylemma}
This Lemma leads to:
\begin{equation}\begin{split}
    & \sum\limits_{k=1}^{K} (\overline{V}_{m,h}^k - \underline{V}_{m,h}^k)(s_h^k) \\
    \leq & (1+\dfrac{1}{H}) \sum\limits_{k=1}^{K} \Big( \overline{V}_{m,h+1}^k-\underline{V}_{m,h+1}^k \Big)(s^k_{h+1})\\
    & \hspace{2em} + 108CH^2\sum\limits_{s\in \mathcal{S}} \sqrt{\myTtil{m}{h}{s}{\mynpr{m}{h}{s}{K}}}\iota^2 + 96H^2\sqrt{SAK\iota} + 6H^2S\iota\\ 
\end{split}\end{equation}
Iterating over $h$, we get:
\begin{equation}\begin{split}
    & \sum\limits_{k=1}^{K} (\overline{V}^k_{m,1} - \underline{V}^k_{m,1})(s_1) \\
    \leq & 108CH^3\max_h\sum\limits_{s\in \mathcal{S}} \sqrt{\myTtil{m}{h}{s}{\mynpr{m}{h}{s}{K}}}\iota^2 + 96H^3\sqrt{SAK\iota} + 6H^3S\iota\\ 
    \lesssim & CH^3\max_h\sum\limits_{s\in \mathcal{S}} \sqrt{\myTtil{m}{h}{s}{\mynpr{m}{h}{s}{K}}}\iota^2 + H^3\sqrt{SAK\iota}\\
    % \lesssim & C\cdot H^3 S\sqrt{\tilde{\mathcal{T}}_{K}}\iota^2 + H^3S\sqrt{A} \sqrt[4]{\tilde{\mathcal{T}}_{K}}\iota + H^3\sqrt{SAK}\iota\\
    % \leq & 32CH^3S \max_{s, h}\min_{\mathcal{L}\in [n]} \Bigg\{ |\mathcal{L}| + \sqrt{\overline{\mathcal{T}}_{m,h}^{\mynpr{m}{h}{s}{K},\mathcal{L}}(s)} \Bigg\}\iota + 16H^3\sqrt{SAK}\iota + 8H^3S \iota\\
\end{split}\end{equation}
Finally, conditioned on the successful event of Lemma~\ref{lem:optskip}, which has probability at least $1-\delta$, the Lemma~\ref{lem:gap} holds.

\subsubsection{Supporting Details}

\begin{proof}[Proof of Lemma \ref{lem:gap_1}]
    Consider any fixed pair $(m,h)$. We inherit the definition of $X_n(s)$ from the proof of Lemma~\ref{lem:comp_1} to denote all episodes where $(h,s)$ is visited and the $n$-th visit of $(h,s)$ is received or been skipped. We then have:
    \begin{equation}
        \begin{aligned}
        I_{3,1} =& H\sum\limits_{k=1}^K \mathbb{I}\Big\{\mysn{k} < d'_{max}(s^k_h)\Big\} = H \sum_{s\in\mathcal{S}} \sum_{x\in X_{0}(s)} \mathbb{I}\Big\{\mysn{x} < d'_{max}(s)\Big\}\\
        \end{aligned}
    \end{equation}
    Notice that the $d_{max}'(s)$-th visit will be received or skipped when the $(2d_{max}'(s)+1)$-th visit happens. So we must have for every $s\in \mathcal{S}$:
    \begin{equation}
        \sum_{x\in X_{0}(s)} \mathbb{I}\Big\{\mysn{x} < d'_{max}(s)\Big\} \leq 2d'_{max}(s)
    \end{equation}
    Consequently, 
    \begin{equation}
        I_{3,1} \leq H \sum_{s\in\mathcal{S}} 2d'_{max}(s) \leq 4H \sum_{s\in\mathcal{S}} \sqrt[4]{\myTtil{m}{h}{s}{\mynpr{m}{h}{s}{K}}} + 2HS
    \end{equation}
    The last line follows Lemma~\ref{lem:regskip_dmax}.
\end{proof}

\begin{proof}[Proof of Lemma \ref{lem:gap_2}]
    Consider any fixed pair $(m,h)$. We have:
    \begin{equation}\begin{split}
        I_{3,2} =& \sum\limits_{k=1}^{K} \sum\limits_{i=1}^{\mysn{k}} \alpha^i_{\mysn{k}}(\overline{V}_{m,h+1}^{k_i} - \underline{V}_{m,h+1}^{k_i})(s_{h+1}^{k_i})\\
        \leq& \sum_{k=1}^K \Big(\overline{V}_{m,h+1}^{k}-\underline{V}_{m,h+1}^{k}\Big)(s^k_{h+1}) \Big[d'_{max}(s^k_h)\alpha_{\mysnpr{k}}+(1+\frac{1}{H})\Big]\\
        \leq& (1+\frac{1}{H}) \sum_{k=1}^K \Big(\overline{V}_{m,h+1}^{k}-\underline{V}_{m,h+1}^{k}\Big)(s^k_{h+1}) + 2H^2\sum_{k=1}^K \frac{\sqrt[4]{4\myT{m}{h}{s^k_h}{\mysnpr{K}}}+1}{\mysnpr{k}}\\
        \leq& (1+\frac{1}{H}) \sum_{k=1}^K \Big(\overline{V}_{m,h+1}^{k}-\underline{V}_{m,h+1}^{k}\Big)(s^k_{h+1}) + 2H^2\sum_{s\in\mathcal{S}}\sum_{i=1}^{\mynpr{m}{h}{s}{K}} \frac{\sqrt[4]{4\myT{m}{h}{s}{\mynpr{m}{h}{s}{K}}}+1}{i}\\
        \leq& (1+\frac{1}{H}) \sum_{k=1}^K \Big(\overline{V}_{m,h+1}^{k}-\underline{V}_{m,h+1}^{k}\Big)(s^k_{h+1}) + 2H^2\sum_{s\in\mathcal{S}}\Big(\sqrt[4]{4\myT{m}{h}{s}{\mynpr{m}{h}{s}{K}}}+1\Big)\sum_{i=1}^{\mynpr{m}{h}{s}{K}} \frac{1}{i}\\
        \leq& (1+\frac{1}{H}) \sum_{k=1}^K \Big(\overline{V}_{m,h+1}^{k}-\underline{V}_{m,h+1}^{k}\Big)(s^k_{h+1}) + 4H^2\sum_{s\in\mathcal{S}}\Big(\sqrt[4]{4\myT{m}{h}{s}{\mynpr{m}{h}{s}{K}}}+1\Big)\iota\\
        \leq& (1+\frac{1}{H}) \sum_{k=1}^K \Big(\overline{V}_{m,h+1}^{k}-\underline{V}_{m,h+1}^{k}\Big)(s^k_{h+1}) + 8H^2\sum_{s\in\mathcal{S}}\sqrt{\myT{m}{h}{s}{\mynpr{m}{h}{s}{K}}}\iota+4H^2S\iota.\\
    \end{split}\end{equation}
    Here the second line follows directly from proof of Lemma~\ref{lem:gap_2}, the third line follows the definition of $d_{max}'(s^k_h)$.
\end{proof}

\begin{proof}[Proof of Lemma \ref{lem:gap_3}]
    Consider any fixed pair of $(m,h)\in [M]\times[H]$. We inherit the definition of $X_n(s)$ from the proof of Lemma~\ref{lem:comp_1} to denote all episodes where $(h,s)$ is visited and the $n$-th visit of $(h,s)$ is received or been skipped. We have:
    
    \begin{equation}\label{eq:gap_3}\begin{split}
        &\sum\limits_{k=1}^{K} \Big(\mybetaOver{m}{h}{s^k_h}{\mysn{k},\mysnpr{k}}+ \mybetaUnd{m}{h}{s^k_h}{\mysn{k}, \mysnpr{k}}+4CH^2\frac{\sqrt{\myT{m}{h}{s^k_h}{\mysnpr{k}}}}{\mysn{k}}\Big) \cdot \mathbb{I}\Big\{\mysn{k} \geq d'_{max}(s^k_h)\Big\}\\
        \leq& 32H^2C \sum_{k=1}^K\sqrt{\dfrac{\myTtil{m}{h}{s^k_h}{\mysnpr{k}}}{\mysn{k}^2}}\iota\cdot \mathbb{I}\Big\{\mysn{k} \geq d'_{max}(s^k_h)\Big\} + 24H^2 \sum_{k=1}^K \sqrt{\dfrac{A}{\mysn{k}}}\iota\cdot \mathbb{I}\Big\{\mysn{k} \geq d'_{max}(s^k_h)\Big\}\\
    \end{split}\end{equation}
    
    For the first term, we have:
    \begin{equation}\begin{split}
        &H^2C \sum_{k=1}^K\sqrt{\dfrac{\myTtil{m}{h}{s^k_h}{\mysnpr{k}}}{\mysn{k}^2}}\iota\cdot \mathbb{I}\Big\{\mysn{k} \geq d'_{max}(s^k_h)\Big\}\\
        \leq & H^2C\iota\sum\limits_{s\in \mathcal{S}} \sqrt{\myTtil{m}{h}{s}{\mynpr{m}{h}{s}{K}}} \cdot \Big[\sum_{x\in X_{d'_{max}(s)}(s)} \frac{1}{\mysn{x}}\Big]\\
        \leq & 3H^2C\sum\limits_{s\in \mathcal{S}} \sqrt{\myTtil{m}{h}{s}{\mynpr{m}{h}{s}{K}}}\iota^2\\
        % \leq & H^2CS\max_{s\in\mathcal{S}}\sqrt{\myTtil{m}{h}{s}{\mynpr{m}{h}{s}{K}}}\iota^2
    \end{split}\end{equation}
    Here the last line is from results in Equation~\eqref{eq:comp_1_3}.

    Then for the second, 
    \begin{equation}\begin{split}
        &H^2 \sum_{k=1}^K \sqrt{\dfrac{A}{\mysn{k}}\iota}\cdot \mathbb{I}\Big\{\mysn{k} \geq d'_{max}(s^k_h)\Big\}\\
        \leq& H^2\sum_{s\in\mathcal{S}} \sqrt{A\iota}\sum_{x\in X_{d'_{max}(s)}(s)} \sqrt{\frac{1}{\mysn{x}}}\\
        \leq& 2H^2\sum_{s\in\mathcal{S}} \sqrt{A\mynpr{m}{h}{s}{K}\iota} + 2H^2S\sqrt{A\iota}\\
        \leq & 4H^2\sqrt{SAK\iota}
    \end{split}\end{equation}
    where the third line follows Equation~\eqref{eq:comp_1_4}.

    Substituting the above results into Equation~\eqref{eq:gap_3}, we have:
    \begin{equation}\begin{split}
        &\sum\limits_{k=1}^{K} \Big(\mybetaOver{m}{h}{s^k_h}{\mysn{k},\mysnpr{k}}+ \mybetaUnd{m}{h}{s^k_h}{\mysn{k}, \mysnpr{k}}+4CH^2\frac{\sqrt{\myT{m}{h}{s^k_h}{\mysnpr{k}}}}{\mysn{k}}\Big) \cdot \mathbb{I}\Big\{\mysn{k} \geq d'_{max}(s^k_h)\Big\}\\
        \leq & 96CH^2\sum\limits_{s\in \mathcal{S}} \sqrt{\myTtil{m}{h}{s}{\mynpr{m}{h}{s}{K}}}\iota^2 + 96H^2\sqrt{SAK\iota}\\        
    \end{split}\end{equation}
\end{proof}

\subsection{\textbf{Step Four:} Proof of Theorem~\ref{thm:compskip}} 
From Lemma~\ref{lem:ignoreset}, we can upper bound $\max_{m,h}\sum_{s\in\mathcal{S}}\sqrt{\myTtil{m}{h}{s}{\mynpr{m}{h}{s}{K}}}$ as follows:
\begin{equation}\begin{split}
    \max_{m,h}\sum_{s\in\mathcal{S}}\sqrt{\myTtil{m}{h}{s}{\mynpr{m}{h}{s}{K}}} \leq& 2 \max_{m,h}\sum_{s\in\mathcal{S}} \min_{\mathcal{L}} \Bigg\{ |\mathcal{L}| + \sqrt{\myTtil{m}{h}{s}{\mynpr{m}{h}{s}{K},\mathcal{L}}}\Bigg\} + 64C^2S\\
    \leq& 2 \max_{m,h}\min_{\mathcal{L}} \Bigg\{S|\mathcal{L}| + \sum_{s\in\mathcal{S}}\sqrt{\myTtil{m}{h}{s}{\mynpr{m}{h}{s}{K},\mathcal{L}}}\Bigg\} + 64C^2S\\
    =& 2 \max_{m,h}\min_{\mathcal{L}} \Bigg\{S|\mathcal{L}| + \sqrt{S\mysTtil{m,h}^{K,\mathcal{L}}}\Bigg\} + 64C^2S,\\
\end{split}\end{equation}
where $\mathcal{L} \subset [K]$. Then from Lemma \ref{lem:gap}, we have:
\begin{equation}\begin{split}
    & \max_m \Big( V_{m,1}^{\dag, \pi_{-m}} - V_{m,1}^{\pi} \Big)(s_1) = \max_m \dfrac{1}{K} \sum_{k=1}^K \Big( V_{m,1}^{\dag, \pi_{-m,k}} - V_{m,1}^{\pi_k} \Big)(s_1)\\
    \lesssim & CH^3 \max_{m,h}\sum_{s\in\mathcal{S}}\sqrt{\frac{\myTtil{m}{h}{s}{\mynpr{m}{h}{s}{K}}}{K^2}}\iota^2 + H^3\sqrt{\frac{SA}{K}\iota}\\
    \lesssim & CH^3 \max_{m,h}\min_{\mathcal{L}} \Bigg\{S|\mathcal{L}| + \sqrt{S\mysTtil{m,h}^{K,\mathcal{L}}}\Bigg\}\iota^2 + H^3\sqrt{\frac{SA}{K}\iota}\\
\end{split}\end{equation}

\subsection{Supporting Details}
\begin{proof}[Proof of Lemma~\ref{lem:ignoreset}]
Consider any fixed pair $(m,h,s)$. When $\mysTtil{n} \leq (8C)^4$, the desired result trivially holds. We now consider $\mysTtil{n} \geq (8C)^4$. Then for any set $\mathcal{L}\subset [K]$, we have:
\begin{equation}
    \mysTtil{n} = \sum\limits_{i=1}^n i - \myse{i} = \sum\limits_{i=1}^n (i - \myse{i})\cdot \mathbb{I}\Big\{e_i\notin\mathcal{L}\Big\} + \sum\limits_{i=1}^n (i - \myse{i})\cdot \mathbb{I}\Big\{e_i\in\mathcal{L}\Big\}
\end{equation}

For the first term, we have:
\begin{equation}
    \sum\limits_{i=1}^n (i - \myse{i})\cdot \mathbb{I}\Big\{e_i\notin\mathcal{L}\Big\} \leq \sum\limits_{i=1}^n (i - \mathop{\arg\min}_{j\notin\mathcal{L}} \Big[\myd{m}{h}{s}{j}+k_j > k_i-1\Big]) = \mysT{n,\mathcal{L}}
\end{equation}
To see why this holds, consider the following cases. If $e_i \notin \mathcal{L}$, then:
\begin{equation}\begin{split}
    &(i - \myse{i})\cdot \mathbb{I}\Big\{e_i\notin\mathcal{L}\Big\} = i - \myse{i}\\
    =&i - \mathop{\arg\min}_{j} \Big[\myd{m}{h}{s}{j}+k_j > k_i-1\Big]\\
    =&i - \mathop{\arg\min}_{j\notin\mathcal{L}} \Big[\myd{m}{h}{s}{j}+k_j > k_i-1\Big]
\end{split}\end{equation}
However, if $e_i\in \mathcal{L}$, then 
\begin{equation}
    (i - \myse{i})\cdot \mathbb{I}\Big\{e_i\notin\mathcal{L}\Big\} = 0 \leq i - \mathop{\arg\min}_{j\notin\mathcal{L}} \Big[\myd{m}{h}{s}{j}+k_j > k_i-1\Big]
\end{equation}
Combining the two cases gives the desired result.

Now for the second term, 
\begin{equation}\begin{split}
    & \sum\limits_{i=1}^n (i - \myse{i})\cdot \mathbb{I}\Big\{e_i\in\mathcal{L}\Big\}\\
    ={}& \sum\limits_{i=1}^n \sum_{j\in\mathcal{L}} (i - j)\cdot \mathbb{I}\Big\{j = e_i\Big\}\\
    ={}& \sum_{j\in\mathcal{L}} \sum_{i=j}^n (i-j) \cdot\mathbb{I} \Big\{j = \myse{i} \Big\}\\
    \leq{} & \sum_{j\in\mathcal{L}} \sum_{i=j}^n (i-j)\cdot \mathbb{I} \Big\{j \in \arg\mysM{k_i} \Big\}\\
    ={} & \sum_{j\in\mathcal{L}} \mysphi{j}{n}.\\
\end{split}\end{equation}
Here the last line follows Equation~\eqref{eq:skip_phidef}. For any visit $j\in \mathcal{O}_{\mysnpr{K}}$, from Lemma~\ref{lem:regskip_dmax}, we have: 
\begin{equation}\begin{split}
    \mysphi{j}{n} \leq& \sqrt{\mysTtil{n}} + \sqrt[4]{4\mysTtil{n}}+1\\
\end{split}\end{equation}
For any visit $j\notin \mathcal{O}_{\mysnpr{K}}$, we have \(\mysphi{j}{n} \leq  \sqrt{\mysTtil{n}}\). Combining the above two cases, we have:
\begin{equation}\begin{split}
    & \sum\limits_{i=1}^n (i - \myse{i})\cdot \mathbb{I}\Big\{e_i\in\mathcal{L}\Big\}\\
    \leq & |\mathcal{L}| \sqrt{\mysTtil{n}} + |\mathcal{O}_{K}| \sqrt[4]{4\mysTtil{n}} + |\mathcal{O}_K|\\
    \leq & |\mathcal{L}| \sqrt{\mysTtil{n}} + 4C (\mysTtil{n})^{3/4},\\
\end{split}\end{equation}
where the last line is due to Lemma~\ref{lem:regskip_omax}.

Finally, combining the first and second term gives:
\begin{equation}\begin{split}
    \mysTtil{n} &\leq \mysT{n,\mathcal{L}} + |\mathcal{L}| \sqrt{\mysTtil{n}} + 4C (\mysTtil{n})^{3/4}\\
\end{split}\end{equation}
Since $\mysTtil{n} \geq (8C)^4$, we have $\mysTtil{n} - 4C(\mysTtil{n})^{3/4} \geq \frac{1}{2} \mysTtil{n}$. This implies:
\begin{equation}\begin{split}
    \mysT{n,\mathcal{L}} \geq \frac{1}{2} \mysTtil{n} - |\mathcal{L}| \sqrt{\mysTtil{n}}
\end{split}\end{equation}
which is equivalent to:
\begin{equation}\begin{split}
    \min_{\mathcal{L}\in [n]} \Bigg\{ |\mathcal{L}| + \sqrt{\mysT{n,\mathcal{L}}} \Bigg\} \geq  \min_{l} \Bigg\{ l + \sqrt{\frac{1}{2} \mysTtil{n} - l \sqrt{\mysTtil{n}}} \Bigg\}
\end{split}\end{equation}
Since the right hand side is concave in $l$, its maximum is reached when $l = \frac{1}{2} \sqrt{\mysTtil{n}}$. So we have:
\begin{equation}\begin{split}
    \frac{1}2{ \sqrt[]{\mysTtil{n}}} \leq \min_{\mathcal{L}\in [n]} \Bigg\{ |\mathcal{L}| + \sqrt{\mysT{n,\mathcal{L}}} \Bigg\}
    % \leq \min_{\mathcal{L}\in [n]} \Bigg\{ |\mathcal{L}| + \sqrt{\overline{\mathcal{T}}_{K}^{\mathcal{L}}} \Bigg\}
\end{split}\end{equation}
\end{proof}

% Recall $\tau_K = \max_{m\in[M], h\in[H], s\in \mathcal{S}} \tau_{m,h,s,K}$ is the largest delay for the whole algorithm. Since $\tilde{\mathcal{T}}_K \leq K \cdot \tau_{K}$, from Lemma \ref{trunc_param_dmax}, we have:
% \begin{equation}\begin{split}
%     \tilde{\mathcal{T}}_{K} \leq 2K \cdot \sqrt[4]{\tilde{\mathcal{T}}_{K}}
% \end{split}\end{equation}
% This implies that $\tilde{\mathcal{T}}_K \leq (2K)^{4/3}$. Consequently, when $K \geq C^6S^3\iota^8 \geq S^3$, we have: 
% \begin{equation}\begin{split}
%     H^3S\sqrt{A} \sqrt[4]{\tilde{\mathcal{T}}_{K}} / K\iota \lesssim H^3S\sqrt{A} K^{-2/3} \iota \leq H^3\sqrt{SA/K}\iota
% \end{split}\end{equation}
% So we have:
% \begin{equation}\begin{split}
%     & \Big( V_{m,1}^{\dag, \pi_{-m}} - V_{m,1}^{\pi} \Big)(s_1) = \dfrac{1}{K} \sum_{k=1}^K \Big( V_{m,1}^{\dag, \pi_{-m,k}} - V_{m,1}^{\pi_k} \Big)(s_1)\\
%     \lesssim & H^3\sqrt{SA/K}\iota + C\cdot H^3 S\iota^2 \min_{\mathcal{L}\in[K]} \Bigg\{ |\mathcal{L}|/K + \sqrt{\overline{\mathcal{T}}_{K}^{\mathcal{L}}/K^2}\Bigg\} + 64C^3\cdot H^3S\iota^4 / K\\    
%     \lesssim & H^3\sqrt{SA/K}\iota + C\cdot H^3 S\iota^2 \min_{\mathcal{L}\in[K]} \Bigg\{ |\mathcal{L}|/K + \sqrt{\overline{\mathcal{T}}_{K}^{\mathcal{L}}/K^2}\Bigg\}\\    
% \end{split}\end{equation}
% The last line is because $K \geq C^6S^3\iota^8 \geq C^6S\iota^8$.

\end{document}